\documentclass[twocolumn,showpacs,preprintnumbers,amsmath,amssymb]{revtex4-2}

\usepackage[colorlinks=true,citecolor=blue,urlcolor=blue,linkcolor=blue,bookmarksopen]{hyperref}
\usepackage{amsfonts}
\usepackage{amsmath}
\usepackage{amssymb}
\usepackage{bbold}
\usepackage{mathrsfs}
\usepackage{graphicx}
\usepackage{dcolumn}
\usepackage{bm}
\usepackage{times}
\usepackage{comment}
\usepackage{latexsym}
\usepackage{epsf}
\usepackage{float}
\usepackage{natbib}
\usepackage{bm}
\usepackage[export]{adjustbox}
\usepackage[makeroom]{cancel}
\usepackage[english]{babel}
\usepackage[T1]{fontenc}
\usepackage{wasysym}
\usepackage[tight, FIGTOPCAP, hang, raggedright, nooneline]{subfigure}
\usepackage[dvipsnames]{xcolor}
\usepackage{epsfig}
\usepackage{lipsum}
\usepackage{enumitem}
\usepackage{scalerel}
\usepackage{ulem}

\begin{document}

\title{Analytical  non-Hermitian description of Photonic Crystals\\
with arbitrary Lateral and Transverse symmetry}

\author{X. Letartre$^1$}
\email{xavier.letartre@ec-lyon.fr}
\author{S. Mazauric$^1$}
\author{S. Cueff$^1$}
\author{T. Benyattou$^1$}
\author{H S. Nguyen$^{1,2}$}
\author{P. Viktorovitch$^1$}

\affiliation{$^1$Univ Lyon, Ecole Centrale de Lyon, INSA Lyon, Universit\'e  Claude Bernard Lyon 1, CPE Lyon, CNRS, INL, UMR5270, Ecully 69130, France} 
\affiliation{$^2$Institut Universitaire de France (IUF)}

\date{\today}	

\begin{abstract}

		We propose a general theoretical approach to the modelling of complex dispersion characteristics of leaky optical modes operating in photonic crystal slab composed of two coupled high-index contrast gratings.  Our analytical model, based on a non-Hermitian Hamiltonian, allows for a unified description of the wide family of optical modes which may be generated within uni-dimensional photonic crystals. Our theory stands for  a variety of illustrative examples relating to the manipulation of bound states in the continuum and exceptional points, and can be used as a powerful enabler for the discovery of novel photonic species.    Finally, as proof-of-concept, we demonstrate experimentally the formation of a Dirac point at the merging of three bound states in the continuum that is the most achieved photonic specie discussed in this work.
\end{abstract}

\maketitle
\section{\label{sec:Introduction} Introduction}
 Taming losses has always been the main challenge for the development of photonic devices in modern history: nurturing lasing emission from leaky channel of an optical cavity\,\cite{Xiao2010}, minimizing attenuation of guided light in integrated circuit\,\cite{McNab03}, confining photons to interact strongly with quantum emitters\,\cite{Press2007}, sharpening photonic resonances for high-sensitivity optical sensing\,\cite{Zhang18}, engineering emission pattern of light-emitting diodes\,\cite{Wierer2009}, to cite a few examples. As a matter of fact, most photonic phenomena are dictated by the complex energy-momentum dispersion characteristic of which the imaginary part corresponds to system losses and the real part corresponds to light frequency. The maturity of nano-fabrication technologies nowadays offers unprecedented degree-of-freedom for dispersion engineering of light via molding periodic arrangement of materials with different permittivity and geometry. More importantly, since the past few years, modern photonics has entered a new paradigm for the research of non-conservative optical systems exploiting non-Hermiticity notions\,\cite{Feng2017,Midya2018,El-Ganainy2019}. In this approach, the complex energy-momentum dispersions, theoretically described by non-Hermitian Hamiltonians, reveal unique features with no Hermitian counterparts. The most famous example is the non-Hermitian extension of topological band theory which is originally built for lossless band structure from condensed matter physics\,\cite{RevModPhys.93.015005,PhysRevB.105.094103,PhysRevA.105.023531}. Two distinctive objects of non-Hermitian topology are Bound states In the Continuum (BICs) and Exceptional Points (EPs). The first one, BICs, are lossless states resulting from destructive interference of coupled lossy photonic resonances\,\cite{PhysRevA.32.3231,Hsu2016}. They are topological charges pinned at polarization vortices in momentum space\,\cite{zhen2014topological,doeleman2018experimental}. Topological manipulations (merging, splitting..) of BIC charges\,\cite{yoda2020generation} propose unique way to modify farfield radiation\cite{Yin2020}, robustness\,\cite{Jin2019} and threshold\,\cite{Hwang2021} of light emitting devices. The second one, EPs, where photonic complex resonances coalesce, are degeneracy points of non-Hermitian physics\,\cite{EPphotonics}. Fundamentally, they take the role of Dirac points as singularities in non-Hermitian topological band theory\,\cite{RevModPhys.93.015005}. For devices applications, they offer novel concepts for making optical sensors with high sensitivities\,\cite{Wiersig2014,Hodaei2017} and lasers with intriguing properties\,\cite{Peng2016,Miao2016,Gao2017}. Interestingly, in light of non-Hermiticity, ``text-book'' dispersion characteristics may gain or evolve with new behaviors: half of photonic band-edges at the centre of the Brillouin zones are BICs in one dimensional lattice\,\cite{LeeMagnusson2019}, Dirac points transform into EP rings in two dimensional lattice\,\cite{Zhen2015} ... Thus strategies used for Hermitian photonics may suggest fruitful scenarii to investigate and study complex dispersion engineering in non-Hermitian context.

We have recently proposed  an analytical approach of One Dimensional (1D) Photonic Crystals (PCs) with broken transverse symmetry, which were shown to provide a new degree of freedom for the design of optical dispersion, and hence for the control of spatial and spectral characteristics of light\,\cite{nguyen2018}. The 1D PCs were composed of two high-index-contrast subwavelength dielectric gratings of same period and in close near field proximity. We demonstrated  that breaking transverse symmetry opens the way to the generation of any local density of photonic states from zero (Dirac cone) to infinity (flatband of zero curvature), as well as any constant density over an adjustable spectral range for the same photonic band. 
At this stage, the considered  photonic modes were assumed to operate inside the light cone and were fully preserved from radiating into the continuum, thus the system were perfectly described by a Hermitian Hamiltonian.

In the present work, we generalize our theoretical approach to the modelling of complex dispersion characteristics of leaky modes operating above the light cone in 1D PCs.  Opening access of wave-guided resonances to free space continuum provides large amount of extra degrees of freedom for mode coupling engineering. Not only can the two gratings communicate via near field coupling, but they are also allowed to couple via the propagating radiated field. A general non-Hermitian Hamiltonian is proposed to capture both coupling schemes. In particular, we show that the lateral and transverse symmetry, both described by phase parameters in the non-Hermitian Hamiltonian, play crucial roles in the radiative coupling processes. Our approach allows for a unified description of the wide family of optical modes which may be generated within an arbitrary 1D PC. Remarkably, through a variety of illustrative examples, we show that our theoretical approach provides a simplified categorization of these modes, and it is also a powerful enabler for the discovery of novel photonic species. Finally, as a proof-of-concept, we demonstrate experimentally the formation of a Dirac point at the merging of three BICs that is the most achieved photonic specie discussed in this work.

\begin{figure}
	\begin{center}
	\includegraphics[width=0.45 \textwidth]{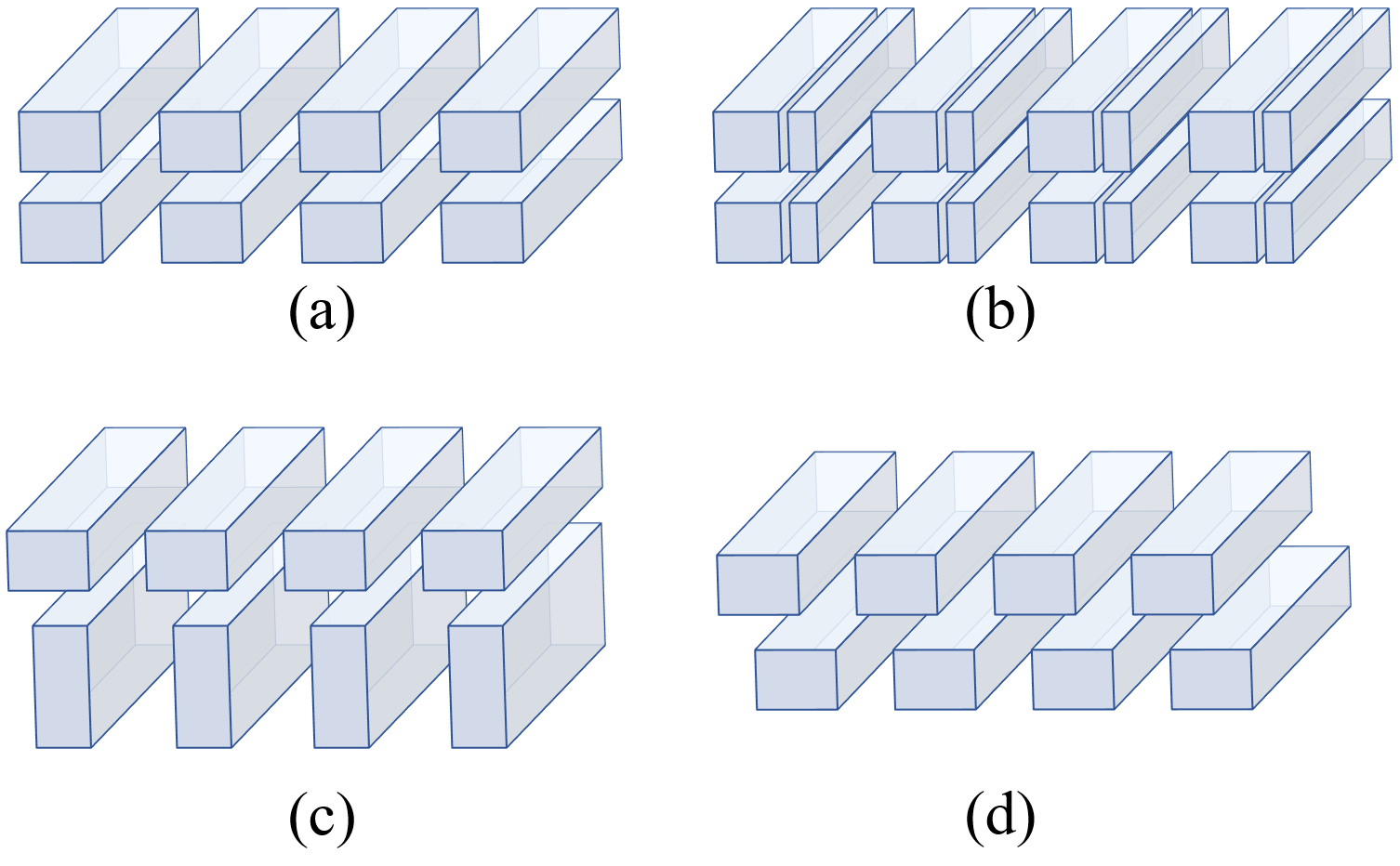}	
\caption{Schematic  view of the different types of coupled grating structures investigated in this paper: with full lateral and transverse symmetry (a), with broken lateral symmetry (b), with broken transverse symmetry (c), ``fish-bone'' structure, a specific case where both the lateral and transverse symmetry are broken (d).}
	\label{fig:1}
\end{center}
\end{figure}

This paper is organized as follows. The next section \ref{sec:TheoryApproach} is devoted to the presentation of the non-Hermitian Hamiltonian $H$. This general Hamiltonian is the back-bone of the present work and makes it possible the derivation of a physically insightful analytical model of the dispersion characteristics. Figure \ref{fig:1} provides a schematic view of the variety of coupled grating structures, which we propose to handle, using our theoretical approach. Since the number of parameters involved for full generality is rather large, it is appropriate to implement simplified versions of the general Hamiltonian, encompassing a wide variety of specific practical cases. 

In sections \ref{sec:aligned_identical} to \ref{sec:transLateralSymBreak}, we select a few exemplifying typical grating configurations to assess the validity and illustrate the effectiveness of our analytical model, whose results are confronted to RCWA (Rigorous Coupled-Wave Analysis) and FEM (Finite Element Method) numerical simulations \footnote{for all simulations the gratings are made in silicon (n=3.5) embedded in silica (n=1.5)}. This selection is directly related to specific symmetry characteristics of coupled grating structures : in sections \ref{sec:aligned_identical}  to \ref{sec:transSymBreak}, we consider  aligned grating structures successively with full lateral and transverse symmetry \ref{sec:aligned_identical} , with broken lateral symmetry \ref{sec:latSymBreak}, and finally with broken transverse symmetry \ref{sec:transSymBreak}. The section \ref{sec:transLateralSymBreak} focuses on a specific case of misaligned grating structures, so called ``fish-bone'' structure formed by two misaligned, identical and fully symmetrical gratings. The ``fish-bone'' is a unique specific case of 1D periodic structure, where both the lateral and transverse symmetries are broken.

We will show that our analytical model is particularly suited as a predictive qualitative design tool for the production of a great variety of photonic states. This includes a simplified and accurate classification of BICs, which have been the matter of intense investigation during recent years\,\cite{plotnik2011experimental, Hsu2013,viktorovitch20103d,chang2012high,letartre2003switching,suh2003displacement,shuai2013coupled,campione2016broken,milord2015engineering,romano2020ultra,Hsu2016,zhen2014topological,doeleman2018experimental,Jin2019,yoda2020generation,kang2021merging,ovcharenko2020bound,Hwang2021}. Our model provides the clues for the generation of BICs, - with adjustable (from very flat to Dirac like) complex dispersion characteristics - as well as the generation of such fascinating photonic species as EPs \cite{EPphotonics} in the complex dispersion curves; particular attention is paid on the physical impact of transverse as well as lateral symmetry of the photonic structures. A specific emphasis will be placed on the demonstration of two unique photonic species, called thereafter (i) ``triple BIC'', exhibiting unprecedented low optical radiation losses over a very large k-vector range of the Brillouin zone and (ii) ``Dirac point at triple BIC'', originating from the degeneracy of one triple and one double BICs at the $\Gamma$ point. Since the latter may be considered as the most achieved photonic specie based on the combination and interaction of BICs, we present our experimental demonstration of ``Dirac point at triple BIC''  in section \ref{sec:exp} as a proof-of-concept. In the last section \ref{sec:conclusion}, we point out the great potential of photonic crystals with controlled lateral and transverse symmetry, as generic building blocks for a range of new practical applications as well as for original physical studies.

Additional specific developments, complementary to the main text, are provided in the Appendix.

\section{\label{sec:TheoryApproach}  Theoretical approach}

\subsection{\label{ssec:NonhermitianH} Non-Hermitian Hamiltonian}
Properties of the eigenmodes in PC slabs are dictated by two symmetry categories\,\cite{Sakoda2001}: the ``transverse symmetry'' (or vertical symmetry), defined by the  z-reflection operation  $\sigma_z$, and the ``lateral symmetry'' (or in-plan symmetry), defined by the x-reflection operation $\sigma_x$ in the case of 1D PC (see Appendix \ref{sec:symmetry_properties} for details of these operators). Leaky optical resonances in PC slabs are generally described as complex eigenvalues of 2$\times$2 non-Hermitian Hamiltonian; this approach has been successfully applied to leaky resonances exhibiting peculiar states such as EP and BIC \cite{Zhen2015,Hsu2016}. However this approach, although attractive in terms of simplified mathematical resolution, provides a rather partial description of the structures. It ignores explicitly the transverse dimension of the structures. The 2$\times$2 Hamiltonian cannot therefore accounts for the impact of a transverse symmetry breaking nor can it describe properly the transverse radiative properties, for example the occurrence of resonance trapped BICs\,\cite{Hsu2013,Kodigala2017}, named transverse BICs thereafter. To fully consider this symmetry, a more complete description with non-Hermitian Hamiltonian of higher dimension is required.

In the case of 1D PC slabs, the transverse dimension and symmetry can be taken into account by using a description based on two coupled grating structures, with no defined transverse dimensions (ie bilayer 1D PC\,\cite{nguyen2018}). This approach might also be considered as a simplified description of a real grating structure, naturally provided with a transverse dimension; however, switching from one single grating, with no transverse dimension, to a pair of coupled gratings allows for taking into account number of factors related to the transverse dimension of a real grating, despite the transverse dimensionless nature of each of the two individual coupled gratings. The reader may find complementary explanation of the rationale and phenomenology of our model in the appendix section \ref{sec:discussion_model}.

\begin{figure}[b]
\begin{center}
\includegraphics[width=0.49 \textwidth]{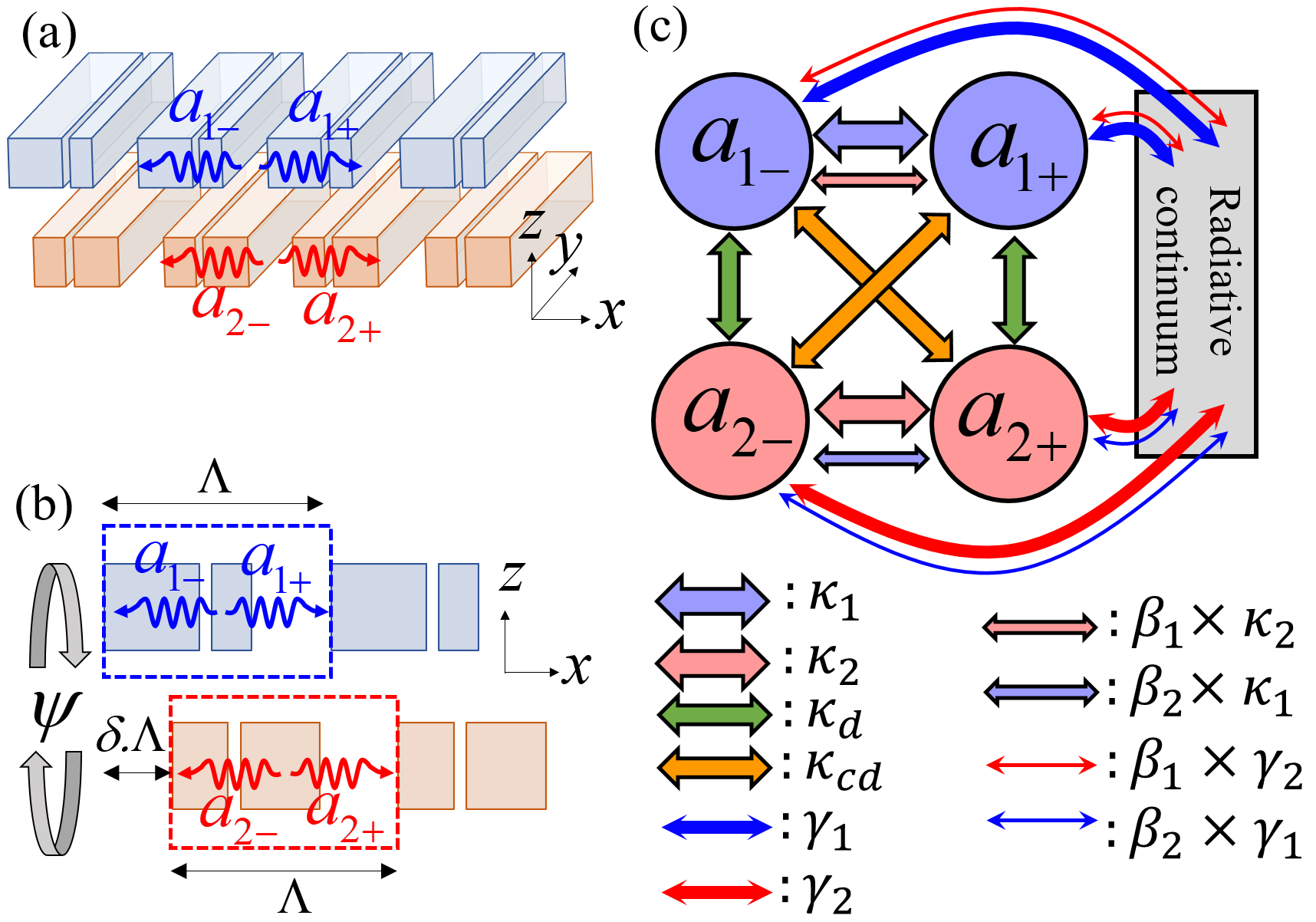}	
\caption{\label{fig:GeneralHamiltonian}Schematic presentation of the coupled grating structure (a) and detailed cross-sectionnal view of the unit cell (b); $\delta$ is the lateral off-set between the two grating of period $\Lambda$. $\psi$ is the transverse phase shift. (c) Different coupling mechanisms between the four guided modes $a_{1\pm}$,$a_{2\pm}$.}
\end{center}
\end{figure}

A sketch of the two coupled gratings (with common period $\Lambda$) is presented in Fig.\ref{fig:GeneralHamiltonian}(a). The gratings are depicted with finite thickness, although this parameter is not explicitly accounted for in the analytical model, nor is explicit the geometrical distance between the two structures. Indeed, these two parameters are merged into a single one, the phase parameter $\psi$, shown in Fig.\ref{fig:GeneralHamiltonian}(b). This phase is the key parameter describing the transverse dimension of the structure, and is further discussed later in the main text (see \ref{ssec:phase}) and in the appendix section \ref{sec:discussion_model}. The transverse symmetry of the system of the bilayer PC is partly defined by the characteristics of each individual grating. 

The lateral symmetry of the system is partly set by the lateral symmetries of the grating unit cells. In addition, both symmetries are also controlled by the lateral offset $\delta\times \Lambda$ between the two gratings [see Fig.\ref{fig:GeneralHamiltonian}(b)]. For the sake of simplifying the writing of the equations of the analytical model, the origin of the lateral x-coordinate is taken at mid lateral offset between the two gratings. Referring the gratings to this origin is straightforward, if both have laterally symmetric unit cells. If this is not the case, it is appropriate to define the anchorage position of each grating at the x-centroid of the lateral distribution of the effective dielectric constant.

In our model, the $\omega(k_x)$ dispersion characteristics are derived from different coupling processes undergone by the forward $\left(a_{1+},a_{2+}\right)$ and backward $\left(a_{1-},a_{2-}\right)$ fundamental zero-order waves of the two planar waveguides of effective refractive index which are assumed to be single-mode in the spectral range of interest. The dispersion engineering is focused in the vicinity of the second Brillouin zone boundary, where second-order diffractive coupling processes between backward and forward wave components as well as first order coupling processes with the radiation continuum (in the vicinity of the $\Gamma$ point) take place. A phenomenological description of coupling processes within the two gratings as well as of cross coupling processes occurring between them can be obtained using the coupled mode theory formalism. In the basis formed by $\left(a_{1+},a_{1-},a_{2+},a_{2-}\right)$, the equations of coupled mode theory end up in an eigenequation of a $4\times4$ non-Hermitian Hamiltonian, given by:
\begin{equation}\label{eq:H}
    H=
    \left(\begin{matrix}
		H_{11} & H_{12} \\ 
		H_{21} & H_{22} \\
	\end{matrix}\right),
\end{equation}
 in which $H_{ij}$ are 2$\times$2 matrices, which describe the optical interactions intra each gratings ($H_{ij=i}$) and inter the two gratings ($H_{ij\ne i}$). The general expressions of $H_{ij}$ are given in the Appendix section \ref{sec:derivation_H}. As pointed out in the introductory section, the Hamiltonian is not meant to be exploited in its full generality, but instead to be broken down into a variety of simplified versions suited to specific practical cases. The parameters chosen to build up the various matrix element expressions (given in the Appendix section \ref{sec:derivation_H}) allow for comprehensive analytical description of leaky 1D PC slabs. They are categorized and scrutinized into different coupling strengths and phases, and will be discussed in the following.

\subsection{\label{ssec:uncoupledModes} Uncoupled modes}
The dispersion characteristics of the uncoupled modes (i.e. $a_{1\pm}$ for the upper grating and $a_{2\pm}$ for the lower grating) are given by the pulsations $\omega_{1,2}$  of the two non-corrugated waveguides at $2\pi/\Lambda$ wave vector, and their corresponding group velocities $v_{1,2}$. 

In the rest of the manuscript, we denote $k$ the variation of the wave vector in the vicinity of the second Brillouin zone boundary (i.e. $2\pi/\Lambda$). The dispersion relationship of uncoupled modes can be written as $\omega_{i\pm}=\omega_i \pm v_i k$ with $i=1,2$.

\subsection{\label{ssec:coupling} Coupling rates} 
The coupling mechanisms between modes in the basis $\left(a_{1+},a_{1-},a_{2+},a_{2-}\right)$, as depicted in Fig.\ref{fig:GeneralHamiltonian}(c), include the coupling between guided modes from the same membrane (intra-layer coupling) or from separated ones (inter-layer coupling), and the coupling between these guided modes with the radiative continuum. The corresponding coupling rates are:

\begin{itemize}
\item For intra-layer diffractive coupling rates (expressed in intra-layer $H_{ii}$ sub-matrices) :
    \begin{itemize}
    \item $\kappa_{1,2}$:diffractive coupling rates between counter-propagating guided waves in the same grating (1 or 2), due to their own corrugation
    \item $\gamma_{1,2}$:diffractive coupling rates between guided waves in grating 1 and 2 and the radiation continuum. The two gratings are supposed to be zero order in the spectral range of interest ($\frac{\Lambda}{\lambda}<1$) around the $\Gamma$ point.
    \item $\beta_{1,2}$:evanescent relative weight of grating 2(1) to the total diffractive coupling rate of grating 1(2). These coefficients quantify the amount of diffractive coupling rates within one membrane layer, induced evanescently by the grating of the other membrane layer. It is assumed that the relative weight is the same, whether it is applied to $\kappa_{1,2}$  or $\gamma_{1,2}$. When the two gratings are far apart, $\beta_{1,2}$  tend to zero, and $H_{ii}$ sub-matrices describe intra-layer interactions within each individual grating, taken alone
    \end{itemize}
\item 	For inter-layer diffractive coupling rates (expressed in inter-layer $H_{ij}$ sub-matrices) :
    \begin{itemize}
    \item $\kappa_d$: evanescent coupling rate between co-directional guided waves from different gratings, with light transfer from grating 1 to grating 2 and reciprocally. 
    \item $\kappa_c$: diffracto-evanescent coupling rate between counter-directional guided waves from different gratings, with light transfer from one grating to the other, then switching the propagation direction.
    \end{itemize}
\end{itemize}

When the two gratings are far apart, $\kappa_d$ and $\kappa_c$ tend to zero; however, the two gratings keep on interacting via radiative coupling, which does not vanish, as expressed by inter-layer $H_{ij}$ sub-matrices. This interaction is delayed by the light time of flight between the two gratings, which is quantified by the transverse phase shift $\psi$ introduced in the upcoming section \ref{ssec:phase} and in Fig.~\ref{fig:GeneralHamiltonian}. This key phase parameter is further commented extensively in this paper (see, for example, section \ref{ssec:H1} and appendix sections \ref{sec:discussion_model} and \ref{sec:derivation_H}). 

The basis is chosen so that $\kappa_{1,2}$ and $\gamma_{1,2}$ are real positive numbers. However, $\kappa_d$ and $\kappa_c$ are complex in general, and they are real numbers if the two gratings are identical.

\subsection{\label{ssec:phase} Coupling phases} 
While most of the coupling rates presented in the previous subsection (\ref{ssec:coupling}) are real numbers, they are associated with different phases when implemented in the Hamiltonian construction: 

\begin{itemize}
    \item $\phi=2\pi\delta$: phase parameter corresponding to different diffraction order phase-shifts occurring in each gratings. It results from	their respective off-set x-coordinate $\pm\frac{\delta}{2}\times\Lambda$ [see Fig.~\ref{fig:GeneralHamiltonian}(b)]. The phase shifts are, respectively, $\pm\phi$ for the second order diffractive coupling between forward and backward guided waves (in the vicinity of the second Brillouin zone boundary), and $\pm\phi/2$ for the first order diffractive coupling between guided waves and waves of radiation continuum (in the vicinity of the $\Gamma$ point). $\phi$ is therefore central in the diffractive coupling processes mentioned above.

    \item $\psi$: transverse phase shift built up in free continuum by radiated photons, along a one way trip between the two gratings. $\psi$ includes the phase shifts accumulated within the gratings and in between; it is related to the effective transverse optical distance $L_{opt}$ between the wave-guided resonances. Therefore, $\psi$ is not solely related to the geometrical distance between the two gratings, but is also impacted by the transverse field distribution. $\psi$ is a central phenomenological parameter of our model, which conveys in itself significant amount of physical significance. An analytical expression of this parameter will be derived later in this paper (Eq.~\eqref{eq:psi} in section \ref{ssec:H1}). 

    \item $\left(\varphi_{1+}, \varphi_{2+}, \varphi_{1-}, \varphi_{2-}\right)$: phases related to the coupling step of forward (+) and backward (-) guided waves to the radiation continuum in gratings 1, 2. If the grating unit cell is laterally symmetric $\varphi_{i+}=\varphi_{i-}$. If the two gratings are identical $\varphi_{1\pm}=\varphi_{2\pm}$.
\end{itemize}

Phases $\psi$ and $\varphi_{i\pm}$ play a major role in the radiative coupling processes, which are accounted for by a phenomenological description of the radiative channel connecting the two gratings. The radiative channel includes successively 3 steps, (i) the extraction of the wave-guided field from one grating to free space continuum, (ii) the propagation of the radiated field in free space continuum and finally (iii) the resonant insertion of the latter into the other wave-guide grating. Phases $\varphi_{i\pm}$ are involved in the first and third steps, while the second step is controlled by the phase $\psi$. Implementation of the different phases in the Hamiltonian construction is given in the appendix section  \ref{sec:derivation_H}.

The reader is invited to refer to this section of the appendix for additional explanation about the derivation of the Hamiltonian matrix elements.

\subsection{\label{ssec:modelApprox} Approximation of the model}

In order to establish a useful and tractable Hamiltonian, two main approximations are made. First of all, as we will see in the upcoming sections, the diagonalization of $H$ gives rise to an implicit equation for the complex eigenvalues $\omega=\omega_R-i\omega_I$ (see e.g. Eqs.~\eqref{eq:omega} to \eqref{eq:complexsplitting}). Indeed, some parameters in $H$ are functions of the phase $\psi$ (see e.g. Eqs.~\eqref{eq:psi}). The latter is related to the propagation of photons in the transverse direction and obviously depends on $\omega$. In the following we will assume that $\psi$ depends on the sole real part $\omega_R$. This remains valid when the time spent by photons for propagating in the transverse direction (called $\tau_{rad}$ in the  appendix \ref{sec:discussion_model}; see also Eq~\eqref{eq:lifetime_rad} in section \ref{sssec:DoubleTransverseBIC}) is much smaller than $1/\omega_I$ . As we will focus in this article on modes with high Q-factors, the condition $\tau_{rad}\ll 1/\omega_I$  is always met.

We also assume that the propagation of light in the vertical direction can be taken into account by the sole dephasing $\psi$ and is not affected by the interfaces between the dielectric layers. It can be shown that the qualitative behaviours described in this article remain valid, even though some corrections must be made on the eigenvalues, mainly on their imaginary parts. 
These approximations are more extensively discussed in the appendix (section \ref{sec:discussion_model}).

\section{\label{sec:aligned_identical} Symmetrical and aligned gratings}

\begin{figure}[b]
\begin{center}
	\includegraphics[width=0.4 \textwidth]{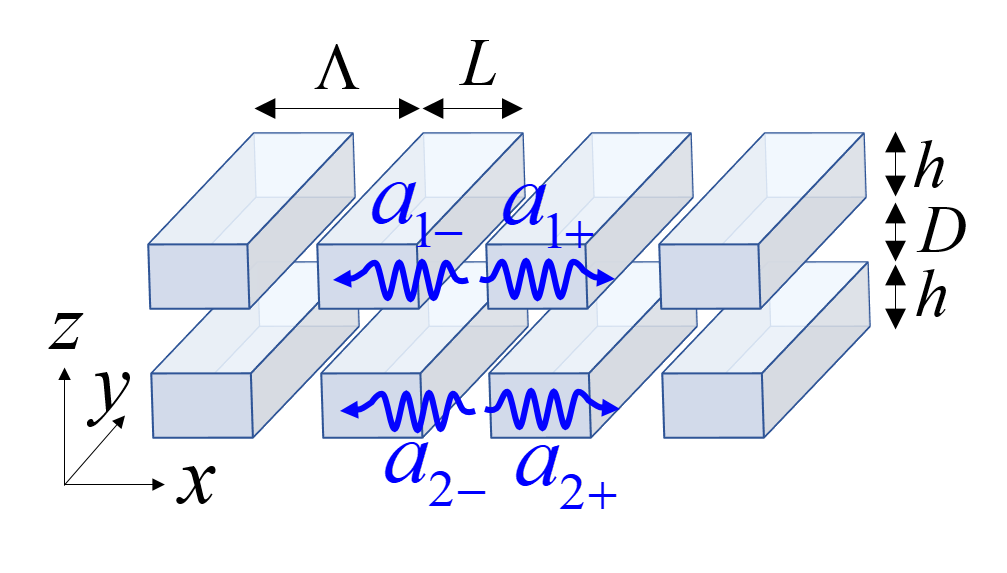}	
\caption{\label{fig:sym_align}Schematic presentation of the fully symmetrical coupled grating structure, along both the transverse and lateral directions.}
\end{center}
\end{figure}

In this section, we will discuss and analyze in detail the configuration of coupled gratings structure which is fully symmetrical along both the transverse and lateral directions (see Fig.~\ref{fig:sym_align}). This is the simplest configuration, yet revealing to be very generic in terms of complex dispersion characteristics, as demonstrated in the following.\\

\subsection{\label{ssec:H1} Analytical expression of Hamiltonian}

 The general Hamiltonian is considerably simplified; first, the phase parameters, except for the phase $\psi$, are no longer relevant: $\phi=0$, since the two gratings are aligned, and $\varphi_{1\pm}=\varphi_{2\pm}$, together with $\varphi_{1,2+}=\varphi_{1,2-}$, since they are identical and formed with laterally symmetric unit cells. Second, the coupling coefficients are identical, $\kappa_{1,2}  =\kappa$, $\gamma_{1,2}=\gamma$, $\beta_{1,2}=\beta$, together with the kinetic parameters, $\omega_{1,2}=\omega_0$ and $v_{1,2}=v_0$. Finally, the coupling parameters $\kappa_{c,d}$ are real. As a result the Hamiltonian can be written along the following simplified version:
 
\begin{widetext}
\begin{equation}\label{eq:H1}
    H(k)=\left(\omega_0 - iG\right)\mathbb{1}_{4} +   \left(
    \begin{matrix}
    v_0k& K - iG &  \kappa_d - iG e^{i\psi} & \kappa_c - iG e^{i\psi} \\
     K - iG &  - v_0k & \kappa_c - iG e^{i\psi} & \kappa_d - iG e^{i\psi} \\
       \kappa_d - iG e^{i\psi} & \kappa_c - iG e^{i\psi}  &   v_0k & K - iG \\
      \kappa_c - iG e^{i\psi} & \kappa_d - iG e^{i\psi}  & K - iG &  - v_0k \\
    \end{matrix}
    \right),
\end{equation}
\end{widetext}
where $K = \kappa\left(1+\beta\right)$ and $G=\gamma\left(1+\sqrt{\beta}\right)^2$. The complex eigenvalues of Eq.~\ref{eq:H1} can be expressed in a general form:
\begin{equation}
    \omega(k)=\omega_R(k) - i\omega_I(k).
\end{equation}
Here $\omega_R(k)$ and $\omega_I(k)$ are given by the real and imaginary part of $\omega(k)$ and represent the optical frequency and the radiative loss respectively. Explicitly, the diagonalization of $H$ provides four complex eigenvalues which can be regrouped into two couples corresponding to two opposite parities $\eta=\pm 1$ of the transverse symmetry:
\begin{equation}
    \omega_{\eta}^\pm = \omega_0 + \eta\kappa_d - i G_\psi \pm \frac{S(k)}{2}  \label{eq:omega}.
\end{equation}
Here the ``complex radiation rate'' $G_\psi$ is given by:
\begin{equation}\label{eq:complex_gamma}
G_\psi = G\left(1+\eta e^{i\psi}\right),
\end{equation}
and the ``complex splitting'' of each couple is determined by: 
\begin{equation}\label{eq:complexsplitting}
    S(k)=2\sqrt{v_0^2k^2 + \left(K + \eta\kappa_c - iG_\psi\right)^2}.
\end{equation}
The first couple of Eq.~\eqref{eq:omega}, given by $\eta=1$, corresponds to two fundamental/even (along the transverse direction) modes. The second one, given by $\eta=-1$, corresponds to two excited/odd (along the transverse direction) modes. We note that the parameter $\kappa_d$ is negative, since the fundamental guided mode of the coupled membranes is obviously even. The even and odd branches of the complex eigenvalues ignore each other, since the corresponding modes have opposite parity. Therefore, possible crossing of these branches is not avoided. \\

As pointed out in the previous section, the phase $\psi$ is central in our model; this parameter expresses the transverse phase shift built up in free continuum by radiated photons within the gratings and in between, along a one way trip between the two gratings. It is related to the effective transverse optical distance $L_{opt}$ between the wave-guided resonances. The analytical expression of the phase $\psi$ is thus simply given by $\psi=\frac{k_{\perp}}{\overline{n}}L_{opt}$ where $k_{\perp}$ is the transverse component of the wave-vector and $\overline{n}$ is the average refractive index perceived by radiated photons along their trip in the continuum. Neglecting the reflection at the membrane interfaces (see sections \ref{ssec:modelApprox} and \ref{sec:discussion_model} ), it can be written:
\begin{equation}
    \psi(k)=\sqrt{\left[\frac{\omega_R(k)}{c}\right]^2 - \left(\frac{k}{\overline{n}}\right)^2}L_{opt}
    \label{eq:psi},
\end{equation}
where $c$ is the light celerity in vacuum.

The phase $\psi$ can be controlled by adjusting the transverse optical distance $L_{opt}$ between the wave-guided resonances in each membrane, which is set mainly by the thickness $h$ of the gratings, and by the distance $D$ between them:
\begin{equation}
    L_{opt}\approx n.h + n_{0}.D
    \label{eq:Lopt}.
\end{equation}
 Here $n$ n is the effective transverse refractive index of the grating and $n_0$ is the refractive index of the spacer.

We point out that the real part of the eigenvalue $\omega$ depends on $\psi$, through Eq.~\eqref{eq:omega}, as well as $\psi$ depends on $\omega$  through Eq.~\eqref{eq:psi}. Therefore different specific values of $\psi$ apply to the Eq.~\eqref{eq:omega}, except when degeneracy of $\omega$ occurs between two eigenvalues. As it is illustrated in the following discussions of this section \ref{sec:aligned_identical}, Eq.~\eqref{eq:psi} is particularly relevant when it comes to provide a quantitative description of the real eigenvalues for symmetrical structures, featuring such distinctive characteristics as, for example, double transverse BICs (section \ref{sssec:DoubleTransverseBIC}) or Dirac point at triple BIC (section \ref{sssec:DiracTripleBIC}). It provides also a faith-full qualitative prediction of the behaviour of imaginary eigenvalues, especially regarding the variations of the imaginary eigenvalues versus momentum $k$ around the $\Gamma$ point.\\ 

Fully analytical resolution of this system of equations is accessible at (close to) the $\Gamma$ point, while numerical assistance is required further away, as this will be illustrated in the following. Numerical resolution procedure will be particularly used, whenever it comes to validate our model against results of RCWA/FEM numerical simulations. 

\subsection{\label{ssec:GenFeaturesEig1} General features of the complex eigenvalues}

\subsubsection{At the $\Gamma$ point: Lateral BIC}
At the $\Gamma$ point (i.e. $k=0$), the expressions given by Eqs.~\eqref{eq:omega},\eqref{eq:complex_gamma} and \eqref{eq:complexsplitting} are greatly simplified. One can easily show that $\omega_\eta^-(k=0)$ for both parities have no imaginary part. They are thus two ``dark modes'' which are free of losses and are systematically observed at the $\Gamma$ point in structures being symmetric along the lateral direction. Both of them belong to the category of ``symmetry protected BICs''\,\cite{Paddon2000,Hsu2016}. This protection arises from the symmetry of the mode which turns to be odd with respect to the unit cell of the grating which is symmetrical along the lateral direction. As a result it cannot radiate at the $\Gamma$ point since coupling to plane waves is prevented, unlike the case of the bright mode which symmetry is even. This type of BIC has been widely documented in the literature\,\cite{Paddon2000,Hsu2016}. In this work, we will refer to it as ``lateral BICs'', since it is controlled by the lateral symmetry of the grating structure. This point will be addressed in more details in the section \ref{sec:latSymBreak}, where it is shown that breaking the lateral symmetry prevents the formation of lateral BICs at the $\Gamma$ point.\\

On the other hand, Eq.~\eqref{eq:omega} at $k=0$ indicates that $\omega_{\eta}^{+}(k=0)$ for both parities are generally “bright modes” with non-zero radiative losses (i.e. imaginary part of the eigenvalues).  From now on we denote « dark branches » the two branches (one per parity) of dispersion characteristics showing a lateral BIC at the $\Gamma$ point and ``bright branches'', the other branches. Note that, for the latter, one may observe more than one branch per parity, since they are solutions of the system of two equations ~\eqref{eq:omega} and ~\eqref{eq:psi}, which are not unique in general. This is particularly the case when $L_{opt}$ is significantly larger than the wavelength, as further discussed in section \ref{sssec:DoubleTransverseBIC} and Fig.~\ref{fig:RCWA_flat_doubleBIC}.

\subsubsection{Out of the $\Gamma$ point: Tranverse BIC}
Equations ~\eqref{eq:omega} to \eqref{eq:complexsplitting} indicate that the four eigenmodes are bright out of the $\Gamma$ point. Lateral symmetry is broken for $k\ne0$, and no more room is left to lateral BICs. It can be shown, from expansion of equations around the $\Gamma$ point, that the imaginary part of the dispersion characteristics of the dark branches varies like $k^2$ (see appendix section \ref{sec:expansion_complex}) .

However, for any $k$ value and for each of the four eigenmodes (but not for all of them at a time, in a given structure), optical losses are inhibited when $\psi=\pi(0)$ (mod $2\pi$) for the two even (odd) branches: the bright mode turns into a BIC, occurring accidentally, as a result of destructive interferences in the transverse direction. Occurrence of a BIC may therefore take place on any of the two couples of above denoted dark/ bright branches of even/odd modes. It can be shown from Eqs. ~\eqref{eq:omega} to \eqref{eq:complexsplitting} that the imaginary part of the dispersion characteristics varies like $(k-k_{BIC})^2$ around the wave-vector $k_{BIC}$ corresponding to the occurrence of the transverse BIC (see appendix section \ref{sec:expansion_imaginary} ). This type of BIC may occur for specific geometrical / optical characteristics of the photonic structure; it is also called resonance trapped BIC \cite{Hsu2013,Kodigala2017}. We classify this BIC within a common family which we name transverse BICs, since it is controlled by the transverse symmetry of the grating structure. This point will be addressed in more details in section \ref{sec:transSymBreak}, where it is shown that breaking the transverse symmetry prevents the formation of transverse BICs at and apart from the $\Gamma$ point.

We may point out at this stage that the simplified classification of BICs into the two categories, transverse and lateral BICs, constitutes an attractive outcome of our model; we will show indeed that this classification naturally applies to a variety of BICs occurring in the generic structure formed by two coupled gratings and revealed by our model, in agreement with results of numerical simulation. Even more, we will show that these two basic BIC building blocks can be combined and result in structures featuring remarkable characteristics (in the subsection \ref{ssec:SymmetricalCases} and appendix \ref{sec:fully_symmetric}).\\

\begin{figure}[hbt]
\begin{center}
	\includegraphics[width=0.48 \textwidth]{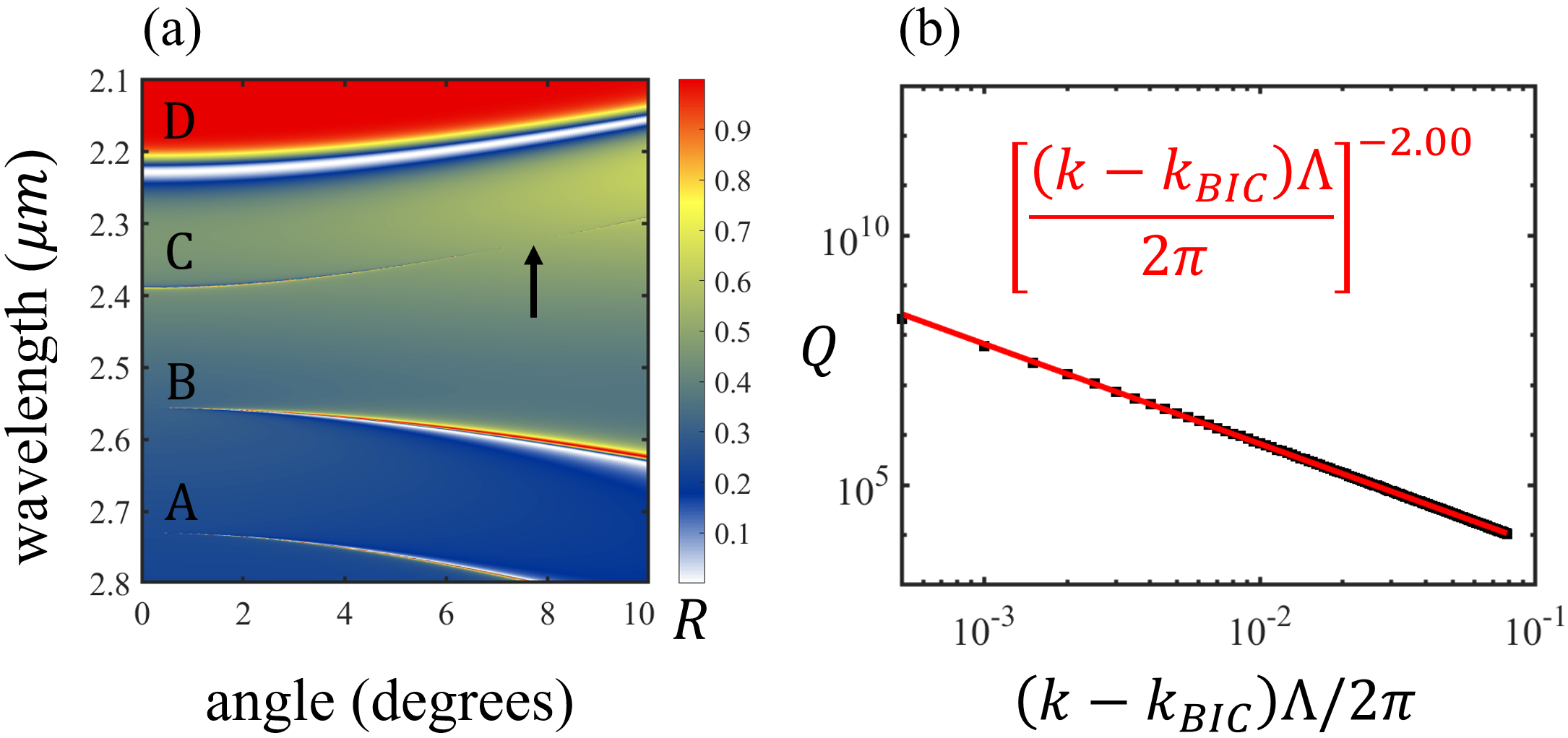}	
\caption{\label{fig:lat_trans_BICs}(a) Angle-resolved reflectivity spectra obtained by RCWA simulations, showing the two couples of dark/bright branches of even/odd modes. The two systematic lateral BICs, even (A) and odd (B), occurring at the $\Gamma$ point can be observed. A transverse BIC (black arrow) is also observed in oblique direction on the bright branch of the even mode (C) ; (b) Log-log plot of the quality factor of the branch C in the vicinity of the wave-vector $k_{BIC}$ corresponding to the occurrence of the transverse BIC in oblique direction; the imaginary part of the eigenfrequency varies like $(k-k_{BIC})^2$. The black squares are numerical results obtained by FEM simulations. The red line is a polynomial fit. The parameters are: $\Lambda=1\ \mu m,\ h=0.3\ \mu m,\ L=0.78\ \mu m$ and $D=0.29\ \mu m$.}
\end{center}
\end{figure}

All the theoretical predictions of the model described above are fully confirmed by RCWA simulations of a fully symmetrical coupled grating structure as reported in Fig.~\ref{fig:lat_trans_BICs}(a) ($\Lambda=1\ \mu m,\ h=0.3\ \mu m,\ L=0.8\ \mu m$ and $D=0.29\ \mu m$). Indeed, from the angle-resolved reflectivity spectra, we distinguish four resonances corresponding to the four branches previously discussed. These are Fano resonances resulting from the coupling between incident plane-wave and the Bloch resonances. Since BICs correspond to a perfect uncoupling of Bloch resonances from the radiative continuum, one may identify easily BICs at local vanishings of these Fano resonances. We observe two lateral BICs at the $\Gamma$ point of A and B branches that are the two dark branches; and a transverse BIC in oblique direction on the bright branch C. More details are written in the caption of Fig.~\ref{fig:lat_trans_BICs}. Finally, Fig.~\ref{fig:lat_trans_BICs}(b) depicts the quality factor of the bright branch C in the vicinity of the transverse BIC, extracted from FEM simulation. We observe indeed a quadratic law as predicted by the analytical theory.

\subsection{\label{ssec:SymmetricalCases} Some selective cases with distinctive characteristics}

\subsubsection{\label{sssec:BandInversion} Band Inversion and Double Exceptional Point}
\begin{figure}[hbt]
\begin{center}
	\includegraphics[width=0.48 \textwidth]{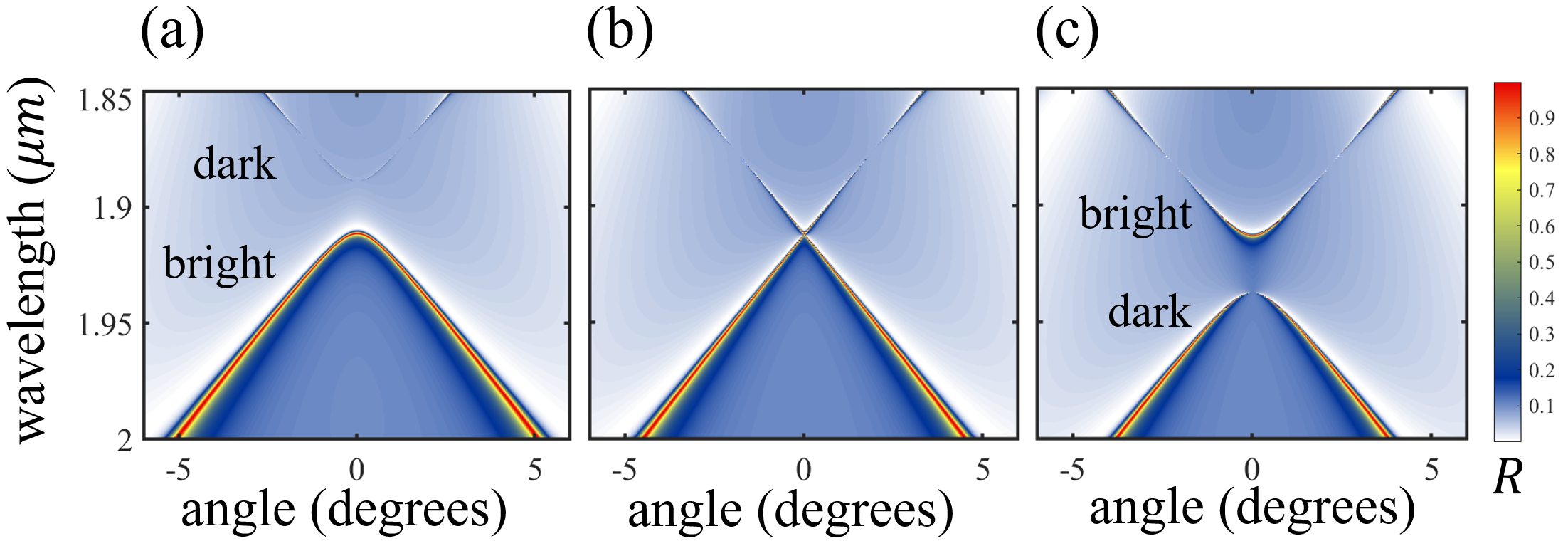}	
\caption{\label{fig:band_inversion}Angle-resolved reflectivity spectra obtained by RCWA simulations, showing band inversion phenomenon resulting from a sign inversion of the factor $K+\kappa_c+G \sin(\psi_0)$. At the onset of the band inversion, the dark branch is flat and degeneracy occurs between the real eigenvalues of the dark and the bright branches. Other than the lateral BIC at $k=0$ of the dark branch, we distinguish also two transverse BIC pinned at $\sim \pm2$ degrees in the upper branch of both configurations.  The band inversion is obtained by varying the filling factor: (a) FF=0.35, (b) FF=0.36 and (c) FF=0.37. The other parameters are $\Lambda=1 \mu m$, $h=0.25 \mu m$ and $D=0.3 \mu m$.}
\end{center}
\end{figure}

Further analytical treatment of the complex eigenvalues of even modes in the vicinity of the $\Gamma$ point or for small values of $k$, where they can be expanded, is given in section \ref{sec:expansion_complex} of the appendix. Results can be summed up as follows. In general, the complex eigenfrequencies are a quadratic function of $k$ in the vicinity of the $\Gamma$ point. The real curvatures change in sign when the quantity $K+\kappa_c+G \sin(\psi_0)$, where $\psi_0=\psi(k=0)$, changes in sign as well, strictly speaking for the dark branch, and in first approximation for the bright branch. For $K+\kappa_c+G \sin(\psi_0)>0$, it is observed that the upper (lower) branch is bright (dark) with upward (downward) curvature and the other way around for $K+\kappa_c+G \sin(\psi_0)<0$. This band inversion behaviour, referred to as band flip in \cite{LeeMagnusson2019}, is shown in RCWA simulations of  Fig.~\ref{fig:band_inversion}. For $K+\kappa_c+G \sin(\psi_0)=0$, degeneracy of the real eigenvalue of the dark and bright branches occurs, which means that the overall diffractive coupling processes between wave-guided resonances cancel, and the dark branch is flat at the $\Gamma$ point (see Eq. \eqref{eq:S25} of the appendix \eqref{sec:expansion_complex}).
RCWA simulations show that the principal controlling parameter of the band inversion phenomena is the filling factor $FF=\frac{L}{\Lambda}$ of the grating structure. Our model does not include explicitly an ab initio parameter to account for FF; the latter is implicitly included through the factor $K+\kappa_{c}+G\sin(\psi_{0})$, which expresses the overall diffractive processes encountered by wave-guided resonances. The reader may find complementary physical insights in the appendix  \ref{sec:Rule_triple_BIC}, which details the practical design rules of a Dirac point at triple BIC.

Conditions for full degeneracy of the complex eigenvalues of the dark and bright branches can be met out of the $\Gamma$ point at double exceptional point for $k=k_{EP}$. Section \ref{sec:Double_EP} of the appendix provides a detailed account of this occurrence, based on our analytical model.
\subsubsection{\label{sssec:DoubleTransverseBIC} Double transverse BIC}
\begin{figure}[hbt]
\begin{center}
	\includegraphics[width=0.45 \textwidth]{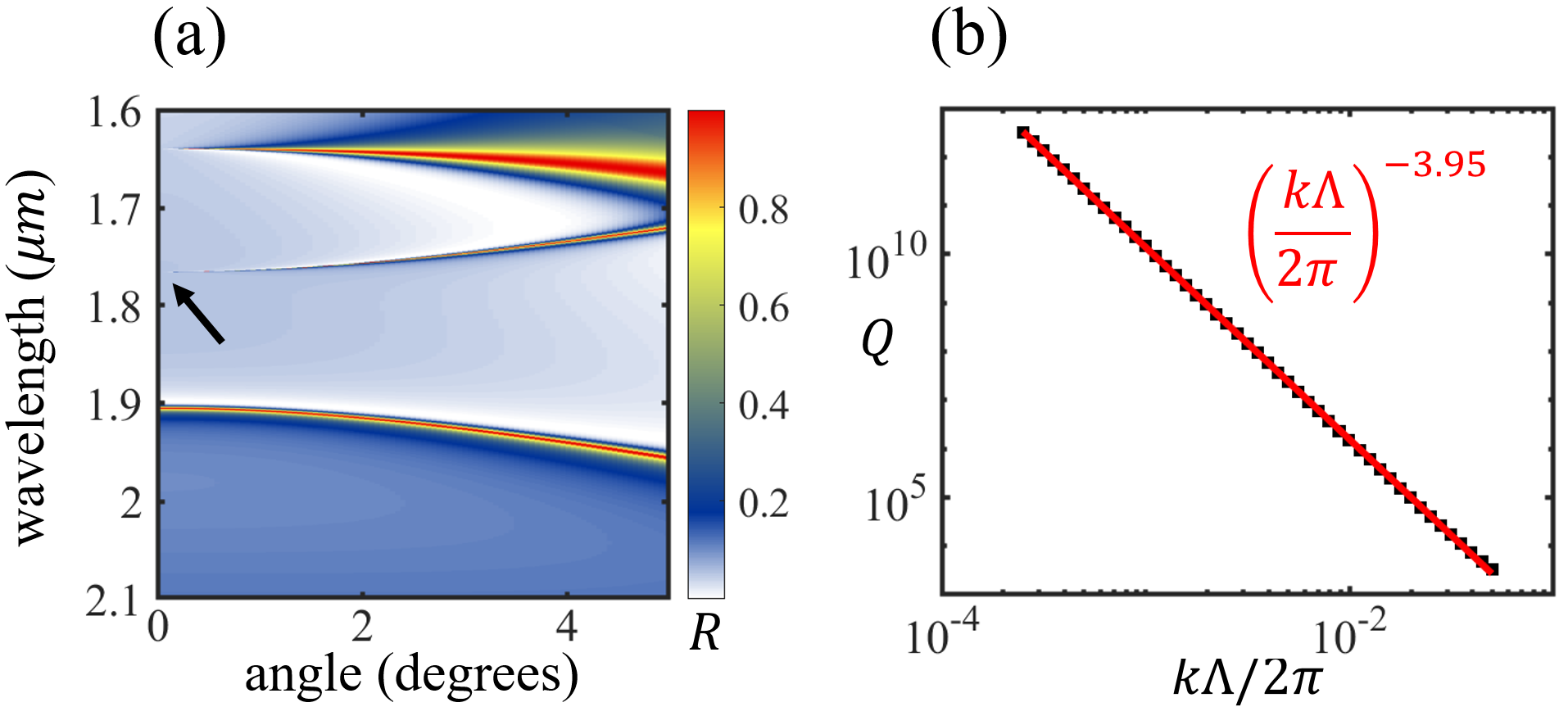}	
\caption{\label{fig:double_transverse_BIC}(a) Angle-resolved reflectivity spectra obtained by RCWA simulations, showing the occurrence of a double transverse BIC at the $\Gamma$ point (black arrow) of the bright branch (here dielectric branch). The ubiquitous lateral BIC shows up at the $\Gamma$ point of the dark branch. (b) Log-log plot of the quality factor of the bright branch versus $k$ vector in the vicinity of the $\Gamma$ point. Black squares are numerical results obtained by FEM simulations. Red line is a polynomial fit. The slope $-3.95$ of the fit confirms the results of our analytical model anticipating variations of the imaginary eigenvalues like $k^4$. The parameters are: $\Lambda=1\ \mu m,\ h=0.25\ \mu m,\ L=0.3\ \mu m$ and $D=0.3\ \mu m$.}
\end{center}
\end{figure}

Let us concentrate on even modes, given that conclusions are qualitatively similar for odd modes. At the $\Gamma$ point of the bright branch, the mode is generally bright except when conditions are met for the formation of a transverse BIC, that is for $\psi=\pi$ (mod $2\pi)$, as explained in the previous section. This particular circumstance corresponds to the merging of two ordinary transverse BICs belonging to the bright branch and occurring at $\pm k_{BIC}$ vectors, when $k_{BIC}\rightarrow0$. We call this particular transverse BIC, double transverse BIC (see appendix sections \ref{sec:expansion_complex} and \ref{sec:expansion_imaginary} for complementary detailed analysis). The generation of a double transverse BIC in fully symmetrical coupled gratings is illustrated in RCWA simulation of Fig.~\ref{fig:double_transverse_BIC} .

Close to the $\Gamma$ point (small $k$), it is possible to derive the complex dispersion characteristic of this double transverse BIC using Eqs.~\eqref{eq:omega}-~\eqref{eq:complexsplitting} where $\eta=1$ and $\psi(k=0)\equiv\psi_0=\pi$ (mod $2\pi)$. The expanded real part $\omega_R$ of the complex dispersion characteristic close to the $\Gamma$ point is given by Eq. \eqref{eq:S21} of appendix \ref{sec:expansion_complex}: 

\begin{equation}\label{eq:real_mode_doubleBIC}
    \omega_{R}(k) = \omega_{R_0} + C_R\frac{k^2}{2},
\end{equation}

where

\begin{equation}\label{eq:omegaR_0}
    \omega_{R_0}=\omega_R(k=0)=\omega_0+\kappa_d+K+\kappa_c,
\end{equation}

and $C_R$ is the curvature of the real dispersion characteristic at the $\Gamma$ point (i.e. the second derivative of $\omega_R(k)$). The curvature $C_R$ can be written:

\begin{equation}\label{eq:real_curvature_doubleBIC}
    C_{R}=\frac{\tau_{wg} C_{wg} + \tau_{rad} C_{rad}} {\tau_{wg} + \tau_{rad}},
\end{equation}

where:

\begin{subequations}
\begin{align}
\tau_{wg}&=\frac{1}{2G} \label{eq:lifetime_wg}\\
    \tau_{rad}&=\frac{L_{opt}}{c} \label{eq:lifetime_rad}
\end{align}
\end{subequations}

and:
\begin{subequations}
\begin{align}
    C_{wg}&=\frac{v^2}{K+\kappa_c} \label{eq:curvature_wg}\\ 
    C_{rad}&=\frac{\left(c/n \right)^2}{\omega_{R_0}} \label{eq:curvature_rad}.  
\end{align}
\end{subequations}

%

$\tau_{wg}$ is the average lifetime of photons in the wave-guided state before being emitted into the continuum and $\tau_{rad}$ is the average lifetime of photons in the radiated state during one way trip between the two gratings. $C_{wg}$ and $C_{rad}$ are the guided and radiated curvature respectively, their contribution to the total curvature is weighted by the relative time spent by photons in the wave-guided state, i.e. $\tau_{wg}/(\tau_{wg}+\tau_{rad})$, and radiated state, i.e. $\tau_{rad}/(\tau_{wg}+\tau_{rad})$, respectively. The expression of the curvature is an eloquent illustration of the hybrid character of this double transverse BIC, which is altogether guided and radiated \cite{Boutami2008}. 

\begin{figure}[hbt]
\begin{center}
	\includegraphics[width=0.45 \textwidth]{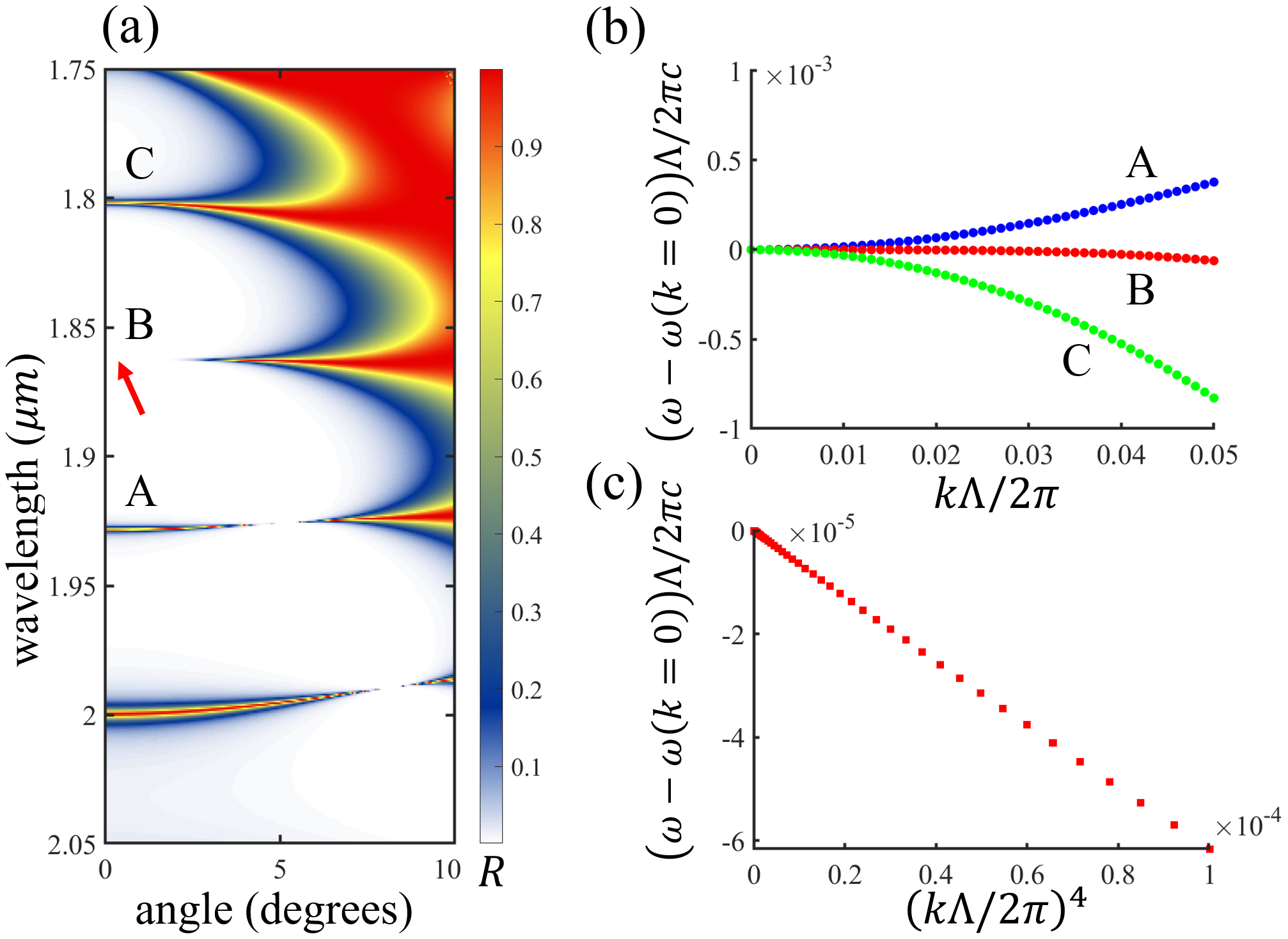}	
\caption{\label{fig:RCWA_flat_doubleBIC}(a) Angle-resolved reflectivity spectra obtained by RCWA simulations, showing a flat double BIC with zero curvature of the real eigenvalue at the $\Gamma$ point (red arrow). (b) FEM simulations of the real eigenvalues of the three branches A, B, C. The real curvature changes from positive value (A) to negative value (C) and vanishes for the intermediate plot (B). Common origin of the energy has been chosen at $\Gamma$ point for the three bands to improve readability of the curvature changes. (c) The real part of the eigenvalue of the flat double BIC is zoomed around the $\Gamma$ point: it shows $k^4$ variation, thus confirming a zero curvature. The optical thickness of the spacer $D$ is about 13$\lambda$.The other parameters are $\Lambda=1\ \mu m,\ h=0.25\ \mu m,\ L=0.3\ \mu m$ and $D=16.4423\ \mu m$.}
\end{center}
\end{figure}

Real dispersion characteristic of double transverse BIC can be made very flat if the condition $C_R=0$ is met. This is shown to be achievable in the supplemental appendix (section \ref{sec:doubleBIC}), provided that $L_{opt}$ is significantly larger than $\lambda$, that is for relatively thick structures. As shown in Fig.~\ref{fig:RCWA_flat_doubleBIC}, RCWA simulation demonstrates indeed that very flat transverse BIC with zero curvature can be achieved provided that the thickness of the structure exceeds a few times $\lambda$. The double transverse BIC with flat band (zero curvature at the $\Gamma$ point ) is observed for a rather thick structure: the gratings are separated by a $13\lambda$ thick spacer.

In that case, as $L_{opt}$ varies rapidly with the wavelength, solving equations \eqref{eq:omega} and \eqref{eq:psi} for the sole even modes gives rise to more than 2 solutions, as this is confirmed by simulations in Fig.~\ref{fig:RCWA_flat_doubleBIC}. All these modes are quasi-Fabry-Pérot modes where the photons share their life time between the guided state in the corrugated membranes (propagation in the lateral direction, see Eq.~\eqref{eq:lifetime_wg}) which act as resonant reflectors, and the radiated state in the spacer (propagation in the transverse direction, see Eq.~\eqref{eq:lifetime_rad}).

As recalled in the introduction, achieving flat band (or zero band curvature) conditions of the real dispersion characteristic provides slow light of zero group velocity with high density of state (DOS) for a broad range of the Brillouin zone, which is an attractive feature for a variety of applications. We propose here a specific approach for that purpose, based on the interplay between radiated and guided hybridized components of a double transverse BIC.\\

The imaginary part $\omega_I$ of the complex dispersion characteristics of the transverse BIC close to the $\Gamma$ point can also be obtained from expansion of Eqs.~\eqref{eq:omega}-\eqref{eq:psi} where $\eta=1$. The general expression of the imaginary part $C_I$ of the complex curvature given by equation \ref{eq:curv_I} of appendix \ref{sec:expansion_complex}, for the even bright branch, indicates that $C_I$ is null when $\psi_0=\pi$, that is the case of a double transverse BIC. Expansion of $\omega_I$ is therefore required to be extended up to the order 4. This means that $\omega_I$, which is null at the $\Gamma$ point, increases like $k^4$. In other word the imaginary part of the complex dispersion characteristics of the double transverse BIC (belonging to the bright branch) is \textit{flat} close to the $\Gamma$ point. This is confirmed by FEM simulations (see Fig.~\ref{fig:double_transverse_BIC}), showing the quality factor of the double transverse BIC to vary like $k^{-4}$ in the vicinity of the $\Gamma$ point.

This result is consistent with our interpretation of the transverse double BIC at the $\Gamma$ point as resulting from the merging of two transverse BICs occurring off the $\Gamma$ point at $k=\pm k_{BIC}$, when $k_{BIC}$ tends to zero momentum. We remind that the imaginary part of the dispersion characteristics varies like $(k-k_{BIC})^2$ for the two transverse BICs with $\pm k_{BIC}$ momentum. For the merging BIC design, that is $k_{BIC}=0$, the imaginary part varies like $(k-k_{BIC})^2(k-k_{BIC})^2\sim k^4$ (see appendix \ref{sec:expansion_imaginary} and equation \eqref{eq:B12} for complementary analysis). This transverse BIC merging scenario, derived from analytical model, is in line with recent interpretation of radiative losses of BICs based on their topological nature \cite{jin2019topological, kang2021merging, Jin2019}.

\subsubsection{\label{sssec:TripleBIC} Triple BIC}
At the $\Gamma$ point of the dark branch, the mode is systematically dark since it is protected by the lateral symmetry. If, in addition, conditions are met for the formation of a double transverse BIC, the eigenmode turns to be doubly protected. We call triple BIC the mode generated in these particular conditions. This specific accidental circumstance corresponds to the merging of two ordinary transverse BICs belonging to the dark branch and occurring at $\pm k_{BIC}$ vectors, when $k_{BIC}$ approaches zero, with the lateral BIC occurring systematically at the $\Gamma$ point. Close to the $\Gamma$ point (small $k$), it is also possible to derive the complex dispersion characteristic of the triple BIC from expansion of Eqs.~\eqref{eq:omega}-\eqref{eq:psi}.\\

The expanded real part $\omega_R$ of the complex dispersion characteristics close to the $\Gamma$ point is given by Eqs.~\eqref{eq:S24} and \eqref{eq:S25} of the appendix \ref{sec:expansion_complex}, for $\psi(k=0)\equiv\psi_0=\pi$ (mod $2\pi)$:

\begin{equation}\label{eq:real_mode_tripleBIC}
    \omega_{R}(k) = \omega_{R_0} -\frac{v^2k^2}{2(K+\kappa_c)},
\end{equation}

where, for the dark branch, $\omega_{R_0}=\omega_0 + \kappa_d -K-\kappa_c$.\\

We may note the fully wave-guiding nature of the real dispersion characteristics, which expresses the efficient protection of the triple BIC from the surrounding continuum. This is clearly unlike the case of the double transverse BIC issued from the bright branch, which hybrid character, both wave-guided and radiated, has been pointed out before in Eq. \eqref{eq:real_curvature_doubleBIC} and  \eqref{eq:real_curvature_doubleBIC_App}. \\

\begin{figure}[hbt]
\begin{center}
	\includegraphics[width=0.45 \textwidth]{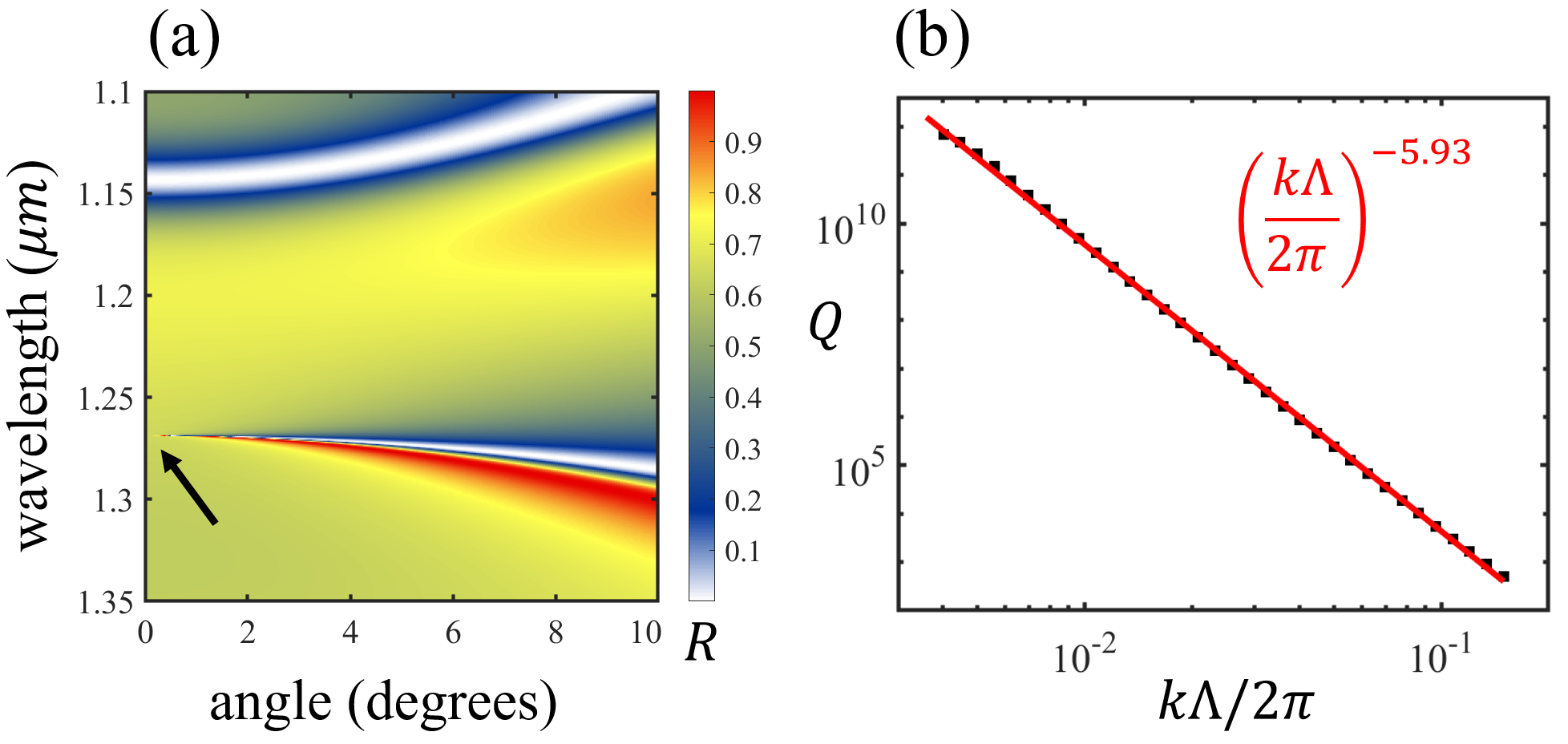}	
\caption{\label{fig:RCWA_ultraflat_tripleBIC} (a) Angle-resolved reflectivity spectra obtained by RCWA simulations, showing the occurrence of a triple BIC at the $\Gamma$ point (black arrow) of the dark branch (here dielectric). (b) log-log plot of the quality factor of the same branch versus $k$ vector in the vicinity of the $\Gamma$ point. Black squares are numerical results obtained by FEM simulations. Red line is a polynomial fit. The slope $-5.93$ of the fit confirms the results of our analytical model anticipating variations of the imaginary eigenvalues like $k^6$. Q factors larger than $10^{12}$ should not be taken into account because they lies beyond the FEM accuracy. The parameters are $\Lambda=0.55\ \mu m,\ h=0.279\ \mu m,\ L=0.275\ \mu m$ and $D=0.$}
\end{center}
\end{figure}

The imaginary part $\omega_I$ of the complex dispersion characteristics of the triple BIC close to the $\Gamma$ point can also be obtained from expansion of Eqs.~\eqref{eq:omega}-\eqref{eq:psi}. Equation \ref{eq:S26} of the appendix \ref{sec:expansion_complex} indicates that the second derivative $d^2\omega_I/dk^2$ is null when $\psi_0=\pi$ (mod $2\pi$). It can be shown that  expansion of $\omega_I$ is required to be extended up to the order 6 (see appendix \ref{sec:expansion_imaginary} for complementary analysis). This means that $\omega_I$, which is null at the $\Gamma$ point, increases like $k^6$. In other word the imaginary part of the complex dispersion characteristics of the triple BIC is \textit{ultra-flat} close to the $\Gamma$ point. This also means that the optical losses of a structure provided with a triple BIC, which are strictly null at the $\Gamma$ point, remain very low off the $\Gamma$ point, up to large lateral $k$ momentum. These predictions of the model are confirmed by RCWA and FEM simulations, illustrated in  Fig.~\ref{fig:RCWA_ultraflat_tripleBIC}. They indicate that, when a triple BIC is generated, the quality factor of the resonance retains very large values (around $10^4$) up to large momentum angle of the $\Gamma$ point ($\pm 10^{\circ})$. FEM data show the quality factor of the triple BIC structure to vary like $k^{-6}$ in the vicinity of the $\Gamma$ point. 

This result is consistent with our interpretation of the triple BIC at the $\Gamma$ point as resulting from the merging of two transverse BICs occurring off the $\Gamma$ point at $k=\pm k_{BIC}$ momentum, with a lateral BIC at the $\Gamma$ point, when $k_{BIC}$ tends to zero momentum. For the triple BIC design, that is $k_{BIC}=0$, the imaginary part varies like $(k-k_{BIC})^2(k-k_{BIC})^2k^2\sim k^6$, given that the lateral BIC imaginary part varies like $k^2$ (see complementary analysis in the appendix section \ref{sec:expansion_imaginary} and Eq.~\eqref{eq:B13}).

\subsubsection{\label{sssec:DiracTripleBIC} Dirac point at triple BIC}
The Dirac point at triple BIC is the central and most outstanding character among all distinctive photonic species described in previous sections, which  combines all the remarkable characteristics at a time. We again concentrate on the two even modes, given that conclusions are qualitatively similar for odd modes. We remind that one of the two even modes is systematically dark at the $\Gamma$ point, being lateral symmetry protected (lateral BIC), while the other is bright in general. 
Let us start with the double exceptional point (see subsection \ref{sssec:BandInversion}) which is further described in the supplemental appendix (section \ref{sec:Double_EP}). We remind that a double exceptional point is formed when conditions for full degeneracy of the complex eigenvalues of the dark and bright branches is achieved \cite{Zhen2015}. It is shown in section \ref{sec:Double_EP} of the appendix (Eqs.~\eqref{eq:kexc} and \eqref{eq:psiexc}) that the double exceptional points show up in the dispersion characteristics at the wave vector $k=k_{EP}=\pm G(1+\cos\psi(k_{EP}))/v$, with the additional condition $K+\kappa_c+G \sin\psi(k_{EP})=0$. If $K+\kappa_c$ turns to be null, $\psi(k_{EP})=\pi $ (mod $2\pi$) (for the even mode) and $k=k_{EP}=0$. The two exceptional points merge at the $\Gamma$ point. It also results obviously that $\psi_0=\psi(k=0)=\pi$ (mod $2\pi$).  Consequently, the bright mode at the $\Gamma$ point turns into a double transverse BIC, which therefore degenerate with a lateral BIC. The double transverse BIC and the lateral BIC having the same real eigenfrequency, we may conclude that the lateral BIC is also a triple BIC, as described in the previous section, since it benefits from both lateral and transverse protections (the condition $\psi_0=\pi$ (mod $2\pi$) holds also for it). While the double transverse BIC and the triple BIC are degenerated at the $\Gamma$ point, the regime of strong coupling between the two eigenmodes applies off the $\Gamma$ point, as soon as the lateral $k$ momentum differs from 0 \footnote{This is in contrast with the behaviour of a double exceptional point, where the weak coupling regime occurs first for $k$ momentum below $k_{EP}$, followed by the strong coupling regime, for $k$ momentum exceeding $k_{EP}$, where the real dispersion characteristics of the dark and bright branches are linear ($\pm vk$) and where both branches equally share the loss rate (see appendix \ref{sec:Double_EP})}. As a result, the real dispersion characteristics follow a Dirac like linear variation. We call therefore Dirac point at triple BIC this specific type of mode, since it features zero-index behaviour at the $\Gamma$ point, with very weak losses, owing to a fine interplay between transverse and lateral BICs. Dirac point at triple BIC characteristics are precisely accounted for by our analytical model, as further detailed below.\\

The expanded real part $\omega_R$ of the complex dispersion characteristics close to the $\Gamma$ point is derived from Eqs.~\eqref{eq:omega}-\eqref{eq:psi}, with $\psi=\psi_0=\pi$  (mod $2\pi$) at the $\Gamma$ point and $K+\kappa_c=0$. It yields the linear relation below:

\begin{equation}\label{eq:real_mode_DiracTripleBIC}
    \omega_{R}(k) = \omega_0+\kappa_d \pm v_Dk,
\end{equation}

where:

\begin{equation}\label{eq:velocity_DiracTripleBIC}
    v_D = \frac{v}{\sqrt{1+\frac{\tau_{rad}}{\tau_{wg}}}}.
\end{equation}

We remind that $\tau_{wg}$ and $\tau_{rad}$, given by Eqs.~\eqref{eq:lifetime_wg} and \eqref{eq:lifetime_rad} respectively, are the average lifetime of photons in the wave-guided state before being emitted into the continuum, and the radiated state during a one way trip between the two gratings. 
The dispersion characteristics include an upper Dirac branch ($-v_Dk$ when $k<0$ and $+v_Dk$ when $k>0$) and a lower Dirac branch  ($+v_Dk$ when $k<0$ and $-v_Dk$ when $k>0$); both branches degenerate at the $\Gamma$ point, where the curvature is infinite. \\

The expanded imaginary part $\omega_I$ of the complex dispersion characteristics close to the $\Gamma$ point can be again derived from Eqs.~\eqref{eq:omega}-\eqref{eq:psi}, with $\psi=\psi_0=\pi$  (mod $2\pi$) at the $\Gamma$ point and $K+\kappa_c=0$.
If $\tau_{wg}\gg\tau_{rad}$, it can be shown that for both Dirac branches, which share the same optical loss rate in strong coupling regime:

\begin{equation}\label{eq:imag_mode_DiracTripleBIC}
    \omega_{I}(k) = -\frac{1}{2} G \left( \frac{L_{opt}}{n_G} \right)^2k^2,
\end{equation}

where $n_G$ is the group index of the non-corrugated grating waveguide. This relation expresses that the wave-guided photons retain the protecting umbrella of the triple BIC, within a domain of the reciprocal space extending over $\Delta k \sim n_G/L_{opt}$, around the $\Gamma$ point. This is another manifestation of the interplay between radiated and wave-guided propagations.\\

\begin{figure}[hbt]
\begin{center}
	\includegraphics[width=0.48 \textwidth]{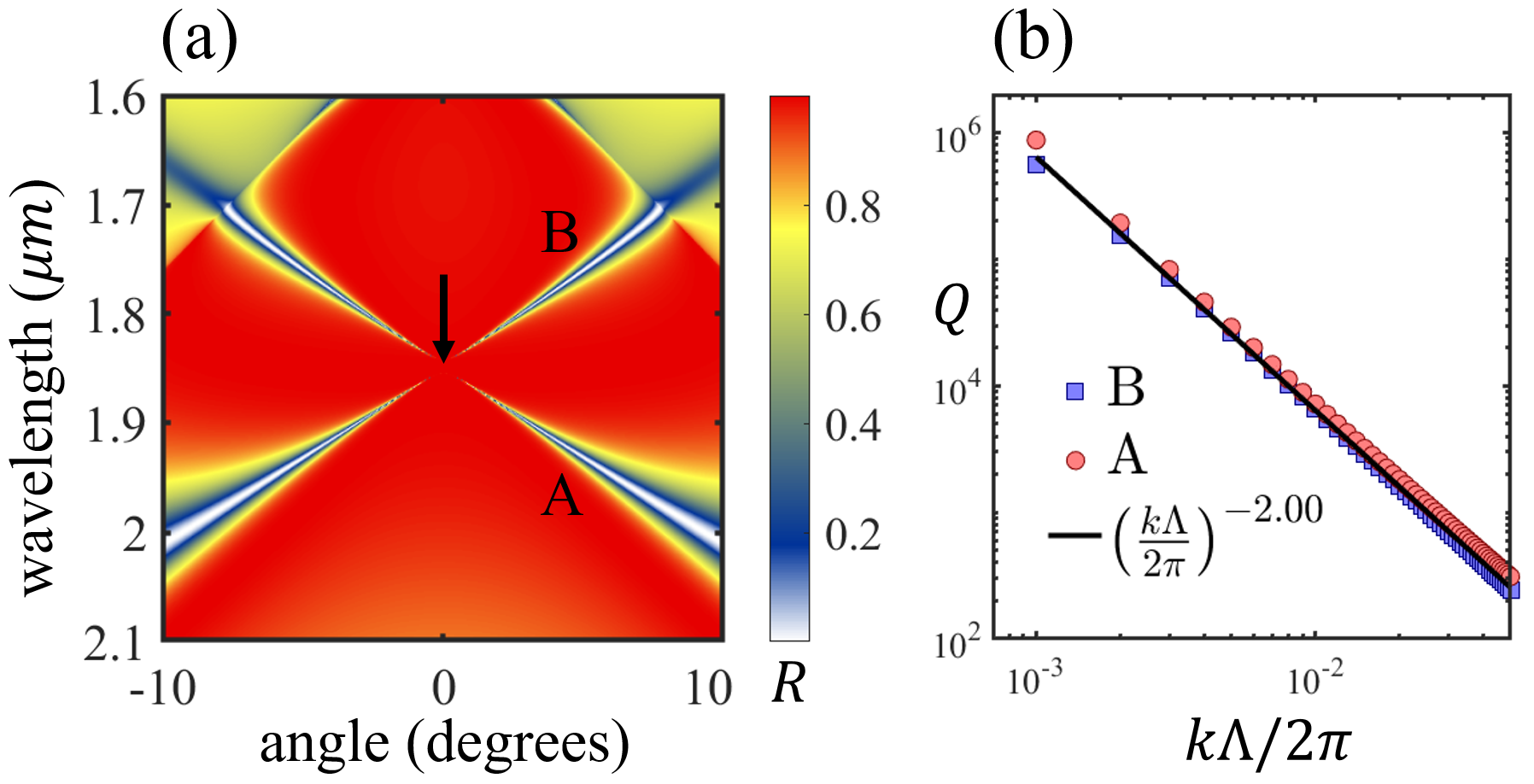}	
\caption{\label{fig:RCWA_Dirac_tripleBIC} (a) Angle-resolved reflectivity spectra obtained by RCWA simulations, showing the occurrence of a Dirac point at triple BIC at the $\Gamma$ point (black arrow). (b) log-log plot of the quality factor of the two branches A and B versus $k$ vector in the vicinity of the $\Gamma$ point. The blue squares and red circles are numerical results obtained by FEM simulations. The black line is a polynomial fit. The slope $-2$ of the fit confirms the results of our analytical model anticipating variations of the imaginary eigenvalues like $k^2$. The parameters are $\Lambda=1\ \mu m,\ h=0.3593\ \mu m,\ L=0.2735\ \mu m$ and $D=0$.}
\end{center}
\end{figure}

In summary, while the real part of dispersion characteristics for the Dirac point at triple BIC is a linear function of $k$, the imaginary part is a quadratic function of $k$, which means that complex Dirac point at triple BIC characteristics retain low optical losses, for $k$ momentum around the $\Gamma$ point. Eq.~\eqref{eq:imag_mode_DiracTripleBIC} shows that optical losses are minimized, if the optical losses of each individual grating and if the transit path of photons in free continuum are kept low. These predictions of the model are confirmed by RCWA and FEM simulations, as illustrated in Fig.~\ref{fig:RCWA_Dirac_tripleBIC}.

The practical design rules of a Dirac point at a triple BIC are given in the appendix \ref{sec:Rule_triple_BIC}.
\begin{figure*}[hbt]
	\includegraphics[width=0.85\textwidth]{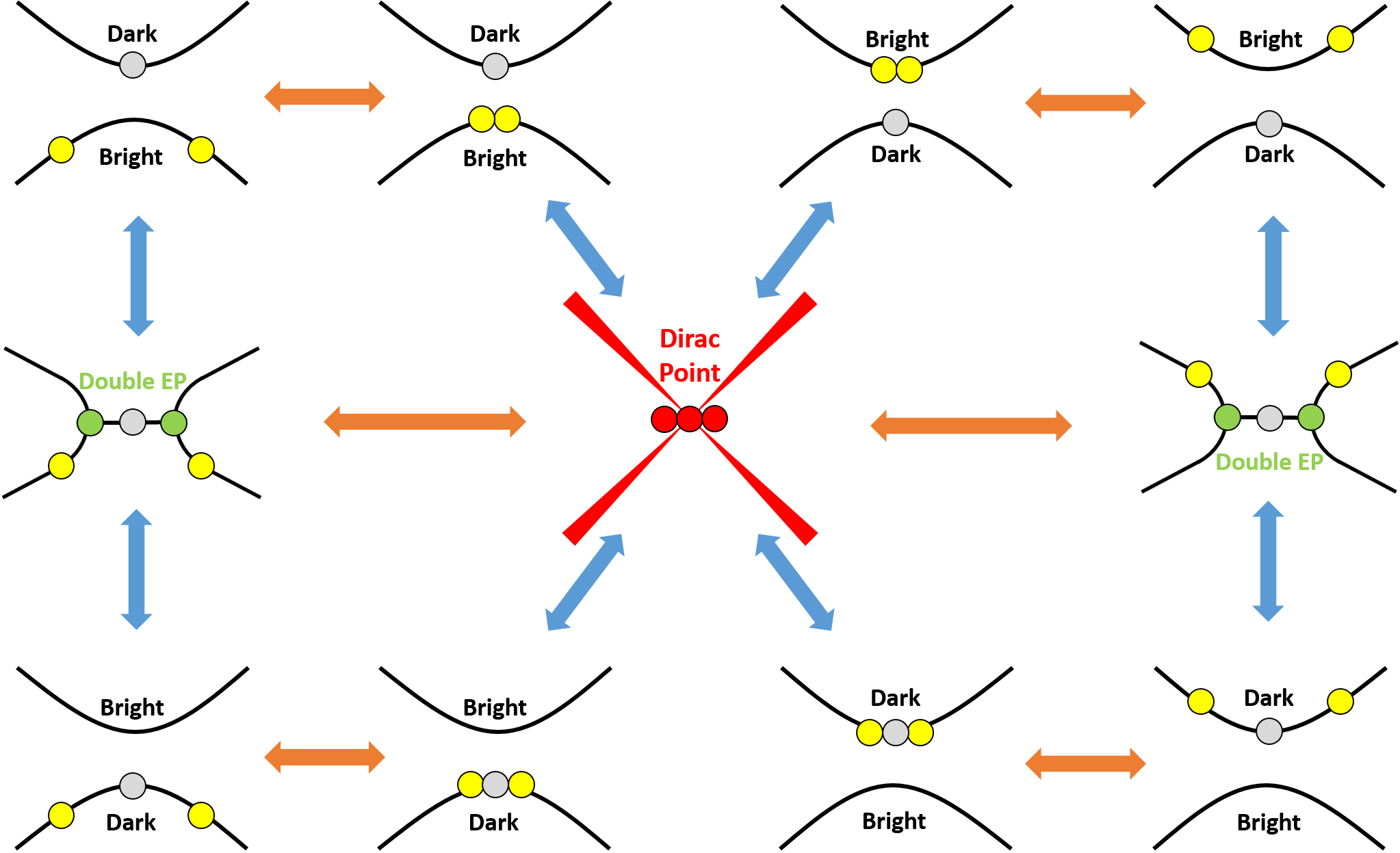}	
\caption{\label{fig:syntheticSummaryDiagram}Synthetic overview of the variety of BIC configurations, which are achievable using ordinary fully symmetrical coupled 1D PC grating structures, and the scenarii for their generation. Here \protect\scalerel*{\includegraphics{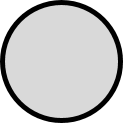}}{B} are lateral BICs, \protect\scalerel*{\includegraphics{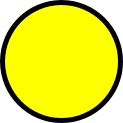}}{B} are transverse BICs, \protect\scalerel*{\includegraphics{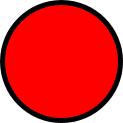}}{B} are triple BIC at Dirac point, \protect\scalerel*{\includegraphics{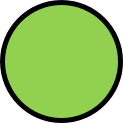}}{B} are EPs, \protect\scalerel*{\includegraphics{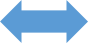}}{B} indicate  opening/closing of real gaps, \protect\scalerel*{\includegraphics{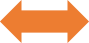}}{B} indicate merging/splitting of tranverse BICs.} 
\end{figure*}
\subsection{\label{ssec:synthethicSummary} Synthetic Summary}
The diagrams of Fig.~\ref{fig:syntheticSummaryDiagram}  thereafter summarize the principal results and conclusions of this section and of the related appendix \ref{sec:fully_symmetric}. The basic BIC building blocks are respectively the lateral BICs (represented by the dark dots) and the transverse BICs (yellow dots). The lateral BICs originate from the lateral symmetry protection, and occur systematically at the $\Gamma$ point on the dark dispersion branch (denoted as “Dark”, in the figure, against “Bright”, which stands for the bright branch, free from lateral BICs). The transverse BICs are “accidental” and may occur anywhere on both the dark and bright branches: their specific occurrences are set by specific opto-geometrical characteristics of the structure.  The green dots represent the exceptional points, where the complex eigenvalues are degenerated. The successive steps of the different scenarii are indicated by the double arrows. The orange double arrows are related to merging-splitting processes of the BICs. The yellow double dots stand for double BICs, originating from the merging at the $\Gamma$ point of two transverse BICs on the bright branch. Triple BICs (yellow-black-yellow triple dots) may also occur from the merging at the $\Gamma$ point of two transverse BICs with a lateral BIC on the dark branch: these triple BICs benefit from both lateral and transverse protections, which help them retaining very low optical losses away from the $\Gamma$ point. The blue double arrows point upon band inversion phenomena, where a dark branch transmute into a bright branch and vice versa. Transmutation event manifests itself by an intermediate state where bands coalesce and show flat dispersion characteristics. Finally, the Dirac characteristic can be viewed as the central character of the plot: it is a particular double exceptional point, where the two exceptional points merge at the $\Gamma$ point and convert into two transverse BICs, which merge in their turn at the $\Gamma$ point with a lateral BIC. In addition, the bright and dark branches degenerate at the $\Gamma$ point. In brief, the Dirac point coincides with two degenerated triple BICs, represented by three red dots. Strong coupling is triggered between the two branches of the Dirac characteristic, right at the $\Gamma$ point, as soon as the lateral $k$ vector deviates from zero, resulting in linear dispersion characteristics. At last, we may recall that the double transverse BIC at the $\Gamma$ point of the bright branch can be made flat (cancellation of the curvature); this is also the case of the lateral BIC at the $\Gamma$ point of the dark branch, in the configuration leading to the band inversion.

\section{\label{sec:latSymBreak} LATERAL SYMMETRY BREAKING: ALIGNED IDENTICAL GRATINGS WITH LATERAL BROKEN SYMMETRY}

\begin{figure}[hbt]
\begin{center}
	\includegraphics[width=0.4 \textwidth]{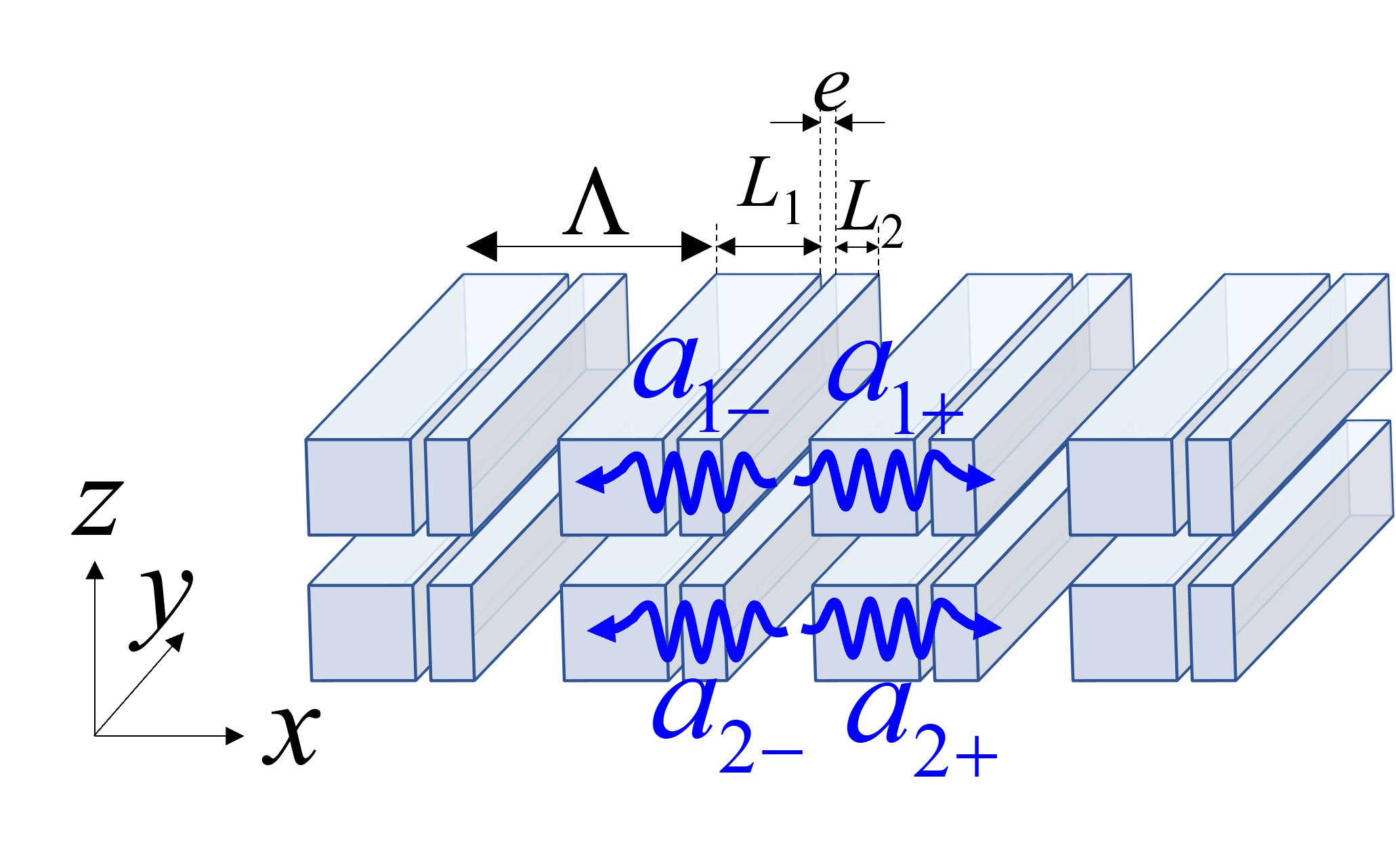}	
\caption{\label{fig:lateralSymmBreakStruct}Schematic view of the simplest coupled grating structure with broken lateral symmetry. The vertical symmetry is preserved.}
\end{center}
\end{figure}

We propose now to analyse in detail the case of a coupled grating structure which is symmetrical along transverse direction and which presents a lateral symmetry breaking (See Fig.~\ref{fig:lateralSymmBreakStruct}). We concentrate on the sole effect on dispersion characteristics of breaking the lateral symmetry of the unit cell.

\subsection{\label{ssec:H2} Analytical expression of Hamiltonian}
The general Hamiltonian is similar to that derived for the fully symmetrical structure, except for the phases $\varphi_{1+}, \varphi_{2+}, \varphi_{1-}, \varphi_{2-}$ related to the coupling step of forward (+) and backward (-) guided waves to the radiation continuum in gratings. The gratings being identical, we have $\varphi_{1\pm}=\varphi_{2\pm}=\varphi_{\pm}$, but the lateral symmetry breaking of the unit cell leads to $\varphi_{+}\ne\varphi_{-}$. As a result, the Hamiltonian can be written along the following modified version, with respect to the fully symmetrical case:

\begin{widetext}
\begin{equation}\label{eq:H2}
    H(k)=\left(\omega_0 - iG\right)\mathbb{1}_{4} +   \left(
    \begin{matrix}
    v_0k& K - iG e^{-i\varphi} &  \kappa_d - iG e^{i\psi} & \kappa_c - iG e^{i\left( \psi-\varphi \right)} \\
     K - iG e^{i\varphi} &  - v_0k & \kappa_c - iG e^{i\left( \psi + \varphi \right)} & \kappa_d - iG e^{i\psi} \\
       \kappa_d - iG e^{i\psi} & \kappa_c - iG e^{i\left( \psi-\varphi \right)}  &   v_0k & K - iG e^{-i\varphi} \\
      \kappa_c - iG e^{i\left( \psi + \varphi \right)} & \kappa_d - iG e^{i\psi}  & K - iG e^{i\varphi} &  - v_0k \\
    \end{matrix}
    \right),
\end{equation}
\end{widetext}

where $\varphi=\varphi_{+}-\varphi_{-}$. All other parameters are the same as in the fully symmetrical case.

From diagonalization of the Hamiltonian, we obtain four complex eigenvalues whose expressions are also given by Eq.~\eqref{eq:omega} (See appendix \ref{sec:broken_lateral_sym} for more details and physics), where the "complex splitting" $S\left(k\right)$ is now written: 

\begin{widetext}
\begin{equation}\label{eq:complexsplitting_lateralsymbreak}
    S(k)=2\sqrt{v_0^2k^2 + \left(K + \eta\kappa_c - iG_\psi e^{-i\varphi}\right)\left(K + \eta\kappa_c - iG_\psi e^{i\varphi}\right)}.
\end{equation}
\end{widetext}

As in the fully symmetrical case, the even and odd branches of the complex eigenvalues ignore each other, since the corresponding modes have opposite parity along the transverse direction. Therefore, possible crossing of these branches is not avoided. The analytical expression of the phase $\psi$ given by Eq.~\eqref{eq:psi} remains valid.
\subsection{\label{ssec:GenFeaturesEig2} General features of the complex eigenvalues}

In the following, we restrict the discussion to the couple of even modes ($\eta=1$ in Eq.~\eqref{eq:complexsplitting_lateralsymbreak}), given that the features of the eigenvalues are qualitatively similar for odd modes ($\eta=-1$). The expressions of the eigenvalues are similar to that applying for fully symmetrical structures, except for the factors $e^{\pm i\varphi}$, whose major impact lies in the disappearance of the lateral BICs, thus confirming the lack of lateral symmetry protection. It is observed, in general, two bright modes for each transverse opposite parities. $\varphi=0$ (mod $2\pi$) or  $\varphi=\pi$ (mod $2\pi$) stand for gratings with symmetrical unit cell along the lateral direction. Note that the eigenvalues are identical if $K+\kappa_c\ge0$ with $\varphi=0$, or if $K+\kappa_c\le0$ with $\varphi=\pi$. Therefore, in general, all situations are accounted for, if one chooses either $K+\kappa_c\ge0$ with $\varphi$ ranging in the interval $[0,\pi]$, or $K+\kappa_c$ spanning positive as well as negative values, with $\varphi$ limited to the interval $[0,\frac{\pi}{2}]$.

When $\varphi\ne0$  (or $\varphi\ne\pi$), room is solely left to the category of accidental transverse BICs. It is indeed possible to generate a transverse BIC for each of the two eigenvalues for a given transverse parity. For example, in the case of even transverse parity ($\eta=1$), this is achieved when the condition $\psi=\pi$ (mod $2\pi$) is met, for any of the two eigenvalues and for two distinct structures having specific opto-geometrical parameters (e.g. the grating thickness). The condition $\psi=\pi$ (mod $2\pi$) can be realized for any $k$ value as well as for any phase $\varphi$.
\begin{figure}[hbt]
\begin{center}
	\includegraphics[width=0.45 \textwidth]{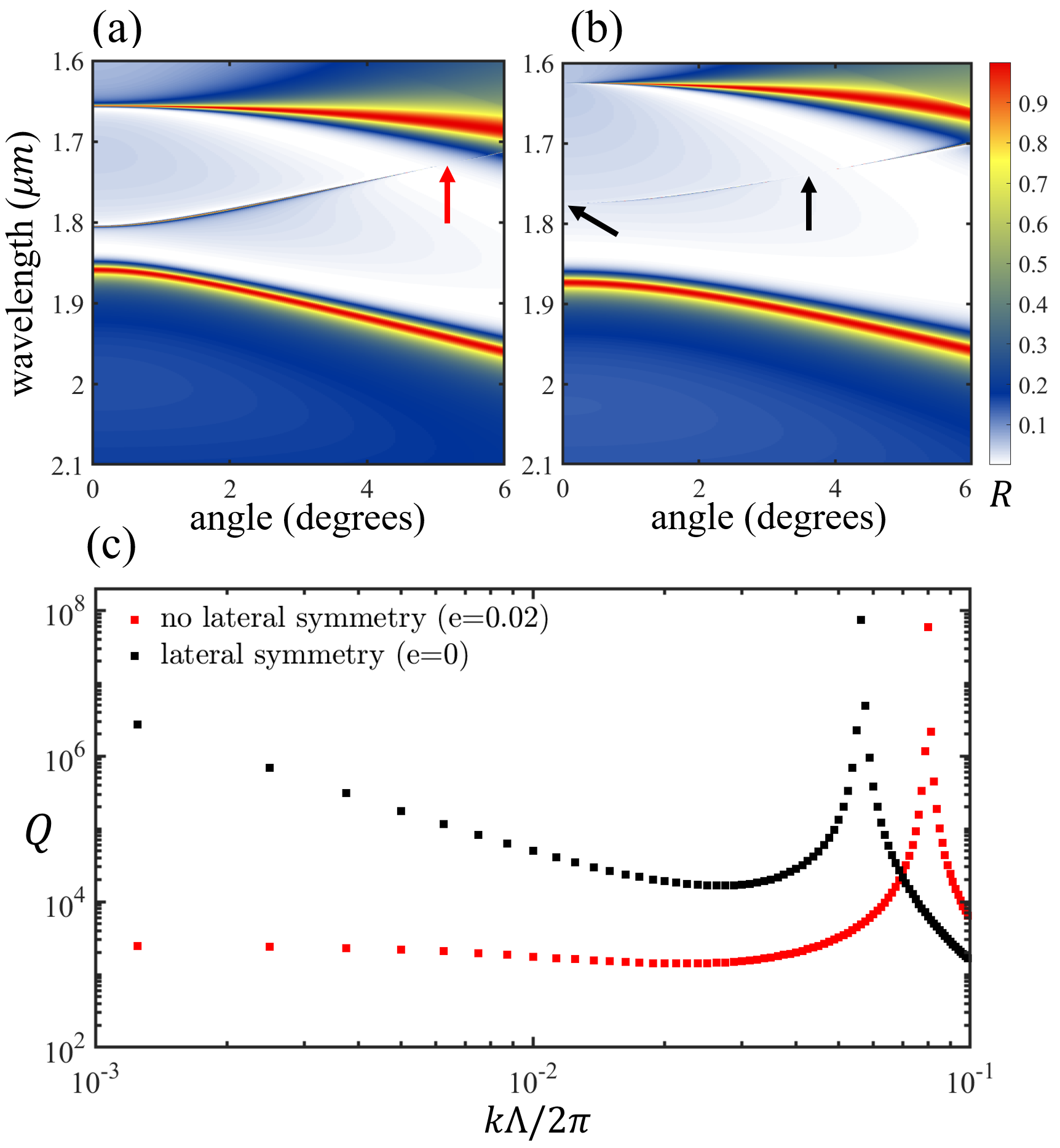}	
\caption{\label{fig:TwoLateralBICs_OneTrasverse_BICs} Angle-resolved reflectivity spectra obtained by RCWA simulations, showing (a) the generation of transverse BICs (red arrow) with on demand angular resolution achievable in lateral symmetry broken structures ($e=0.02$), where they are solely observed, lateral BICs at the $\Gamma$ point being forbidden ; (b) in structures with preserved lateral symmetry ($e=0$), transverse BICs in oblique (right black arrow) have to cohabit with unavoidable lateral BICs at the gamma point (left black arrow). (c) log-log plot of the quality factor of the corresponding branches versus $k$ vector, confirming the sole presence of a transverse BIC in the structure with broken lateral symmetry, while, in the structure with preserved lateral symmetry, both the transverse BIC and lateral BIC coexist. Here the red and black squares are numerical results obtained from FEM simulations. The others parameters are $\Lambda=1\ \mu m,\ h=0.25\ \mu m,\ L_1=0.1\ \mu m, \ L_e2=0.2\ \mu m$ and $D=0.25\ \mu m$.}
\end{center}
\end{figure}
The dispersion characteristics around the $\Gamma$ point (small $k$ values) of the two double transverse BICs are obtained from expansion of Eq.~\eqref{eq:omega}, where $S(k)$ is now given by Eq.~\eqref{eq:complexsplitting_lateralsymbreak}. The two double transverse BICs are referred to as BIC$^{+}$ and BIC$^{-}$, below. The mathematical process is the same as in the case of fully symmetrical structures, and leads to the real dispersion characteristics, in the case of even modes for example ($\eta=1$).

\begin{equation}\label{eq:doubleBICs_lateralsymbreak}
    \omega_{R}^{\pm}(k)\approx\omega_{0} + \kappa_{d} \pm K+\kappa_c + C_{R}^{\pm} \times \frac{k^2}{2},
\end{equation}

where:

\begin{equation}\label{eq:real_curv}
    C_{R}^{\pm}=\frac{\pm\tau_{wg}^{\pm}(\varphi) C_{wg} + \tau_{rad} C_{rad}} {\tau_{wg}^{\pm}(\varphi) + \tau_{rad}},
\end{equation}

\begin{equation}\label{eq:lifetime_wg2}
    \tau_{wg}^{\pm}(\varphi)=\frac{1}{G \left( 1 {\pm} \cos \varphi \right)},
\end{equation}

and where $\tau_{rad}$, $C_{wg}$ and $C_{rad}$ are given by Eqs.~\eqref{eq:lifetime_rad}, \eqref{eq:curvature_wg} and \eqref{eq:curvature_rad} respectively.

In the case of fully symmetrical structure ($\varphi=0$), we find, as expected, that the BIC$^{+}$ coincides with the double transverse BIC issued from the bright branch mode given by Eq.~\eqref{eq:real_mode_doubleBIC} and that the BIC$^{-}$ turns into a triple BIC given by Eq.~\eqref{eq:real_mode_tripleBIC}.
As to the imaginary eigenvalue $\omega_I$, which is null at the $\Gamma$ point (transverse BIC), it can be shown easily that it increases like $k^4$, for both the double transverse BIC$^{+}$ and the double transverse BIC$^{-}$ (if $\varphi\ne0$).

\subsection{\label{ssec:SpecificLateralSymBreaking} Principal asset of structures with lateral symmetry breaking: angular resolved BICs}

Breaking some of the chains imposed upon fully symmetrical structures by symmetry rules is the main asset of structures with broken lateral symmetry. For example, it is possible to get full resolution of the angular characteristics of BICs which can be formed in broken lateral symmetry structures. This is particularly the case at the $\Gamma$ point where the presence of a BIC can be decided or avoided on demand, unlike in fully symmetrical structures, where lateral BIC is systematically present. This particular aspect, accounted for by our analytical model, is illustrated in RCWA simulations showed in Fig.~\ref{fig:TwoLateralBICs_OneTrasverse_BICs} in the case of a lateral symmetry broken structure where $\Lambda=1 \mu m$, $h=0.25 \mu m$, $D=0.25 \mu m$, $L_1=0.1 \mu m$, $L_2=0.2 \mu m$ and $e=0.02$.

\section{\label{sec:transSymBreak} Transverse symmetry breaking: aligned and different symmetrical gratings}

\begin{figure}[hbt!]
\begin{center}
	\includegraphics[width=0.3 \textwidth]{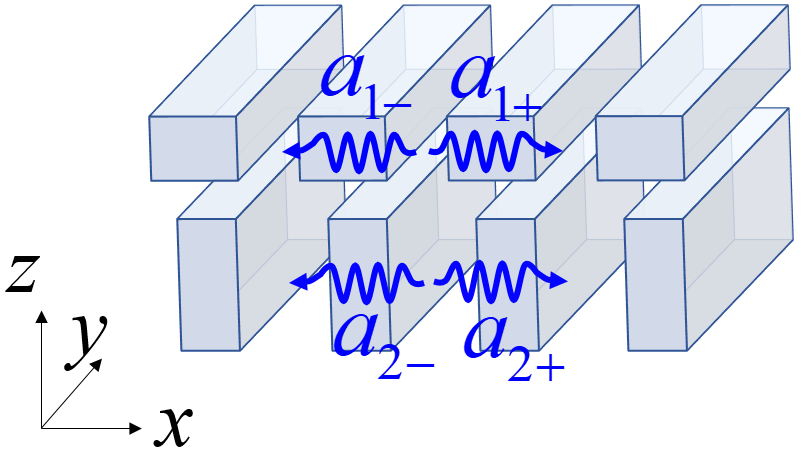}	
\caption{\label{fig:transvSymmBreak} Schematic view of the general coupled grating structure, with broken transverse symmetry. The lateral symmetry is preserved.}
\end{center}
\end{figure}

In this section, we analyse the impact of breaking the transverse symmetry of 1D PC wave-guiding structures on the complex dispersion characteristics. Fig.~\ref{fig:transvSymmBreak} shows the schematic view of a coupled grating structure, where the two gratings are aligned and different: this is the general configuration allowing for the analysis of the sole effect on dispersion characteristics of breaking the transverse symmetry.

\subsection{\label{ssec:H3} Analytical expression of Hamiltonian }

With regard to the general Hamiltonian presented in section \ref{ssec:NonhermitianH} and in appendix \ref{sec:derivation_H}, the only specific simplified characteristics of the 4×4 Hamiltonian are: $\phi=0$, since the two gratings are aligned, and $\varphi_{1,2+}=\varphi_{1,2-}$, since they are formed with laterally symmetric unit cells (no lateral symmetry breaking) and since the two gratings are different, $\varphi=\varphi_1-\varphi_2\ne0$.

Writing and diagonalizing the Hamiltonian in the base formed by base vectors $(a_{1+},a_{1-},a_{2+},a_{2-})$ results in heavy mathematical wording, which is detrimental to the physical readability. The latter is considerably improved if the Hamiltonian is written in the new base $(a_{1+}+a_{1-},a_{2+}+a_{2-},a_{1+}-a_{1-},a_{2+}-a_{2-})$, as below:

\begin{widetext}
\begin{equation}\label{eq:H3}
    H=
    \left(\begin{matrix}
		\omega_1+K_{12}-iG_{12} & K_{+}-i\sqrt{G_{12} G_{21}} e^{i(\psi-\varphi)} & v_1 k & 0 \\ 
		K_{+}^*-i\sqrt{G_{12} G_{21}} e^{i(\psi+\varphi)}  & \omega_2+K_{21}-iG_{21} & 0 & v_2 k \\
		v_1 k & 0 & \omega_1-K_{12} & K_{-}^*\\
		0 & v_2 k & K_{-} & \omega_2-K_{21}
	\end{matrix}\right),
\end{equation}
\end{widetext}

where $K_{ij}=\kappa_i+\beta_i \kappa_j$,  $G_{ij}=(\sqrt{\gamma_i}+\sqrt{\beta_i\gamma_j})^2$ and $K_{\pm}=\kappa_d \pm \kappa_c$.

\subsection{\label{ssec:GenFeaturesEig3} General features of the complex eigenvalues }

\begin{figure}[hbt]
\begin{center}
	\includegraphics[width=0.4 \textwidth]{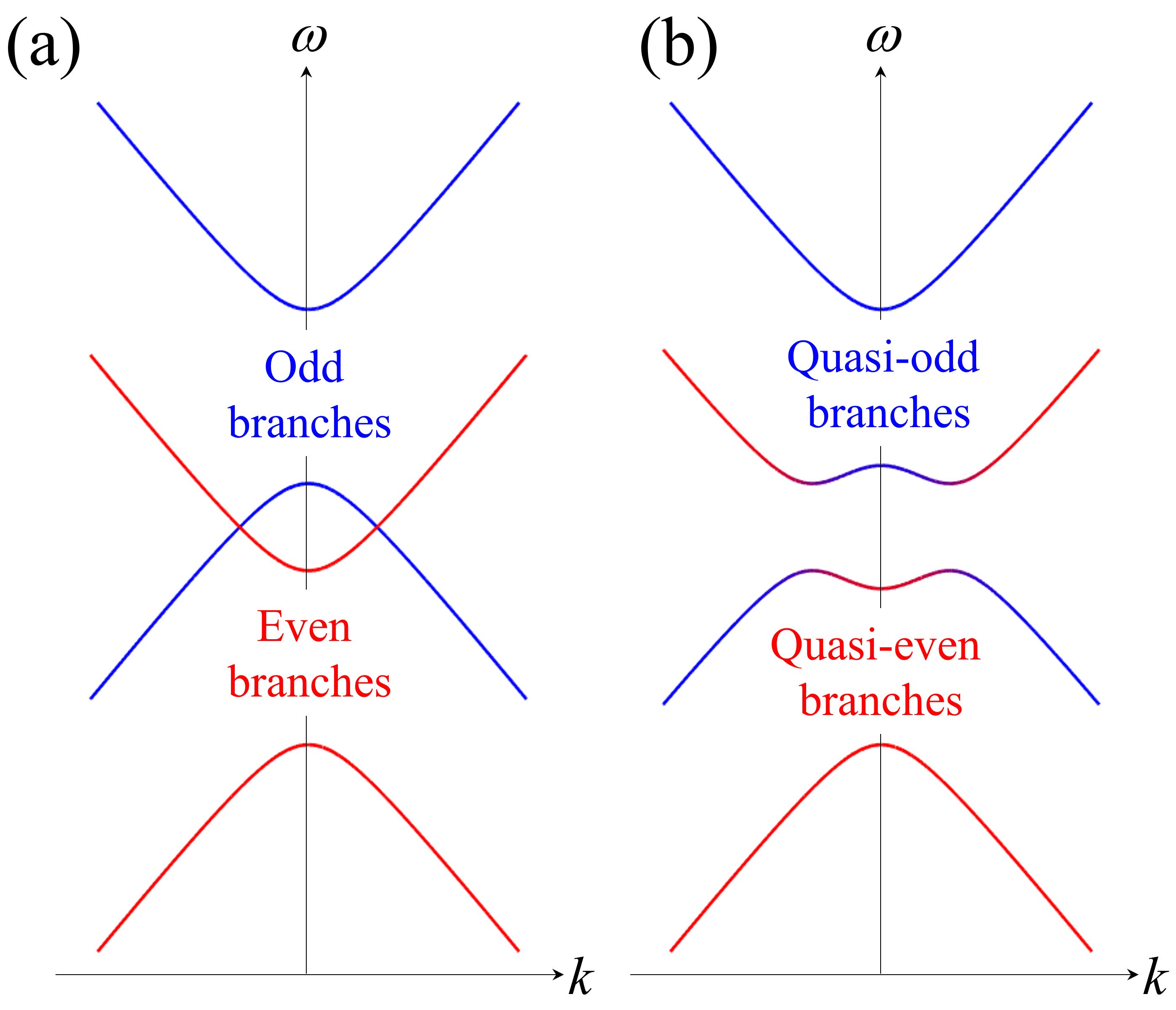}	
\caption{\label{fig:schematicRepres1} Schematic representation of real dispersion characteristics of structures with preserved transverse symmetry (a) and with broken transverse symmetry (b). For the latter, crossing of branches with opposite quasi-parities is avoided and modes are fully hybridized in the anti-crossing areas}
\end{center}
\end{figure}

As a result of broken transverse symmetry, it is no more possible to separate the four eigenvalues into two couples of eigenvalues of opposite parity along the transverse direction, which would ignore each other and whose crossing would be allowed, as in the case of symmetric structures. The 4 eigenvalues can be now classified into two couples of eigenvalues with quasi-even and quasi-odd parities at $k=0$, respectively. This is schematically illustrated in Fig.~\ref{fig:schematicRepres1}. 

\subsection{\label{ssec:SpecificPropertiesEig3} Specific properties of eigenvalues at the $\Gamma$ point}

At the $\Gamma$ point, the Hamiltonian is a diagonal block matrix, and the 4 eigenvalues can be easily determined through the diagonalization of the two 2×2 matrices being on the diagonal, resulting respectively in two eigenvalues which are systematically complex for the first and purely real for the second. The latter correspond to 2 lateral BICs which are systematically observed, as a result of the lateral symmetry, while the former correspond to bright modes which cannot turn into transverse BICs, owing to the broken transverse symmetry. It can be easily deduced that the non-existence of transverse BICs results from the finite phase $\varphi$, the two aligned gratings being different. This fact is confirmed by results of numerical simulations, which do not reveal the presence of transverse BICs, unlike the case of fully symmetrical structures where $\varphi=0$, and where transverse BICs are present for $\psi=0$ (mod $\pi$). 
Complementary physical discussion concerning the generation of quasi transverse BICs at the $\Gamma$ point is given in the appendix \ref{sec:quasi_transverse_BIC}.

\subsection{\label{ssec:PrincipalAssets3} Principal asset of structures with transverse symmetry breaking: flat BIC }

Lateral BICs can be made flat under specific conditions at the $\Gamma$ point: this category of flat BIC has not the same nature as the flat transverse double BICs observed in structures with preserved transverse symmetry and presented in sections \ref{sec:aligned_identical} and \ref{sec:latSymBreak}. For the latter, flat BICs result from the interplay between the radiated and wave-guided components of transverse BICs. In structures with broken transverse symmetry, flat lateral BICs may result from the interplay between eigenmodes with quasi-even and quasi-odd symmetries. This is unlike the case of structures with preserved transverse symmetry, where such an interplay does not exist. It is noteworthy that this category of flat lateral BIC is the counterpart of the loss-less flat band characteristics exhibited in structures with broken transverse symmetry operating under the protection of the light cone, around the first Brillouin zone boundary \cite{nguyen2018}. This category of flat BIC is also present in the ``fish-bone'' structures, as exposed in details in subsection \ref{ssec:flatBIC}.

\subsection{\label{ssec:PracticalImpl} Practical implementation of transverse symmetry breaking }

\begin{figure}[hbt]
\begin{center}
	\includegraphics[width=0.45 \textwidth]{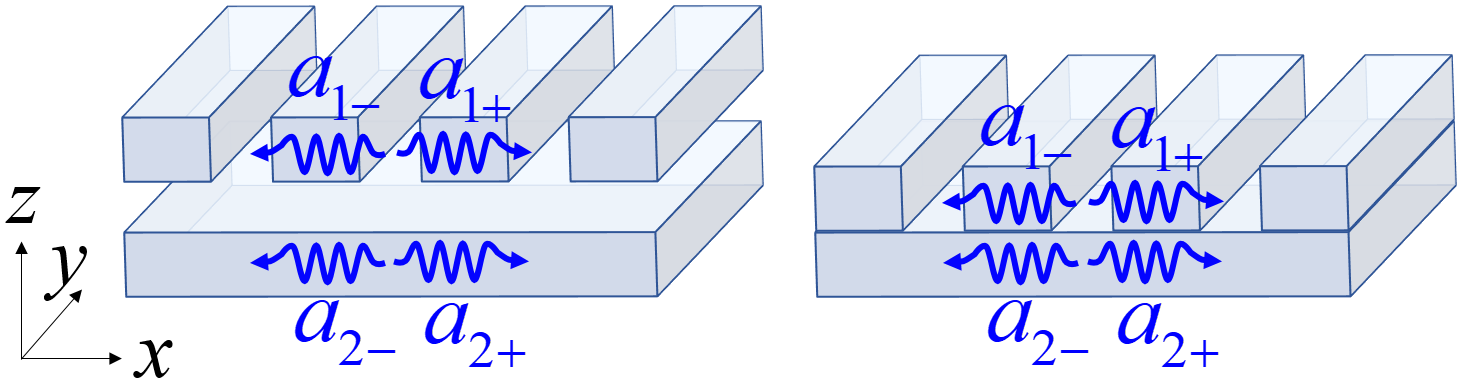}	
\caption{\label{fig:schematicViewComb} Schematic views of a “Comb” structure. The right view shows a more standard configuration, where the two superimposed “gratings” are not separated}
\end{center}
\end{figure}

A practical implementation of structures with broken transverse symmetry is schematically represented in Fig.~\ref{fig:schematicViewComb}. 

It consists in coupling a symmetrical grating with a plain membrane, which can be considered as a non-corrugated grating. The right view is representative of a more standard configuration, where the two parts of the structure are not separated. It consists of an asymmetric grating, which is viewed as two superimposed non separated symmetric gratings, the second “grating" being non corrugated. A straightforward “joy-stick" to tune the vertical symmetry breaking is the etch depth ratio $\epsilon$ (the etch depth is equal to $\epsilon h$, where $h$ is the total thickness of the grating), which may span the range 0 to 1. This configuration is very little demanding in terms of technological constraints, yet featuring properties quite similar to those presented in this section. Also, the mathematical treatment of the analytical mode is significantly simplified, since such parameters as $\kappa_2$, $\gamma_2$, in addition to the phase $\phi$, vanish.

\section{\label{sec:transLateralSymBreak} Transverse and lateral symmetry breaking: misaligned gratings}

When the two coupled gratings are misaligned, both the lateral and transverse symmetry are broken and a very wide variety of configurations can be considered and handled using the general Hamiltonian presented in the theoretical approach in section \ref{sec:TheoryApproach}. However, as noted in the introduction, the number of parameters involved for full generality is rather large and it is appropriate to implement simplified versions of the general Hamiltonian, encompassing a wide variety of specific practical cases. In this section we concentrate on a simple case of misaligned structure where the two gratings are identical and symmetrical: we call this structure a ``fish-bone'' structure. In this particular configuration, both the lateral and the transverse symmetry are broken. Fig.~\ref{fig:schematicViewFishBone} shows a schematic view of a ``fish-bone'' structure.

\subsection{\label{ssec:H4} Analytical expression of Hamiltonian }

The general $4\times4$ Hamiltonian can be written along the following simplified version:

\begin{widetext}
\begin{equation}\label{eq:H4}
    H=
    \left(\begin{matrix}
		\omega_0+vk-i\sqrt{G G^*} & K^*-iG^* & \kappa_d-i e^{i\psi}G^* & \kappa_c-i e^{i\psi}\sqrt{G G^*} 
		\\ 
		K-iG & \omega_0-vk-i\sqrt{G G^*} & \kappa_c-i e^{i\psi}\sqrt{G G^*} & \kappa_d-i e^{i\psi}G
		\\
		\kappa_d-i e^{i\psi}G & \kappa_c-i e^{i\psi}\sqrt{G G^*} & \omega_0+vk-i\sqrt{G G^*} & K-iG
		\\
		\kappa_c-i e^{i\psi}\sqrt{G G^*} & \kappa_d-i e^{i\psi}G^* & K^*-iG^* & \omega_0-vk-i\sqrt{G G^*}
		
	\end{matrix}\right).
\end{equation}
\end{widetext}

\begin{figure}[hbt]
\begin{center}
	\includegraphics[width=0.4 \textwidth]{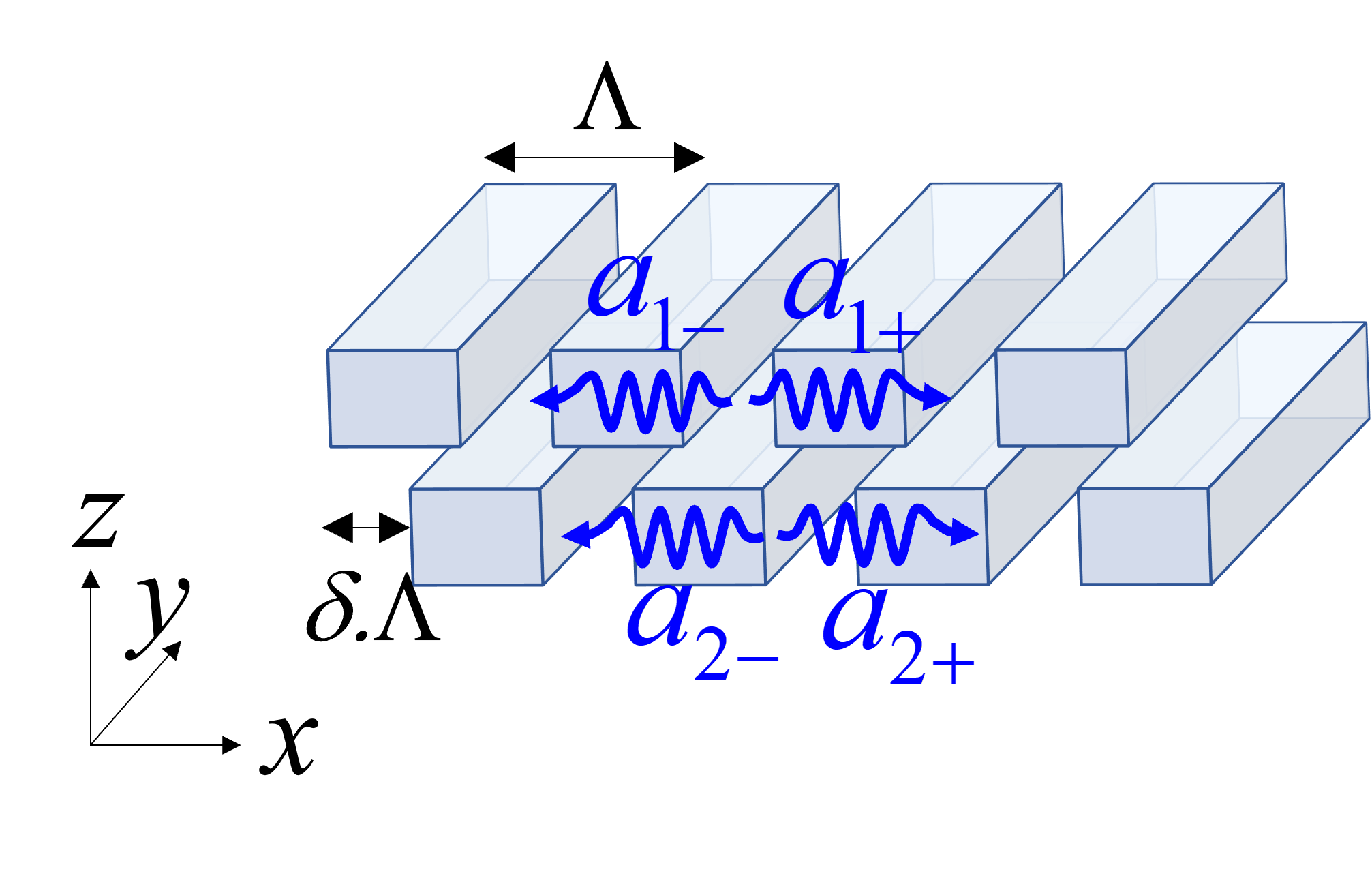}	
\caption{\label{fig:schematicViewFishBone} Schematic  view of a ``fish-bone'' structure. It consists in two identical gratings, super-imposed and misaligned along the lateral direction. The misalignment is given by the lateral off-set $\pm\delta/2\times\Lambda$ with $\delta\in[-0.5,0.5]$ of the upper and lower grating respectively. This corresponds to a relative displacement $\delta\times\Lambda$ between the two gratings.}
\end{center}
\end{figure}

The Hamiltonian is very similar to the case of a fully symmetrical structure: the only difference lies in the phase parameter $\phi$, related to the respective x-coordinate off-set $\pm\delta/2\times\Lambda$ of the two gratings (see  Fig.~\ref{fig:schematicViewFishBone}). One may show that this phase parameter is given by:
\begin{equation}\label{eq:phi_delta}
\phi=2\pi\delta\,(mod\,2\pi)
\end{equation}

It results that the coupling parameters $G$ and $K$ are complex in general:
\begin{equation}\label{eq:sqrt_G}
    \sqrt{G}=\sqrt{\gamma} \left( e^{i\phi/2} +\sqrt{\beta} e^{-i\phi/2} \right),
\end{equation}

\begin{equation}\label{eq:sqrt_K}
    K=\kappa \left( e^{i\phi} +\beta e^{-i\phi} \right).
\end{equation}

The misalignment, characterized by the relative displacement $\delta\times\Lambda$ of the two gratings results in breaking both the transverse and lateral symmetry of the structure, except for two  cases: $\delta=0$ (perfect alignment) and $\delta=\pm 0.5$ (half period misalignment). As discussed in details in the appendix section \ref{sec:symmetry_properties}, one may show that both lateral and transverse symmetries of eigenmodes are preserved in these two cases (the lateral symmetry requires in plus an operation at the $\Gamma$ point). Thus the aligned gratings and half period misalignment are equivalent in term of symmetry for the complex eigenmodes. Therefore transverse BICs and lateral BICs can be obtained in both configurations.  
	
The equivalence of the two cases $\delta=0$ and  $\delta=\pm 0.5$ is also found in the resolution of complex eigenvalues. Indeed,  these two cases correspond to the value of phase parameter  $\phi=0$ (mod $2\pi$) and $\phi=\pi$ (mod $2\pi$) respectively. This leads to real values of coupling parameters $G$ and $K$ given by Eqs.\,\eqref{eq:sqrt_G},\eqref{eq:sqrt_K}. Thus, the Hamiltonian given by Eq.\,\eqref{eq:H4} is simplified into the one given by Eq.\,\eqref{eq:H1}. In other words, the two configurations share the same Hamiltonian description of aligned gratings. Also, as noted in the appendix section \ref{sec:Fishbone1} [eq. \eqref{eq:E1}], the formal compact expressions of the Hamiltonian for aligned gratings and half period misalignment are identical. This means that for these two limit cases, the properties of the eigenvalues are formally identical. For example, when $\phi=\pi$ (mod $2\pi$), in the same way as when $\phi=0$ (mod $2\pi$), the four eigenvalues consists also in two couples of fundamental and excited branches, which ignore each other, respectively, and whose crossing is not avoided (see appendix section \ref{sec:Fishbone1}, Eqs\,\eqref{eq:E2},\eqref{eq:E3}). Consequently, the general as well as specific features of the fully symmetrical structures described in sections \ref{ssec:GenFeaturesEig1} and \ref{ssec:SymmetricalCases}, apply to the case of structures with half period lateral off-set.\\

In structures with arbitrary misalignment, physical readability and exploitation of the Hamiltonian is significantly improved if one choose the alternative base ($a_{1+} + a_{2-} , a_{2+} + a_{1-} , a_{1+} - a_{2-} , a_{2+} - a_{1-}$). In this base, the Hamiltonian can be written as below:

\begin{widetext}
\begin{equation}\label{eq:H5}
    H=
    \left(\begin{matrix}
		\omega_0 + \kappa_c - i\sqrt{GG^*} (1+e^{i\psi}) & \kappa_d + K^{*} - i G^* (1+e^{i\psi}) & v k & 0 
		\\
		\kappa_d + K - i G (1+e^{i\psi})  & \omega_0 + \kappa_c - i\sqrt{GG^*} (1+e^{i\psi}) & 0 & v k
		\\
		v k & 0 & \omega_0 - \kappa_c - i\sqrt{GG^*} (1-e^{i\psi}) & \kappa_d - K^* - i G^* (1-e^{i\psi})
		\\
		0 & v k & \kappa_d - K - i G(1-e^{i\psi}) & \omega_0 - \kappa_c - i\sqrt{GG^*} (1-e^{i\psi})
	\end{matrix}\right).
\end{equation}
\end{widetext}

\subsection{\label{ssec:GenFeaturesEig4} General features of the complex eigenvalues }

\begin{figure*}[hbt]
	\includegraphics[width=\textwidth]{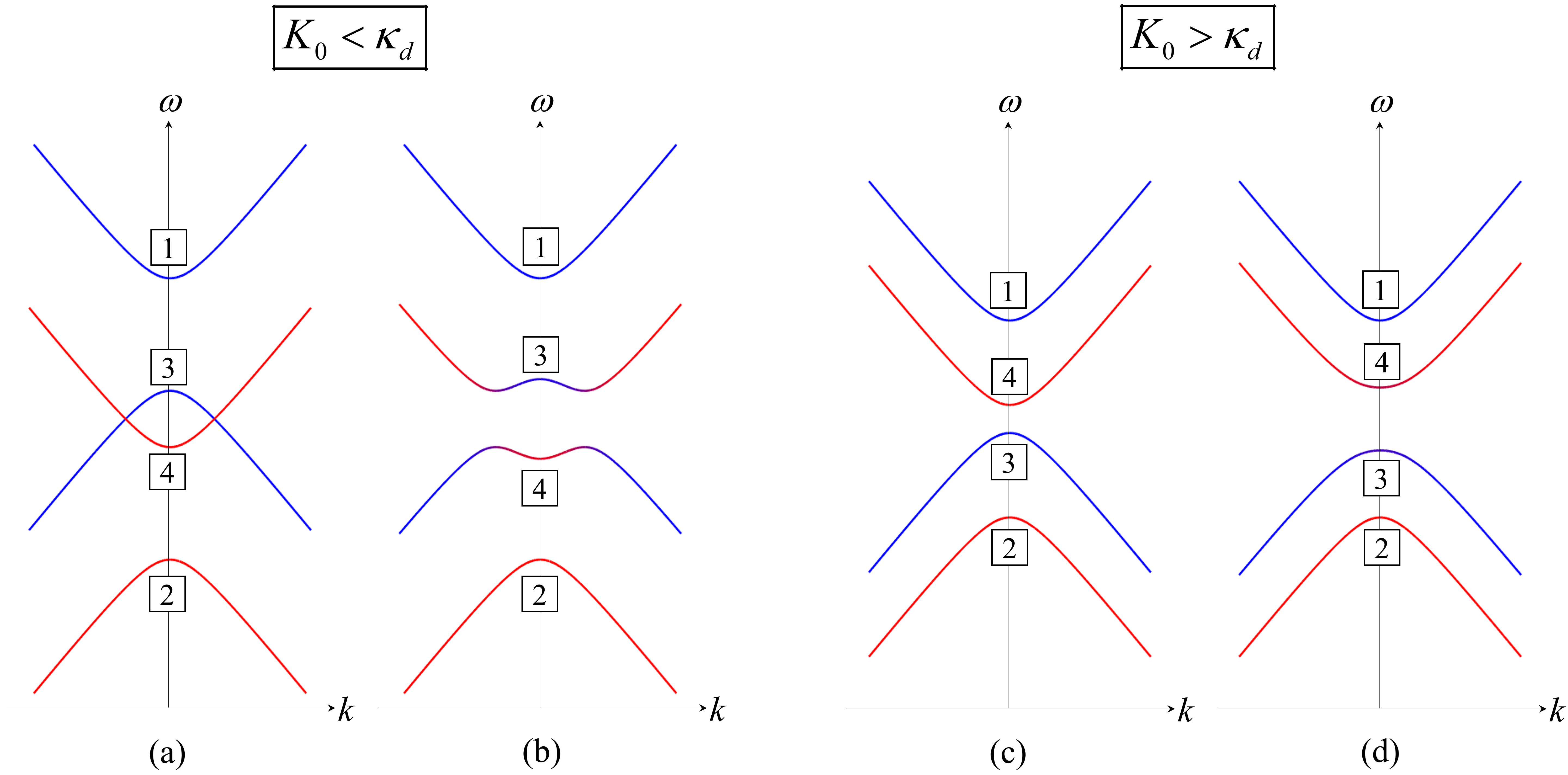}	
\caption{\label{fig:schematicView2} Schematic representation of real dispersion characteristics of aligned grating structures, i.e. where $\phi=0$ [(a), (c)], and misaligned grating structures, i.e. where $\phi\ne0$ [(b), (d)]. Blue (red) curves represent respectively odd (even) branches for the former and quasi-odd (quasi-even) branches for the latter. For the misaligned structures, the eigenvalues are fully hybridized in the area of dispersion characteristics [(b), (d)] coloured in purple. There are two possible configurations depending on whether $K_0=\kappa(1+\beta)=K(\phi=0)$ is smaller [(a), (b)] or larger [(c), (d)] than $|\kappa_d|$ (remind that $\kappa_d<0$). The numbering 1 to 4 of eigenvalues at the $\Gamma$ point corresponds to the numbering of Eqs.~\eqref{eq:omega1} to \eqref{eq:omega4}.}
\end{figure*}

As a result of broken transverse symmetry, it is no more possible to separate the four eigenvalues into two couples of eigenvalues of opposite parity along the transverse direction, which would ignore each other and whose crossing would be allowed, as in the case of symmetric structures. The 4 eigenvalues can be now classified into two couples of eigenvalues with quasi-even and quasi-odd parities, respectively. This is schematically illustrated in Fig.~\ref{fig:schematicView2}.

The two possible configurations shown in Fig.~\ref{fig:schematicView2}, depending on whether $K_0=\kappa(1+\beta)=K(\phi=0)$ is smaller [(a) and (b)] or larger [(c) and (d)] than $|\kappa_d|$, will be discussed further, later on in this paper, in connection with the topic of flat BICs at the $\Gamma$ point.

\subsection{\label{ssec:SpecificPropertiesEig4} Specific properties of eigenvalues at the $\Gamma$ point}

At the $\Gamma$ point, the Hamiltonian is a diagonal block matrix, and the 4 eigenvalues can be easily determined through the diagonalization of the 2×2 Hamiltonians, resulting in 4 eigenvalues, numbered 1 to 4 below: 

\begin{widetext}

\begin{equation}\label{eq:omega1}
    \omega_1=\omega_0-\kappa_c-i\sqrt{GG^*}(1-e^{i\psi})+\sqrt{[\kappa_d-K+iG(1-e^{i\psi})][\kappa_d-K^*+iG^*(1-e^{i\psi})]}),
\end{equation}

\begin{equation}\label{eq:omega2}
    \omega_2=\omega_0-\kappa_c-i\sqrt{GG^*}(1-e^{i\psi})-\sqrt{[\kappa_d-K+iG(1-e^{i\psi})][\kappa_d-K^*+iG^*(1-e^{i\psi})]}),
\end{equation}

\begin{equation}\label{eq:omega3}
    \omega_3=\omega_0+\kappa_c-i\sqrt{GG^*}(1+e^{i\psi})+\sqrt{[\kappa_d+K-iG(1+e^{i\psi})][\kappa_d+K^*-iG^*(1+e^{i\psi})]}),
\end{equation}

\begin{equation}\label{eq:omega4}
    \omega_4=\omega_0+\kappa_c-i\sqrt{GG^*}(1+e^{i\psi})-\sqrt{[\kappa_d+K-iG(1+e^{i\psi})][\kappa_d+K^*-iG^*(1+e^{i\psi})]}).
\end{equation}

\end{widetext}

In general, the 4 eigenvalues are complex and correspond to bright modes; in absence of lateral symmetry protection, lateral BICs are forbidden. On the other hand, for $\psi=0$ or $\psi=\pi$, that is for opto-geometrical parameters of the structures (e.g. the grating thickness and/or separation distance) specific to each of the 4 considered eigenvalues, 4 transverse BICs can be formed (one single transverse BIC at a time, for a given coupled grating structure):

\begin{equation}\label{eq:omega_psi_0}
    \omega_{1,2}(\psi=0)=\omega_{R\pm}(\psi=0)=\omega_0-\kappa_c\pm|\kappa_d-K|,
\end{equation}

\begin{equation}\label{eq:omega_psi_pi}
    \omega_{3,4}(\psi=\pi)=\omega_{R\pm}(\psi=\pi)=\omega_0+\kappa_c\pm|\kappa_d+K|.
\end{equation}

Note that for $\phi=0$ mod($2\pi$) or $\phi=\pi$  mod($2\pi$) (structure with preserved lateral symmetry), two of these transverse BICs coincide with lateral BICs, the two other being transverse double BICs.

Eqs.~\eqref{eq:omega_psi_0} and \eqref{eq:omega_psi_pi} show that the eigenfrequency values of the BICs depend on the parameter $\phi$, that is on the lateral off-set between the two gratings. Hence, if $\phi$ is varied, the effective transverse optical distance $L_{opt}$ between the grating wave-guided resonances has to be adjusted to meet the condition required for the formation of a transverse BIC ($\psi=0$ or $\psi=\pi$). In summary, a fine interplay of the lateral and transverse shifts between the two gratings may be implemented to control the strength of the transverse eigenresonance.

At this stage, it must be emphasized that the analytical model does not anticipate the presence of transverse BICs in ``fish-bone'' structures off the $\Gamma$ point in such a formal and straightforward way as in the case of aligned or half period shifted gratings. This is confirmed by RCWA simulations, which show the presence of transverse BICs solely at the $\Gamma$ point, when $\phi\neq 0,\pi$ mod($2\pi$) . Interestingly, since the transverse symmetry is broken, this transverse BIC is not inherent to the transverse symmetry but another symmetry called ``reversal symmetry'' which is preserved for any misalignment but requires operation at the $\Gamma$ point. This symmetry is fully described in the Appendix section\,\ref{sec:symmetry_properties}. Here the accidental BIC at the $\Gamma$ point goes by the name transverse symmetry only because its formation relies on the value of the transverse phase shift $\psi$. As a result, BICs numbered 1 to 4 are single transverse BICs and not double transverse BICs as in aligned grating structures or with half-period offset: they do not result, indeed, from the merging of two transverse oblique BICs. In summary, fish-bone structures of arbitrary misalignment can accommodate in general only single transverse BICs at the $\Gamma$ point, with lateral BICs at the $\Gamma$ point and transverse BICs off the $\Gamma$ point being excluded by lateral symmetry breaking and transverse symmetry breaking respectively.

As a final comment about the behaviour of eigenvalues at the $\Gamma$ point, we draw the attention of the reader to the remarks below:
\begin{itemize}
    \item Since, in general, that is for $\phi\ne 0,\pi$ mod($2\pi$) , one single BIC belonging to the sole category of transverse BICs can be formed, it results that the generation of triple BIC is prohibited, in absence of lateral BICs. We remind indeed that they may occur only in aligned ($\phi=0$ mod($2\pi$) or half-period shifted ($\phi=\pi$ mod($2\pi$) gratings, where lateral and transverse (or reversal) symmetry protection is made possible. 

    \item The generation of Dirac point at triple BICs is also prohibited, in general. They can be formed only in aligned [$\phi=0$ mod($2\pi$)] or half-period shifted [$\phi=\pi$ mod($2\pi$)] gratings, when degeneracy of a double transverse BIC and a triple BIC is achieved.
    
\end{itemize}
\begin{figure}[hbt]
\begin{center}
	\includegraphics[width=0.5 \textwidth]{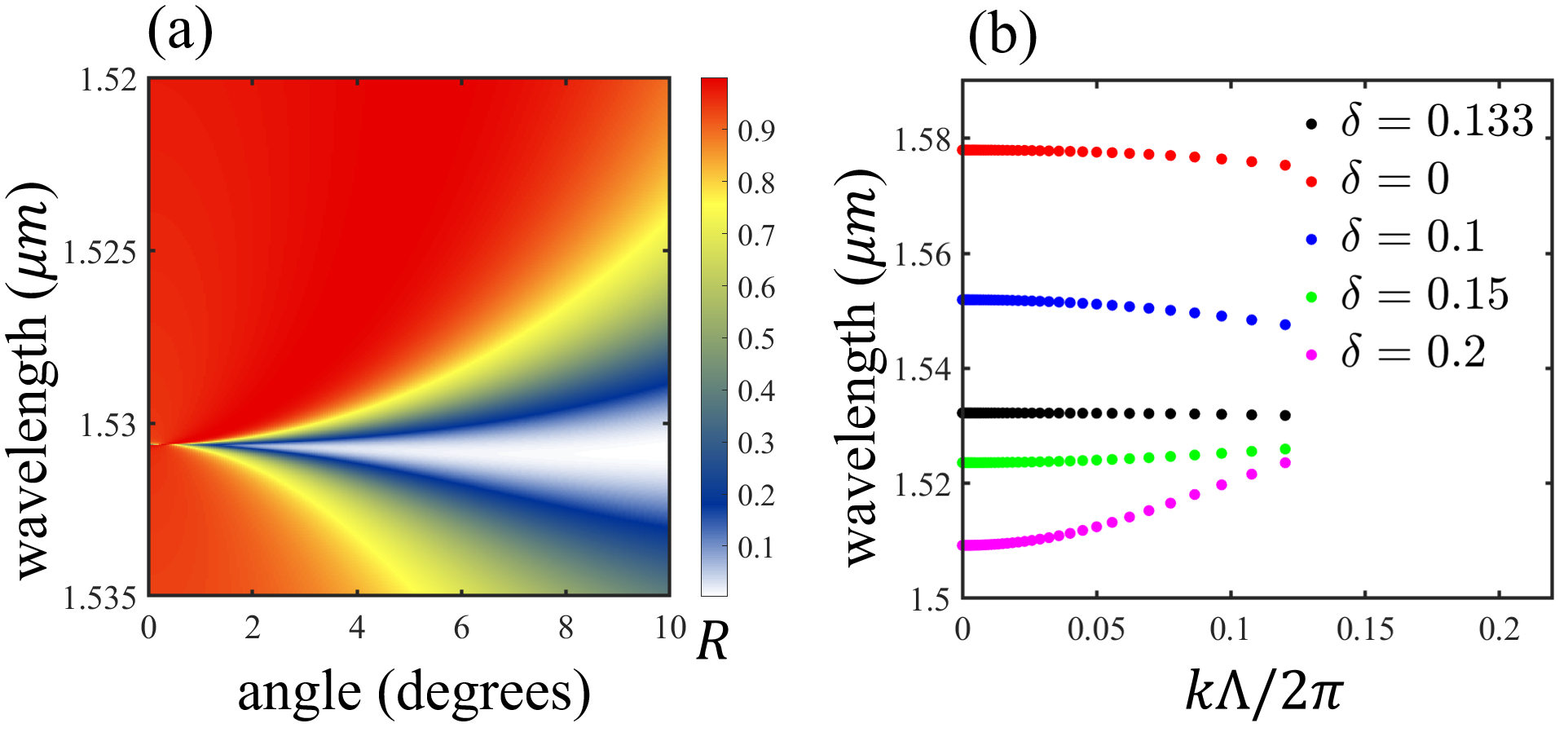}	
\caption{\label{fig:RCWA_flat_BIC} (a) Angle-resolved reflectivity spectra obtained by RCWA simulations, showing a flat BIC for a lateral offset $\delta=0.133$, with zero curvature of the real eigenvalue at the $\Gamma$ point. The protocol adopted for the generation of this flat BIC is based on the hybridization of eigenmodes of opposite parities, as a result of transverse symmetry breaking. The curvature of the dispersion characteristic is efficiently controlled by the lateral off-set $\delta=\phi/2\pi$ of the two gratings: it vanishes for $K_0 \cos\phi\approx|\kappa_d|$, it is positive for $K_0 \cos\phi>|\kappa_d|$ and negative for $K_0 \cos\phi<|\kappa_d|$. (b) FEM simulations of real eigenvalues for several values of lateral  offset $\delta$. The others parameters are $\Lambda=1\ \mu m,\ h=0.25\ \mu m,\ L=0.5\ \mu m$ and $D=0$.}
\end{center}
\end{figure}
\subsection{\label{ssec:flatBIC} Flat BIC at the $\Gamma$ point}

Apart from the $\Gamma$ point, the analytical resolution of equations is, in practice, either a bit heavy in its close vicinity (small $k$ momentum), or impossible further away in the $k$ momentum space. For the latter, numerical resolution of the analytical equations is mandatory. We concentrate in the present work on analysis in the vicinity of the $\Gamma$ point. See complementary information in appendix section \ref{sec:FB2}.

For small $k$ momentum, it is possible to derive the analytical expression of the dispersion characteristics by expanding Eqs.~\eqref{eq:S53} and \eqref{eq:psi}. We concentrate specifically on the study of flat BICs, which can be considered as being among the most attractive features of ``fish-bone'' structures. We therefore concentrate on the analysis of the curvature or second derivative $C_R$ of dispersion characteristics at the $\Gamma$ point. In subsection \ref{sssec:DoubleTransverseBIC}, we have shown that it is possible to design flat double transverse BICs in fully symmetrical structures, by engineering a fine interplay between the radiated and wave-guided components of the hybrid double transverse BIC. We apply here a quite different protocol to design flat BICs at the $\Gamma$ point: it is based on the hybridization of eigenmodes of opposite parities, as a result of transverse symmetry breaking. Therefore, the relevant eigenmodes lending themselves to efficient hybridization, are related to eigenvalues numbered 3 and 4, given to their spectral proximity. The general expression of $C_R$ is a bit heavy. We give below a version of the expression of $C_R$ whose physical significance is made fully readable; it is in line with the expression of $C_R$ obtained for double transverse BICs, in fully symmetrical structures (see Eq.~\eqref{eq:real_curvature_doubleBIC}):  

\begin{equation}\label{eq:real_curv_flat_BIC}
    C_{R}=\frac{\tau_{wg} C_{wg} + \tau_{rad} C_{rad}} {\tau_{wg} + \tau_{rad}},
\end{equation}

where $\tau_{rad}$ and $C_{rad}$ are given by Eq.~\eqref{eq:lifetime_rad} and \eqref{eq:curvature_rad} respectively, and:

\begin{equation}\label{eq:lifetime_wg_flat_BIC}
    \tau_{wg}=\frac{1}{(1+\alpha)|G|},
\end{equation}

\begin{equation}\label{eq:curv_wg_flat_BIC}
    C_{wg}=\frac{v^2}{K_{eff}}.
\end{equation}

The expression of the parameters $\alpha$ and $1/K_{eff}$  are given in the appendix \ref{sec:FB2} (Eqs.~\eqref{eq:E7}  and \eqref{eq:E8}). The module of $\alpha$ is smaller than 1. The parameter $K_{eff}$ includes all the diffractive processes in action within the structure. In the limit case of aligned gratings ($\phi=0$), we have $\alpha=1$ and $K_{eff}=K+\kappa_{c}$. We remind that $\tau_{wg}$  and $\tau_{rad}$ are respectively the average lifetime of photons in the wave-guided state before being emitted into the continuum and in the radiated state during a one way trip between the two gratings, and that $C_{wg}$ and $C_{rad}$ are the guided and radiated curvatures respectively. 

The similarity of the formal expressions of transverse BIC curvatures in aligned ($\phi=0$) and misaligned ($\phi\ne0$) structures should not hide the major difference introduced by the lateral off-set $\delta$ between the two gratings. This difference lies in the guided curvature $C_{wg}$, which is heavily dependent of the parameter $\phi=2\pi\delta$  (mod $2\pi$). $C_{wg}$ can be controlled, \textit{ad libitum} and continuously, from negative to positive values by varying the lateral offset $\delta$. A detailed analysis of the parameter $1/K_{eff}$  (see Eq.~\eqref{eq:E8}) indicates that this is indeed possible provided that $K_0=\kappa (1+\beta)=K(\phi=0)$ is larger than |$\kappa_{d}$|. This condition corresponds to the dispersion characteristic configuration shown in Fig.~\ref{fig:schematicView2}, (c) and (d). In the limit of negligible losses (small $G$), it can be shown, from the expression of $1/K_{eff}$ , that the condition for the production of a flat BIC is reduced to the simple relation: $K_0\cos\phi\approx|\kappa_d|$.

In summary the lateral offset $\delta$ offers a very efficient joystick for the production of flat transverse BICs in coupled grating structures, with no request for the thickness of the coupled grating structure to be large and to exceed a few times $\lambda$, as this is the case in the design scenario specific to aligned gratings (see section \ref{sssec:DoubleTransverseBIC}). These conclusions are faithfully confirmed by the results of RCWA numerical simulations presented in Fig.~\ref{fig:RCWA_flat_BIC}.

\section{\label{sec:exp} Experimental Demonstration}

Although the present paper is essentially devoted to theoretical work, we chose however to include a proof of concept of the Dirac point at triple BIC in the  current experimental section, since it may be considered as the most achieved photonic specie based on the combination and interaction of BICs.

\begin{figure}[hbt]
\begin{center}
	\includegraphics[width=0.48 \textwidth]{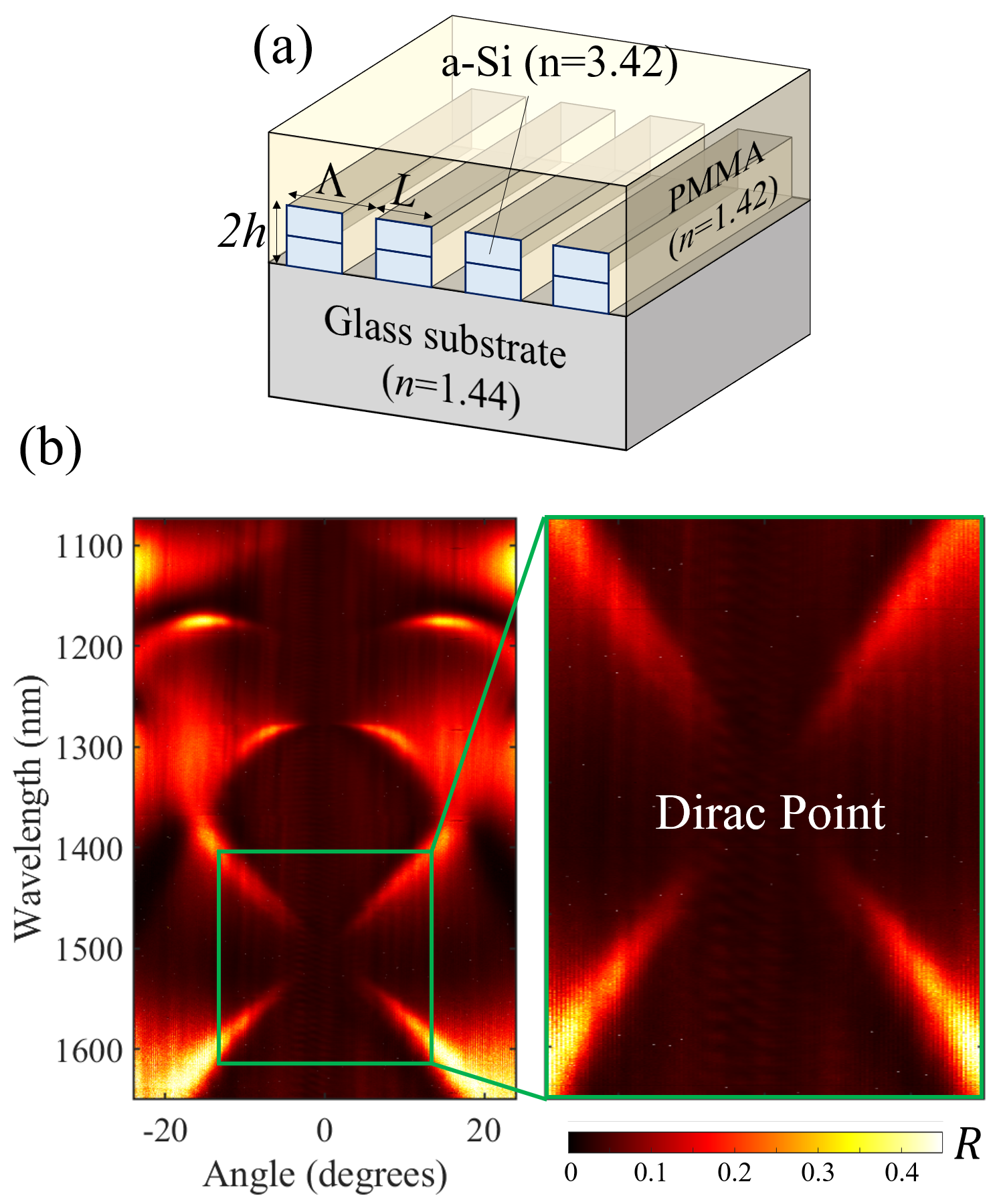}	
\caption{\label{fig:Exp_Dirac_Point_Triple_BIC}(a) Sketch of the fabricated structure, (b) Experimental angle-resolved reflectivity spectra of the sample with a close-up view of the Dirac point at triple BIC. Parameters are $\Lambda=0.83\ \mu m,\ h=0.3\ \mu m,\ L=0.23\ \mu m$ and $D=0$.}
\end{center}
\end{figure}

The fabrication of the sample includes the different steps described thereafter, starting from a commercial quartz substrate on which a 590 nm-thick amorphous silicon film is deposited by plasma-enhanced chemical vapor deposition (PECVD) using SiH$_4$ as a precursor and helium as the plasma gas. The substrate temperature is kept at 300 $^{\circ}$C and the pressure in the chamber is 2 Torr. The plasma is set by an RF signal at 25W. A 100 nm-thick hydrogen silsesquioxane (HSQ) resist is spun on the sample and baked at 80 $^{\circ}$C for 4 minutes. The resist is then exposed by electron-beam lithography and developed with a solution of TMAH. The patterns are subsequently transferred to the a-Si by inductively-coupled reactive ion etching (ICP-RIE) using a mixture of Cl$_2$ and O$_2$. To finalize the device, 828 nm of PMMA is spun directly on top of the patterns (see  Fig.~\ref{fig:Exp_Dirac_Point_Triple_BIC}(a)). 

Devices are characterized using a home-made setup that measures the angle-resolved reflectivity. A broadband white light source (Halogen) is focused on the sample through a microscope objective (NA=0.42) and the reflectivity of the device is collected via the same objective. The back-focal plane of the objective is imaged using a lens, focused at the entrance slit of a spectrograph and then collected on an InGaAs camera sensor. With this configuration, we obtain a direct measurement of the experimental energy- and momentum-resolved dispersion of samples. A polarizer is placed between the Fourier lens and the focusing lens to select the measured polarization.

Figure.~\ref{fig:Exp_Dirac_Point_Triple_BIC}(b) depict the result of angle-resolved reflectivity measurement. The Dirac point at triple BIC dispersion is experimentally demonstrated around a wavelength of 1.5 $\mu m$ in TE-polarization. 

\section{\label{sec:conclusion} Prospects}

The theoretical approach for analytical modelling of complex dispersion characteristics of optical modes operating in PCs proposed in this work is, all at the same time, extremely concise and very generic. It shows an unprecedented degree of completeness and generality, owing to its capability to provide a faith-full description of the photonic properties of an unlimited number of photonic structures. It reveals in particular that apparently simple photonic structures, featuring plain fully symmetrical 1D PCs, may offer an unexpected wealth of remarkable complex dispersion characteristics. It proves, in addition, to be a powerful enabler of the design tool-box in order to predict, control and assess the impact of breaking the transverse and lateral symmetry of grating structures, on their complex dispersion characteristics. Among the variety of conceptual and practical outcomes provided by our approach, we wish to emphasize that all remarkable photonic species being generated in 1D PCs rely fully or, at least to an essential extent, on two basic building blocks so-named lateral and transverse (accidental) BICs in this paper, and explicitly unveiled by our modelling approach. It results that this theoretical analytical approach provides very efficient support to numerical simulations which, taken alone, may not be able to predict the underlying physics, when it comes to design photonic structures with remarkable desired properties. Along this design track, we selected, for experimental demonstration, such a remarkable structure as the so-named ``Dirac point at triple BIC'', which results from the merging / degeneracy of two couples of transverse BICs and one lateral BIC.  
This work opens a unique playground for both exotic Dirac and flatband physics. In particular, it will be possible to study the light delocalization when a dispersion is gradually transformed from flat dispersion (exceptional localized states and ultra-sensitive to disorders \cite{baboux2016bosonic,vicencio2015observation, mukherjee2015observation,faggiani2016lower}) to Dirac cones (very robust versus disorder effect of Anderson localization  \cite{deng2015transverse}). This gradual transformation may be naturally implemented in moir\'e lattices, where, for example, two grating structures with slightly different periods  \cite{nguyen2021magic} or two identical gratings twisted around a common axis  \cite{salakhova2021} would be superimposed and coupled to each-other with an adjustable coupling rate, thus resulting in a fish-bone like moir\'e. 

More generally, the conceptual and design approach proposed in the present work offers a promising practical route for on demand tailoring of the local density of optical states and processing of light at the nanometer scale  \cite{cueff2019}, which are paramount for applications in optoelectronic devices. At this stage, plenty of room is left for an extra wide variety of configurations, with, among other prospects, the extension to 2D PC slab, which naturally lend themselves to additional degrees of freedom in 3D manipulation of light. Indeed, extending these concept to 2D PC slab would benefit richer in-plane symmetry (reflections and rotations) as well as the possibility to harness diffractive couplings between different polarization guided modes. In terms of non-Hermitian features, one may expect more complicated scenario for merging/splitting BICs, engineering  exceptional rings/lines instead of EPs, studying high-order EPs. 

\section*{Acknowledgements}
This work was partly funded by the the IDEXLYON from Université de Lyon, Scientific Breakthrough project TORE within the Programme Investissements d’Avenir (ANR-19-IDEX-0005). 
\bibliography{main.bib}

\appendix
\onecolumngrid



	\section{Symmetry properties of bilayer structures made of two identical symmetric gratings}\label{sec:symmetry_properties}
	
	The lateral and transverse symmetries are at the heart of the Hamiltonian construction and the properties of its complex eigenmodes. In this section, we will discuss in details the symmetry operators and their interplay with the $4\times4$ Hamiltonian of two identical symmetric gratings. The two gratings can be eventually misaligned to make a ``fishbone'' structure as discussed in section \ref{sec:transLateralSymBreak} of the main text, with a lateral offset $\pm\frac{\delta}{2}\times\Lambda$ for the upper and lower grating respectively (see Fig.\,\ref{fig:schematicViewFishBone}). Note that the case of aligned symmetric gratings of section \ref{sec:aligned_identical} is a particular case of ``fishbone'' structure in which $\delta=0$. 
	
	\subsubsection{Lateral symmetry of ``fishbone''}
	Since each grating is symmetric, the lateral symmetry is dictated by the x-reflection operator $\sigma_x$ that swaps the forwards and backwards modes:
	\begin{subequations}
		\begin{align}
			a_{1\pm} &\xrightarrow{\sigma_x} a_{1\mp}\\
			a_{2\pm} &\xrightarrow{\sigma_x} a_{2\mp} \label{eq:sigma_x_action}
		\end{align}
	\end{subequations}
	In the basis formed by $(a_{1+},a_{1-},a_{2+},a_{2-})$, the corresponding matrix of the operator $\sigma_x$ is therefore given by:
	\begin{equation}\label{eq:Mx}
		M_x= \left(
		\begin{matrix}
			0 & 1 &  0 & 0 \\
			1 & 0 &  0 & 0 \\
			0 & 0 &  0 & 1 \\
			0 & 0 &  1 & 0 \\
		\end{matrix}
		\right).
	\end{equation}
	Moreover, the same operator would inverse the wavevector and the lateral off-set:
	\begin{equation}\label{eq:sigma_x_action2}
		(k,\delta)\xrightarrow{\sigma_x} (-k,-\delta),
	\end{equation}
	Thus the Hamiltonian $H(k,\delta)$ needs to satisfy the symmetry condition:
	\begin{equation}\label{eq:fishbone_lateralsym_condition}
		H(k,\delta) = M_x.H(-k,-\delta).M_x
	\end{equation}
	It is straightforward to demonstrate that the Hamiltonian in Eq.\,\eqref{eq:H4} verifies nicely the condition above.\\
	
	We now look for the sets of $(k,\delta)$ for which the eigenmodes of the system preserve the lateral symmetry. These ``high symmetry points'' require the commutation relationship $[H(k,\delta),M_x]=0$. From Eq.\,\eqref{eq:fishbone_lateralsym_condition}, this requirement is equivalent to $H(k,\delta) = H(-k,-\delta)$. Therefore a trivial configuration is with $k=0$ and $\delta=0$. In other words, the lateral symmetry is preserved at the $\Gamma$ point with perfectly aligned gratings.\\ 
	
	Interestingly, the Hamiltonian in Eq.\,\eqref{eq:H4} shows that the requirement $H(k,\delta) = H(-k,-\delta)$ is also fulfilled when $k=0$ and $\delta=\pm 0.5$, i.e. half period misalignment. This is due to the fact that the phase parameters $\phi=\pi$ (mod $2\pi$) and $\phi=-\pi$ (mod $2\pi$) are strictly equivalent. As a consequence, the lateral symmetry is also preserved at the $\Gamma$ point with gratings of half period misalignment.
	
	\subsubsection{Transverse symmetry  of ``fishbone''}
	Since the two gratings are identical, the transverse symmetry is dictated by the z-reflection operator $\sigma_z$ that swaps the co-propagating modes from different layers:
	\begin{subequations}
		\begin{align}
			a_{1\pm} &\xrightarrow{\sigma_z} a_{2\pm}\\
			a_{2\pm} &\xrightarrow{\sigma_z} a_{1\pm} \label{eq:sigma_z_action}
		\end{align}
	\end{subequations}
	In the basis formed by $(a_{1+},a_{1-},a_{2+},a_{2-})$, the corresponding matrix of the operator $\sigma_z$ is therefore given by:
	\begin{equation}\label{eq:Mz}
		M_z= \left(
		\begin{matrix}
			0 & 0 &  1 & 0 \\
			0 & 0 &  0 & 1 \\
			1 & 0 &  0 & 0 \\
			0 & 1 &  0 & 0 \\
		\end{matrix}
		\right).
	\end{equation}
	Moreover, the same operator would inverse the lateral off-set but keeping the wavevector unchanged:
	\begin{equation}\label{eq:sigma_z_action2}
		(k,\delta)\xrightarrow{\sigma_z} (k,-\delta),
	\end{equation}
	Thus the Hamiltonian $H(k,\delta)$ needs to satisfy the symmetry condition:
	\begin{equation}\label{eq:fishbone_transversesym_condition}
		H(k,\delta) = M_z.H(k,-\delta).M_z
	\end{equation}
	Again, one may easily show  that the Hamiltonian in Eq.\,\eqref{eq:H4} verifies nicely the condition above.\\
	
	In the same fashion as the lateral symmetry in previous section, we now look for the sets of $(k,\delta)$ for which the eigenmodes of the system preserve the transverse symmetry. These ``high symmetry points'' require the commutation relationship $[H(k,\delta),M_z]=0$. From Eq.\,\eqref{eq:fishbone_transversesym_condition}, this requirement is equivalent to $H(k,\delta) = H(k,-\delta)$. Following the discussion from the lateral symmetry, we deduce that there are two configurations: the first one is the trivial case is with $\delta=0$ (perfectly aligned gratings), and the second one is with $\delta=\pm 0.5$ (half period misalignment). Unlike the ``high symmetry points'' of the lateral symmetry, here both configurations are valid for any value of the wavevector $k$. 
	
	\subsubsection{Reversal symmetry  of ``fishbone''}
	Other than lateral and transverse symmetry, the group symmetry of ``fishbone'' structure also exhibits another symmetry inherited from the two previous ones. This symmetry is dictated by the operator $\sigma_r$ that is defined by executing successively the  z-reflection operator $\sigma_z$ and x-reflection operator $\sigma_x$ or vice versa. Such operator, that is called here ``reversal symmetry'',  swaps the counter-propagating modes from different layers:
	\begin{subequations}
		\begin{align}
			a_{1\pm} &\xrightarrow{\sigma_{r}} a_{2\mp}\\
			a_{2\pm} &\xrightarrow{\sigma_{r}} a_{1\mp} \label{eq:sigma_r_action}
		\end{align}
	\end{subequations}
	In the basis formed by $(a_{1+},a_{1-},a_{2+},a_{2-})$, the corresponding matrix of the operator $\sigma_{r}$ is therefore given by:
	\begin{equation}\label{eq:Mz}
		M_r= \left(
		\begin{matrix}
			0 & 0 &  0 & 1 \\
			0 & 0 &  1 & 0 \\
			0 & 1 &  0 & 0 \\
			1 & 0 &  0 & 0 \\
		\end{matrix}
		\right).
	\end{equation}
	One may verify that $M_r=M_x.M_z=M_z.M_x$. 
	 
	Interestingly, since $\sigma_x$ inverses both the wavevector and the lateral off-set while $\sigma_z$ only inverses the lateral off-set, $\sigma_{r}$ would only inverse the wavevector but keeping the lateral off-set unchanged:
	\begin{equation}\label{eq:sigma_r_action2}
		(k,\delta)\xrightarrow{\sigma_{r}} (-k,\delta),
	\end{equation}
	This action corresponds to the time reversal symmetry, thus explains the name ``reversal symmetry''.

	As a consequence, the Hamiltonian $H(k,\delta)$ needs to satisfy the symmetry condition:
	\begin{equation}\label{eq:fishbone_reversalsym_condition}
		H(k,\delta) = M_r.H(-k,\delta).M_r
	\end{equation}
	Again, one may easily show  that the Hamiltonian in Eq.\,\eqref{eq:H4} verifies nicely the condition above.\\
	
	The reversal symmetry is preserved in eigenmodes of the system if $H(k,\delta) = H(-k,\delta)$. That limits operation at the $\Gamma$ point in a similar way as the lateral symmetry. However, unlike the lateral and transverse symmetry which require $\delta=0$ or $\pm 0.5$, the reversal symmetry is preserved for all lateral off-set. Thus this symmetry is the only one that is preserved for any arbitrary misalignment. 
	
	\subsubsection{Symmetry and dispersion of eigenmodes in ``fishbone'' structures}
	As discussed in  previous subsections, the lateral and transverse symmetry are preserved in eigenmodes only for the cases of perfect alignment or half period misalignment. Moreover, the lateral symmetry requires an operation at $\Gamma$ point. When both symmetries are preserved, the four operator $H$, $M_x$, $M_z$ and $M_r$ commute one to each other and having the same four eigenmodes,  given by:
	\begin{subequations}
		\begin{align}
			\Psi_1 &= \frac{a_{1+}-a_{1-}}{2} + \frac{a_{2+}-a_{2-}}{2}
				=\left(\begin{matrix} 1 \\-1 \\ 1 \\-1\\ \end{matrix}\right) \\
			\Psi_2 &= \frac{a_{1+}-a_{1-}}{2} - \frac{a_{2+}-a_{2-}}{2} 
				=\left(\begin{matrix} 1 \\-1 \\ -1 \\1\\ \end{matrix}\right) \\
			\Psi_3 &= \frac{a_{1+}+a_{1-}}{2} + \frac{a_{2+}+a_{2-}}{2}
				=\left(\begin{matrix} 1 \\1 \\ 1 \\1\\ \end{matrix}\right) \\
			\Psi_4 &= \frac{a_{1+}+a_{1-}}{2} - \frac{a_{2+}+a_{2-}}{2} 
		     	=\left(\begin{matrix} 1 \\1 \\ -1 \\-1\\ \end{matrix}\right) 
		\end{align}	
	\end{subequations}
	Here $\Psi_{1,2}$ and $\Psi_{3,4}$ are eigenvectors of $M_x$ with corresponding eigenvalues $1$ (even mode) and $-1$ (odd mode) respectively. Therefore $\Psi_{3,4}$ are lateral BICs which are protected by the lateral symmetry. The photonic bands having  $\Psi_{1,2}$ and  $\Psi_{3,4}$ at the $\Gamma$ points are one called bright and dark branches respectively in the section\,\ref{ssec:GenFeaturesEig1}.\\
	
	Moreover,  $\Psi_{1,3}$ and $\Psi_{2,4}$ are eigenvectors of $M_z$ with corresponding eigenvalues $1$ (even mode) and $-1$ (odd mode) respectively. Since the transverse symmetry is preserved even at out of the $\Gamma$ point, the photonic modes $\tilde{\Psi}_n(k)$ ($n=1..4$) that are  $\Psi_n$  at the $\Gamma$ points preserve the transverse parity of $\Psi_n$. The eigenmodes are divided into two groups of opposite transverse symmetry and the dispersions of two modes from different groups do not avoid crossings (see Fig.\,\ref{fig:schematicRepres1}(a)). Most importantly, transverse BICs can take place at any value of wavevector $k$ as already discussed in section\,\ref{ssec:GenFeaturesEig1}.\\
	
	Finally, for an arbitrary misalignment having $\delta\neq 0,\pm 0.5$, both lateral and and transverse symmetry are broken. However, the reversal symmetry is still preserved as long as we stay at the $\Gamma$ point. Therefore, it is still possible to obtain accidental BIC at $\Gamma$ point if the transverse phase shift $\psi$ is equal to 0 or $\pi$ (mod 2$\pi$) (see discussions in section \ref{ssec:GenFeaturesEig4}). Interestingly, although this accidental BIC is inherent to the reversal symmetry and the transverse symmetry is broken, its formation is still dictacted by the transverse phase shift. Therefore we still call it by the name of transverse BIC.

\section{Complemental information on the theoretical approach} \label{sec:complemental_info}

\subsection{Discussion on the approximations of the model}\label{sec:discussion_model}
\subsubsection{Complements on the rationale and phenomenology of the model }\label{rationale}

\begin{figure*}[hbt]
	\includegraphics[width=\textwidth]{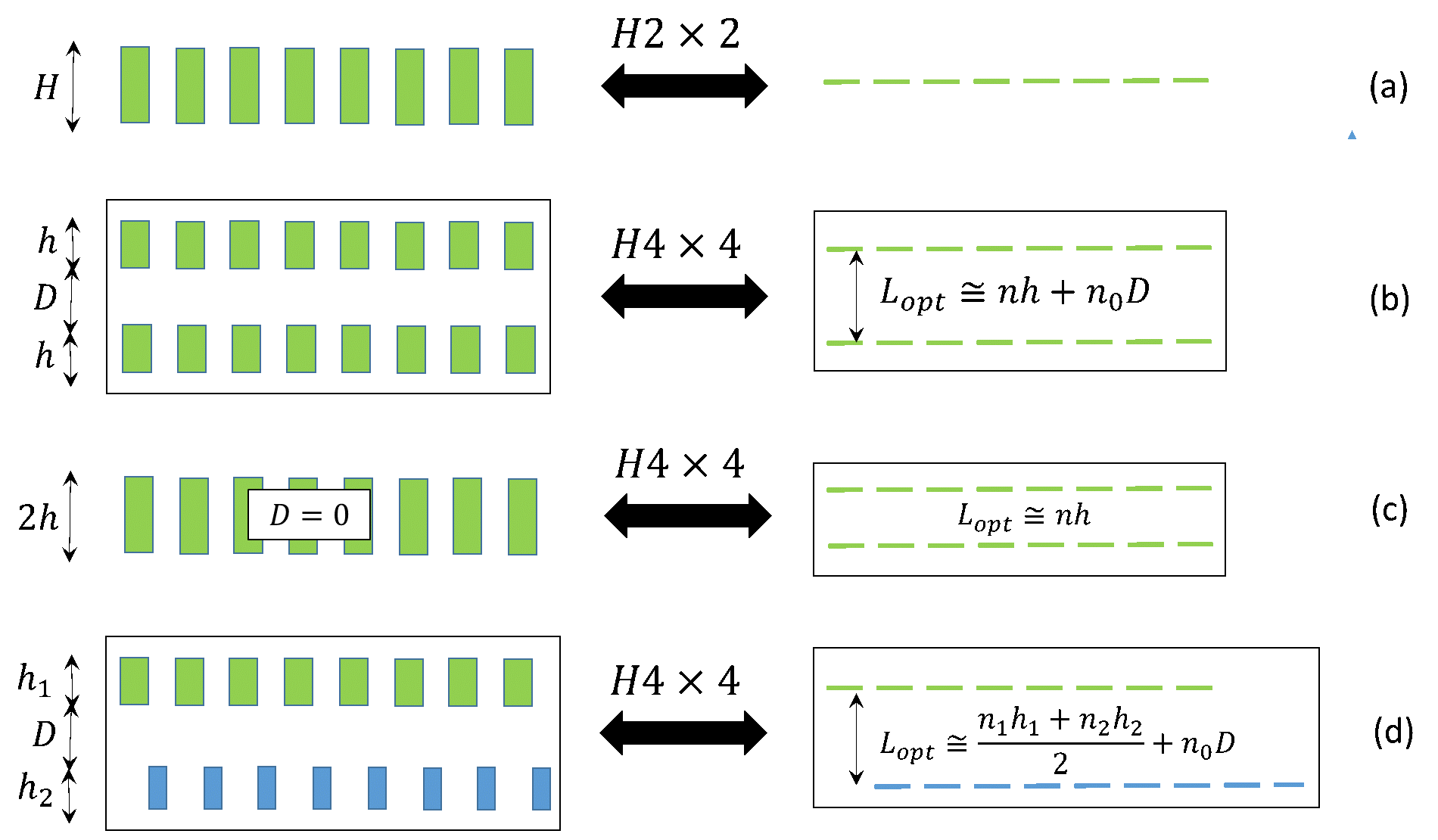}	
\caption{\label{fig:figure20} rationale of the H4x4 versus the H2x2 approach}
\end{figure*}

As recalled in section \ref{ssec:NonhermitianH}, leaky optical resonances in PC slabs are generally described as complex eigenvalues of 2×2 non-Hermitian Hamiltonian. However this H2×2 approach, although attractive in terms of simplified mathematical resolution, provides a rather partial description of the structures. In particular, it ignores explicitly the transverse dimension of the structures, as illustrated in Fig.~\ref{fig:figure20}(a): along the H2×2 approach, a 1D grating is viewed as a flat zero thickness periodic structure, able to accommodate wave-guided leaky resonances, which are, by design, restricted to be mono-modal in the transverse direction; also the H2×2 approach cannot account for events encountered by the leaky light when travelling across the grating structure along the transverse direction.

Let us now proceed one step further by considering the case of two identical and aligned h thick gratings, with are apart by the distance D, as shown in Fig.~\ref{fig:figure20}(b) left. They may communicate via the near field, as well as via the radiated field. In Fig.~\ref{fig:figure20}(b) right, it is proposed a simplified view of this coupled grating structure as two coupled flat zero thickness periodic structures, which are considered as being strictly mono-modal. Coupling of the two single mode grating structures results in the occurrence of additional modes of opposite (even and odd) parity along the transverse direction, which are the photonic expression of the transverse dimension of the structure.  It is then required to move to the H4×4 formalism proposed in this work for a proper analytical description of the real grating structure. The parameter $L_{opt}$ is the effective optical transverse distance between the wave-guided resonances accommodated by each of the two periodic structures.  $L_{opt}$ encompasses transverse optical paths across the two gratings as well as across the spacer and may be written along the following phenomenological relation: $L_{opt}=nh+n_{0}D$, where $n$($n_{0}$) is the refractive index of the effective (surrounding) medium. Our modelling approach applies to the case of two gratings in contact ($D=0$, Fig.~\ref{fig:figure20}(c)), at least for a physical trustworthy description of the physics, although the small perturbation approximation of the coupled mode formalism may not strictly apply. For example this modelling approach provides a faithful account of the variety of transverse BICs occurring in real grating structures, as fully demonstrated in the present work.

One must also emphasize that, unlike in the H2x2 approach, our model can account for the impact of breaking the lateral as well transverse symmetry of the coupled grating structure, as illustrated in Fig.~\ref{fig:figure20}(d) (different gratings of same period with arbitrary lateral offsets), and as thoroughly developed in the main text of this article.

As explained in the main text (section \ref{ssec:modelApprox}), in order to get a tractable Hamiltonian, two main approximations are made. These assumptions will be discussed in the following two subsections  and exemplified in the case of two identical gratings. 

\subsubsection{Neglecting the imaginary part of $\psi$}\label{sec:discussion_psi}
In the case of two identical gratings, the implicit equation for the complex eigenvalue $\omega$ at $\Gamma$-point for the bright branch can be written (see Eq. \eqref{eq:omega} and following):
\begin{equation}\label{eq:omega_psi1}
    \omega=\omega_{R}+i\omega_i = \omega_b-i\gamma_be^{i\psi},
\end{equation}
where $\omega_b$ ($\gamma_b$) is a complex (real) constant.
Rigorously speaking $\psi=\omega\frac{L_{opt}}{c}$, whereas we have assumed  $\psi=\omega_R\frac{L_{opt}}{c}$.
This leads to an error which can be quantified on the term:
\begin{equation}\label{eq:omega_psi2}
    |\omega-\omega_{b}| = \gamma_be^{-\psi_i}.
\end{equation}
Neglecting $\psi_i$ results in a relative error:
\begin{equation}\label{eq:omega_psi3}
    1-e^{-\psi_i}=1-e^{-\omega_i\frac{L_{opt}}{c}},
\end{equation}
which is small if:
\begin{equation}\label{eq:omega_psi4}
    \omega_i\frac{L_{opt}}{c}=\frac{\tau_{rad}}{\tau}<<1,
\end{equation}
where $\tau$ is the lifetime of the eigenmode.
In this wortk we always consider high quality factor modes, which valids this approximation.

\subsubsection{Neglecting the reflectivity of the interfaces}\label{sec:discussion_interfaces}
From first principle Coupled Mode Theory \cite{Suh2004} and taking into account the effect of the interfaces on the transverse propagation of photons, a rigorous Hamiltonian can be derived in the case of two identical gratings. Following this approach, new implicit equations are obtained for the eigenvalues. For example Eq.~\eqref{eq:omega} in the main text becomes (for $\eta$=+1):
\begin{equation}\label{eq:omega_interface}
    \omega=\omega_{R}+i\omega_{i} = \omega_0+\kappa_{d}-iG(1+\mu)\pm\sqrt{v^2k^2+\left(K+\kappa_{c}-iG(1+\mu)\right)^2},
\end{equation}
where:
\begin{equation}\label{eq:S11b}
\mu=\frac{r_m+t_m}{1-r_me^{i\theta}}e^{i\theta},
\end{equation}
with $r_m$ and $t_m$ the complex reflection and transmission coefficients of one grating at frequencies far from resonance. These coefficients can be easily expressed by considering the propagation across a plain (non patterned) slab. $\theta$ is the phase acquired by photons during their travel in the spacer layer between the 2 gratings.
The term $\mu$ accounts for the interferential process experimented by photons in the transverse stack.
Obviously, if $r_m=0$ (which is our approximation):
\begin{equation}\label{eq:S11c}
\mu=e^{i(\theta+\theta_m)}=e^{i\psi},
\end{equation}
where $\theta_m$ is the phase corresponding to the transmission across the plain slab ($t_m=|t_m|e^{i\theta_m}$). In other words, the Hamiltonian in the main text is strictly valid if the slab thickness is half-wavelength (for the eigenfrequency). On the contrary, the error is maximum if the slab thickness is quarter-wavelength.
A numerical solving of Eq.~\eqref{eq:omega_interface}, for usual double gratings, shows that this error can be significant for the imaginary part of the eigenfrequencies, while being negligible for the real one. More importantly, Eq.~\eqref{eq:omega_interface} leads to the same qualitative results which are exposed in the main text (especially $k^n$ dependencies of the dispersion). As an illustration, in the case of the Dirac point at triple BIC (section \ref{sssec:DiracTripleBIC}), simple algebra allows to get the dispersion in the vicinity of the $\gamma$-point:
\begin{equation}\label{eq:disp}
\frac{d\omega_R}{dk}=\frac{\pm v}{\sqrt{1+\frac{2G}{1-|r_m|^2}\frac{\partial\psi}{\partial\omega_R}}}.
\end{equation}
Out of resonance, the stack can be consider as a simple Fabry-Pérot cavity closed by mirrors with reflection coefficient $|r_m |^2$. The lifetime, $\tau_{rad}$, inside this cavity is therefore (see e.g.  \cite{Sauvan2005}):
\begin{equation}\label{eq:taurad}
\tau_{rad}=\frac{1}{1-|r_m|^2}\frac{\partial\psi}{\partial\omega_R},
\end{equation}
and the dispersion obtained in the main text (Eqs.~\eqref{eq:real_mode_DiracTripleBIC}  and \eqref{eq:velocity_DiracTripleBIC}) is found again:
\begin{equation}\label{eq:wr_taurad}
\frac{d\omega_R}{dk}=\frac{\pm v}{\sqrt{1+\frac{\tau_{rad}}{\tau_{wg}}}}.
\end{equation}

\subsection{Derivation of the Hamiltonian matrix elements}\label{sec:derivation_H}
As explained in section \ref{sec:TheoryApproach} (Theoretical approach), the general 4$\times$4 Hamiltonian is written:
\begin{equation}\label{eq:H_general}
      H=
    \left(\begin{matrix}
		H_{11} & H_{21}\\
		H_{12} & H_{22}
	\end{matrix}\right),
\end{equation}
where $H_{ij}$ are 2$\times$2 matrices which describe the optical interactions intra each of the two gratings $(H_{ij=i})$, and inter the 2 gratings $(H_{ij\neq i})$, with $i=1,2$.

The general expressions of $H_{ij}$ matrices are given below:
\begin{equation}\label{eq:H11}
      H_{11}=
    \left(\begin{matrix}
		\omega_1+v_1k-i\sqrt{G_1G_1^*} & K_1^*-iG_1^*e^{-i(\varphi_{1+}-\varphi_{1-})}\\
		K_1-iG_1e^{i(\varphi_{1+}-\varphi_{1-})} & \omega_1-v_1k-i\sqrt{G_1G_1^*}
	\end{matrix}\right),
\end{equation}
\begin{equation}\label{eq:H22}
      H_{22}=
    \left(\begin{matrix}
		\omega_2+v_2k-i\sqrt{G_2G_2^*} & K_2^*-iG_2^*e^{-i(\varphi_{2+}-\varphi_{2-})}\\
		K_2-iG_2e^{i(\varphi_{2+}-\varphi_{2-})} & \omega_1-v_2k-i\sqrt{G_2G_2^*}
	\end{matrix}\right),
\end{equation}
\begin{equation}\label{eq:H12}
      H_{12}=
    \left(\begin{matrix}
		\kappa_{d}-ie^{i\psi}\sqrt{G_1G_2^*}e^{i(\varphi_{1+}-\varphi_{2+})} & \kappa_{c}^*-ie^{i\psi}\sqrt{G_1^*G_2^*}e^{i(\varphi_{1-}-\varphi_{2+}})\\
		\kappa_{c}-ie^{i\psi}\sqrt{G_1G_2}e^{i(\varphi_{1+}-\varphi_{2-})} & \kappa_{d}-ie^{i\psi}\sqrt{G_1^*G_2}e^{i(\varphi_{1-}-\varphi_{2-})}
	\end{matrix}\right),
\end{equation}
\begin{equation}\label{eq:H21}
      H_{21}=
    \left(\begin{matrix}
		\kappa_{d}^*-ie^{i\psi}\sqrt{G_1^*G_2}e^{-i(\varphi_{1+}-\varphi_{2+})} & \kappa_{c}^*-ie^{i\psi}\sqrt{G_1^*G_2^*}e^{-i(\varphi_{1+}-\varphi_{2-})}\\
		\kappa_{c}-ie^{i\psi}\sqrt{G_1G_2}e^{-i(\varphi_{1-}-\varphi_{2+})} & \kappa_{d}^*-ie^{i\psi}\sqrt{G_1G_2^*}e^{-i(\varphi_{1-}-\varphi_{2-})}
	\end{matrix}\right),
\end{equation}
where:
\begin{equation}\label{eq:rac_gamma1}
    \sqrt{G_1}=\sqrt{\gamma_1}exp^{i\frac{\phi}{2}}+\sqrt{\beta_1\gamma_2}exp^{-i\frac{\phi}{2}},
\end{equation}
\begin{equation}\label{eq:rac_gamma2}
    \sqrt{G_2}=\sqrt{\gamma_2}exp^{-i\frac{\phi}{2}}+\sqrt{\beta_2\gamma_1}exp^{i\frac{\phi}{2}},
\end{equation}
\begin{equation}\label{eq:kappa1}
    K_{1}=\kappa_1e^{i\phi}+\beta_1\kappa_2e^{-i\phi},
\end{equation}
\begin{equation}\label{eq:kappa2}
    K_{2}=\kappa_2e^{-i\phi}+\beta_2\kappa_1e^{i\phi}.
\end{equation}
An approximate expression of $\kappa_{c}$ may be derived from classical coupling rate evaluation based on the spatial overlap of the field distribution within the two coupled gratings.
\begin{equation}\label{eq:kappa_BW}
    \kappa_{c}=\sqrt{\beta_1}\kappa_2e^{-i\phi}+\sqrt{\beta_2}\kappa_1e^{i\phi}.
\end{equation}
This relation indicates that $\kappa_{c}$ is real when the two gratings are aligned ($\phi=0$), or for any lateral off-set if they are identical.
We give below the clues for establishing the expressions of the Hamiltonian matrix elements through a few illustrative exemplifying cases. Let us concentrate on the elements of the first column of $H$ (first columns of $H_{11}$ and $H_{12}$), which express the optical field transfers supplied by the forward wave of grating 1 $a_{1+}$, to  all waves $a_{i\pm}$ (i=1,2).
\begin{equation}\label{eq:firstcolumn_1}
    \left(\begin{matrix}
		\omega_1+v_1k-i\color{red}(\sqrt{\gamma_1}e^{i\frac{\phi}{2}}+\sqrt{\beta_1\gamma_2}e^{-i\frac{\phi}{2}})e^{i\varphi_{1+}}(\sqrt{\gamma_1}e^{-i\frac{\phi}{2}}+\sqrt{\beta_1\gamma_2}e^{i\frac{\phi}{2}})e^{-i\varphi_{1+}}\\
		 \kappa_1e^{i\phi}+\beta_1\kappa_2e^{-i\phi}-i(\sqrt{\gamma_1}e^{i\frac{\phi}{2}}+\sqrt{\beta_1\gamma_2}e^{-i\frac{\phi}{2}})e^{i\varphi_{1+}}(\sqrt{\gamma_1}e^{i\frac{\phi}{2}}+\sqrt{\beta_1\gamma_2}e^{-i\frac{\phi}{2}})e^{-i\varphi_{1-}}\\
		 \kappa_{d}-ie^{i\psi}(\sqrt{\gamma_1}e^{i\frac{\phi}{2}}+\sqrt{\beta_1\gamma_2}e^{-i\frac{\phi}{2}})e^{i\varphi_{1+}}(\sqrt{\beta_2\gamma_1}e^{-i\frac{\phi}{2}}+\sqrt{\gamma_2}e^{i\frac{\phi}{2}})e^{-i\varphi_{2+}}\\
		 \kappa_{c}-ie^{i\psi}(\sqrt{\gamma_1}e^{i\frac{\phi}{2}}+\sqrt{\beta_1\gamma_2}e^{-i\frac{\phi}{2}})e^{i\varphi_{1+}}(\sqrt{\beta_2\gamma_1}e^{i\frac{\phi}{2}}+\sqrt{\gamma_2}e^{-i\frac{\phi}{2}})e^{-i\varphi_{2-}}
	\end{matrix}\right).
\end{equation}
The matrix element in red expresses the optical field transfer from $a_{1+}$ to $a_{1+}$, mediated by the $1^{st}$ order diffraction processes at the $ \Gamma $ point occurring between the guided wave $a_{1+}$ and the radiation continuum. This transfer takes place in two successive steps, as illustrated in the figure below:

\begin{figure*}[hbt]
\begin{center}
	\includegraphics[width=0.8 \textwidth]{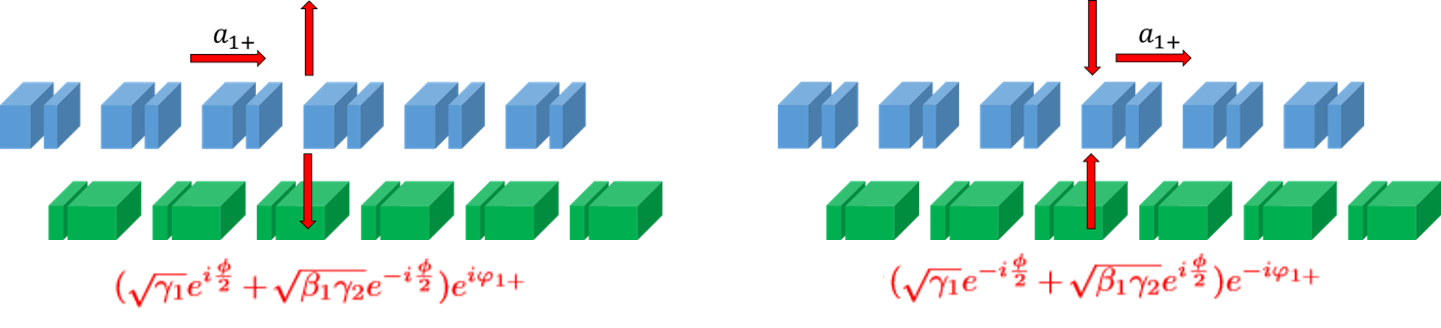}	
\caption{$a_{1+} / a_{1+}$ radiative coupling}
\end{center}
\end{figure*}

Diffraction of $a_{1+}$ is induced by the corrugation of grating 1 (factor $\sqrt{\gamma_1})$ and by the corrugation of grating 2, the action of the latter being limited to the evanescent tail of the guided wave $a_{1+}$ (factor $\gamma_2$, weighted by the parameter $\beta_1$). The coefficients $e^{\pm i\frac{\phi}{2}}$, with $\phi=2\pi\delta$, correspond to the $1^{st}$ order diffraction phase-shift at $\Gamma$ point between the two gratings due to the lateral offset $\delta$. The coefficients $e^{\pm i\varphi_{1+}}$ and $e^{\pm i\varphi_{2+}}$ account for the phase shifts resulting from the diffraction of $a_{1+}$ to (+) or from (-) the radiation continuum, induced by gratings 1 and 2 respectively. If the grating unit cell is laterally symmetric, 
$\varphi_{i+}=\varphi_{i-}$. If the two gratings are identical, $\varphi_{1\pm}=\varphi_{2\pm}$. When the two gratings are far apart, this matrix element is reduced to the parameter $\gamma_1$.
\begin{equation}\label{eq:firstcolumn_2}
    \left(\begin{matrix}
		\omega_1+v_1k-i(\sqrt{\gamma_1}e^{i\frac{\phi}{2}}+\sqrt{\beta_1\gamma_2}e^{-i\frac{\phi}{2}})e^{i\varphi_{1+}}(\sqrt{\gamma_1}e^{-i\frac{\phi}{2}}+\sqrt{\beta_1\gamma_2}e^{i\frac{\phi}{2}})e^{-i\varphi_{1+}}\\
		 \color{red}\kappa_1e^{i\phi}+\beta_1\kappa_2e^{-i\phi}\color{black}-i(\sqrt{\gamma_1}e^{i\frac{\phi}{2}}+\sqrt{\beta_1\gamma_2}e^{-i\frac{\phi}{2}})e^{i\varphi_{1+}}(\sqrt{\gamma_1}e^{i\frac{\phi}{2}}+\sqrt{\beta_1\gamma_2}e^{-i\frac{\phi}{2}})e^{-i\varphi_{1-}}\\
		 \kappa_{d}-ie^{i\psi}(\sqrt{\gamma_1}e^{i\frac{\phi}{2}}+\sqrt{\beta_1\gamma_2}e^{-i\frac{\phi}{2}})e^{i\varphi_{1+}}(\sqrt{\beta_2\gamma_1}e^{-i\frac{\phi}{2}}+\sqrt{\gamma_2}e^{i\frac{\phi}{2}})e^{-i\varphi_{2+}}\\
		 \kappa_{c}-ie^{i\psi}(\sqrt{\gamma_1}e^{i\frac{\phi}{2}}+\sqrt{\beta_1\gamma_2}e^{-i\frac{\phi}{2}})e^{i\varphi_{1+}}(\sqrt{\beta_2\gamma_1}e^{i\frac{\phi}{2}}+\sqrt{\gamma_2}e^{-i\frac{\phi}{2}})e^{-i\varphi_{2-}}
	\end{matrix}\right).
\end{equation}
The matrix element in red accounts for the diffractive coupling between $a_{1+}$ and $a_{1-}$ (optical field transfer from $a_{1+}$ to  $a_{1-}$) at the 2nd Brillouin zone boundary induced by the corrugation of grating 1  (factor $\kappa_1$) and by the corrugation of grating 2 (factor $\kappa_2$, weighted by the parameter $\beta_1$). The coefficients $e^{\pm i\phi}$ correspond to the 2nd-order diffraction phase-shift at the 2nd Brillouin zone boundary between the two gratings due to the lateral offset $\delta$.  
\begin{equation}\label{eq:firstcolumn_3}
    \left(\begin{matrix}
		\omega_1+v_1k-i(\sqrt{\gamma_1}e^{i\frac{\phi}{2}}+\sqrt{\beta_1\gamma_2}e^{-i\frac{\phi}{2}})e^{i\varphi_{1+}}(\sqrt{\gamma_1}e^{-i\frac{\phi}{2}}+\sqrt{\beta_1\gamma_2}e^{i\frac{\phi}{2}})e^{-i\varphi_{1+}}\\
		 \kappa_1e^{i\phi}+\beta_1\kappa_2e^{-i\phi}-i\color{red}(\sqrt{\gamma_1}e^{i\frac{\phi}{2}}+\sqrt{\beta_1\gamma_2}e^{-i\frac{\phi}{2}})e^{i\varphi_{1+}}(\sqrt{\gamma_1}e^{i\frac{\phi}{2}}+\sqrt{\beta_1\gamma_2}e^{-i\frac{\phi}{2}})e^{-i\varphi_{1-}}\\
		 \kappa_{d}-ie^{i\psi}(\sqrt{\gamma_1}e^{i\frac{\phi}{2}}+\sqrt{\beta_1\gamma_2}e^{-i\frac{\phi}{2}})e^{i\varphi_{1+}}(\sqrt{\beta_2\gamma_1}e^{-i\frac{\phi}{2}}+\sqrt{\gamma_2}e^{i\frac{\phi}{2}})e^{-i\varphi_{2+}}\\
		 \kappa_{c}-ie^{i\psi}(\sqrt{\gamma_1}e^{i\frac{\phi}{2}}+\sqrt{\beta_1\gamma_2}e^{-i\frac{\phi}{2}})e^{i\varphi_{1+}}(\sqrt{\beta_2\gamma_1}e^{i\frac{\phi}{2}}+\sqrt{\gamma_2}e^{-i\frac{\phi}{2}})e^{-i\varphi_{2-}}
	\end{matrix}\right).
\end{equation}
The matrix element in red expresses the optical field transfer from $a_{1+}$ to $a_{1-}$, mediated by the $1^{st}$ order diffraction processes at the $\Gamma$ point occurring between the guided waves $a_{1+}$ and $a_{1-}$, via the radiation continuum. This transfer takes place in two successive steps, as illustrated in the figure below. The 1st step is identical to the previous case; for the $2^{nd}$ step, insertion of the radiated light in the backward direction is accounted for in the different phase parameters.

\begin{figure*}[hbt]
\begin{center}
	\includegraphics[width=0.8 \textwidth]{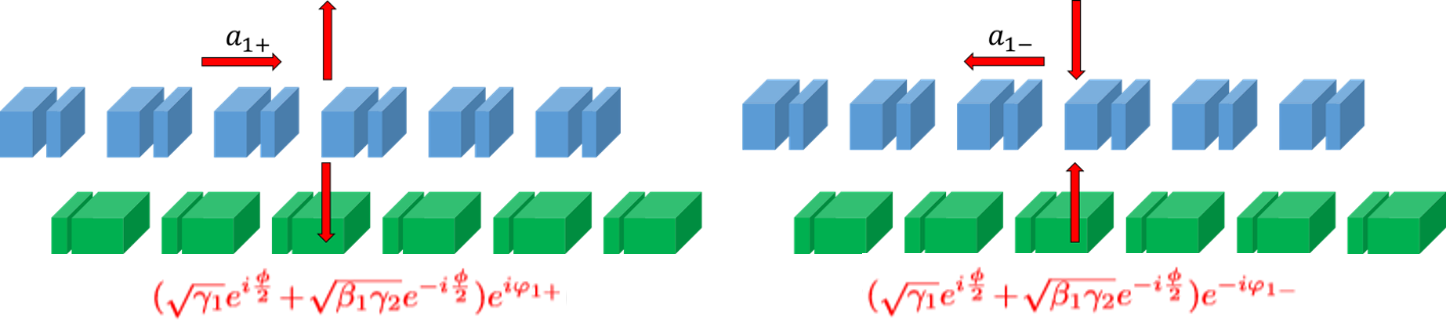}	
\caption{$a_{1+} / a_{1-}$ radiative coupling}
\end{center}
\end{figure*}
\begin{equation}\label{eq:firstcolumn_3_2}
    \left(\begin{matrix}
		\omega_1+v_1k-i(\sqrt{\gamma_1}e^{i\frac{\phi}{2}}+\sqrt{\beta_1\gamma_2}e^{-i\frac{\phi}{2}})e^{i\varphi_{1+}}(\sqrt{\gamma_1}e^{-i\frac{\phi}{2}}+\sqrt{\beta_1\gamma_2}e^{i\frac{\phi}{2}})e^{-i\varphi_{1+}}\\
		 \kappa_1e^{i\phi}+\beta_1\kappa_2e^{-i\phi}-i(\sqrt{\gamma_1}e^{i\frac{\phi}{2}}+\sqrt{\beta_1\gamma_2}e^{-i\frac{\phi}{2}})e^{i\varphi_{1+}}(\sqrt{\gamma_1}e^{i\frac{\phi}{2}}+\sqrt{\beta_1\gamma_2}e^{-i\frac{\phi}{2}})e^{-i\varphi_{1-}}\\
		 \kappa_{d}-\color{red}ie^{i\psi}(\sqrt{\gamma_1}e^{i\frac{\phi}{2}}+\sqrt{\beta_1\gamma_2}e^{-i\frac{\phi}{2}})e^{i\varphi_{1+}}(\sqrt{\beta_2\gamma_1}e^{-i\frac{\phi}{2}}+\sqrt{\gamma_2}e^{i\frac{\phi}{2}})e^{-i\varphi_{2+}}\\
		 \kappa_{c}-ie^{i\psi}(\sqrt{\gamma_1}e^{i\frac{\phi}{2}}+\sqrt{\beta_1\gamma_2}e^{-i\frac{\phi}{2}})e^{i\varphi_{1+}}(\sqrt{\beta_2\gamma_1}e^{i\frac{\phi}{2}}+\sqrt{\gamma_2}e^{-i\frac{\phi}{2}})e^{-i\varphi_{2-}}
	\end{matrix}\right).
\end{equation}
The matrix element in red expresses the optical field transfer from $a_{1+}$ to $a_{2+}$, mediated by the $1^{st}$ order diffraction processes at the $\Gamma$ point occurring between the guided waves $a_{1+}$ and $a_{2+}$, via the radiation continuum. This transfer takes place in three successive steps, as illustrated in the figure below. The $1^{st}$ step is identical to the previous cases; the $2^{nd}$ step consists in the radiative flying of plane waves from grating 1 to grating 2, resulting in the phase shift $\psi$ ; the $3^{rd}$ step corresponds to light insertion in grating 2 in the forward direction.

\begin{figure*}[hbt]
\begin{center}
	\includegraphics[width=0.5 \textwidth]{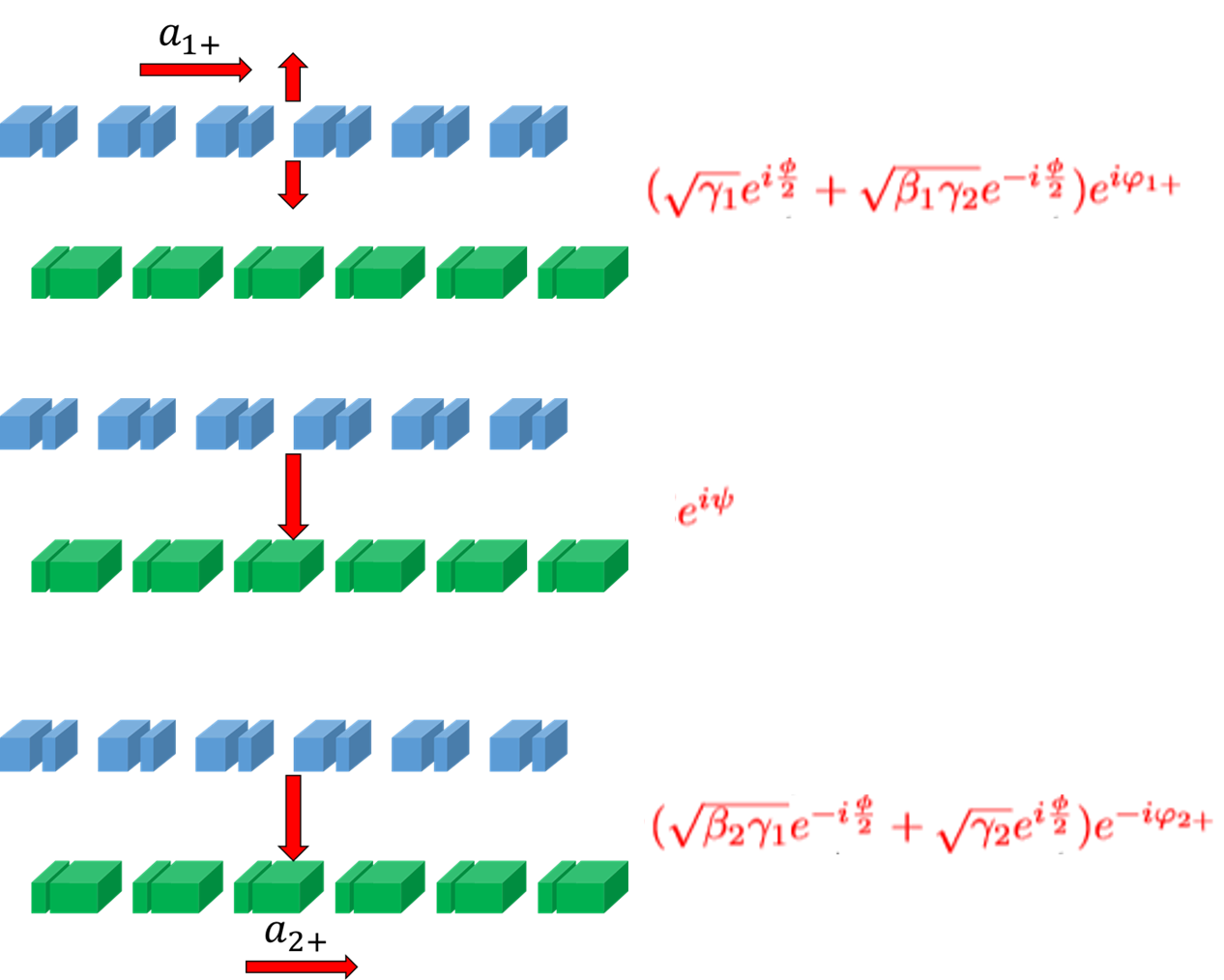}	
\caption{\label{fig:S3} $a_{1+} / a_{2+}$ radiative coupling}
\end{center}
\end{figure*}

\begin{equation}\label{eq:firstcolumn_3_3}
    \left(\begin{matrix}
		\omega_1+v_1k-i(\sqrt{\gamma_1}e^{i\frac{\phi}{2}}+\sqrt{\beta_1\gamma_2}e^{-i\frac{\phi}{2}})e^{i\varphi_{1+}}(\sqrt{\gamma_1}e^{-i\frac{\phi}{2}}+\sqrt{\beta_1\gamma_2}e^{i\frac{\phi}{2}})e^{-i\varphi_{1+}}\\
		 \kappa_1e^{i\phi}+\beta_1\kappa_2e^{-i\phi}-i(\sqrt{\gamma_1}e^{i\frac{\phi}{2}}+\sqrt{\beta_1\gamma_2}e^{-i\frac{\phi}{2}})e^{i\varphi_{1+}}(\sqrt{\gamma_1}e^{i\frac{\phi}{2}}+\sqrt{\beta_1\gamma_2}e^{-i\frac{\phi}{2}})e^{-i\varphi_{1-}}\\
		 \kappa_{d}-ie^{i\psi}(\sqrt{\gamma_1}e^{i\frac{\phi}{2}}+\sqrt{\beta_1\gamma_2}e^{-i\frac{\phi}{2}})e^{i\varphi_{1+}}(\sqrt{\beta_2\gamma_1}e^{-i\frac{\phi}{2}}+\sqrt{\gamma_2}e^{i\frac{\phi}{2}})e^{-i\varphi_{2+}}\\
		 \kappa_{c}-\color{red}ie^{i\psi}(\sqrt{\gamma_1}e^{i\frac{\phi}{2}}+\sqrt{\beta_1\gamma_2}e^{-i\frac{\phi}{2}})e^{i\varphi_{1+}}(\sqrt{\beta_2\gamma_1}e^{i\frac{\phi}{2}}+\sqrt{\gamma_2}e^{-i\frac{\phi}{2}})e^{-i\varphi_{2-}}
	\end{matrix}\right).
\end{equation}
The matrix element in red expresses the optical field transfer from $a_{1+}$ to $a_{2-}$, mediated by the $1^{st}$ order diffraction processes at the $\Gamma$  point occurring between the guided waves $a_{1+}$ and $a_{2-}$, via the radiation continuum. This transfer takes place also in three steps and is similar to the previous case; the only difference concerns the $3^{rd}$ step, where light insertion in grating 2 occurs along in the backward direction, which is accounted for with the various phase parameters.

\section{Fully symmetrical structures}\label{sec:fully_symmetric}
\subsection{Eigenvalues}\label{sec:eigenvalues}
Eigenvalues can be obtained from diagonalization of the Hamiltonian given in section \ref{ssec:H1} and expressed in the base formed by  vectors $(a_{1+},a_{1-},a_{2+},a_{2-})$. The Hamiltonian can be rewritten in a new base formed by even and odd (along the transverse direction), forward and backward  vectors $(a_{e+},a_{e-},a_{o+},a_{o-})$ with $a_{e\pm}=\frac{1}{\sqrt{2}}(a_{1\pm}+a_{2\pm})$ and $a_{o\pm}=\frac{1}{\sqrt{2}}(a_{1\pm}-a_{2\pm})$. Since the eigenmodes are also even and odd (the coupled grating being symmetrical transversally), this writing of equations results in a new Hamitonien which is a priori both more physically readable and mathematically tractable. The new 4×4 Hamiltonien is as below,
\begin{equation}\label{eq:Hevenodd}
      H=
    \left(\begin{matrix}
		\omega_0+vk+\kappa_{d}-iG(1+e^{i\psi}) & K+\kappa_{c}-iG(1+e^{i\psi}) & 0 & 0\\
		K+\kappa_{c}-iG(1+e^{i\psi}) & \omega_0-vk+\kappa_{d}-iG(1+e^{i\psi}) & 0 & 0\\ 
		0 & 0 & \omega_0+vk-\kappa_{d}-iG(1-e^{i\psi}) & 
		K-\kappa_{c}-iG(1-e^{i\psi})\\
		0 & 0 & K-\kappa_{c}-iG(1-e^{i\psi}) & \omega_0-vk-\kappa_{d}-iG(1-e^{i\psi})
	\end{matrix}\right),
\end{equation}
and can be rewritten as :
\begin{equation}\label{eq:Hevenoddcompact}
      H=
    \left(\begin{matrix}
    H_e & 0 \\ 0 & H_o
    \end{matrix}\right),
\end{equation}
where $H_e$, $H_o$ are 2×2 Hamiltonians whose diagonalization provides, separately, the couples of even and odd eigenvalues. It can be easily shown that for both couples of even and odd eigenvalues, one eigenvalue is systematically dark (called lateral BIC in the main text) at the $\Gamma$ point ($k=0$), while the other is bright in general except when $\psi=\pi(0) mod2\pi$, for the even (odd) mode (in the transverse direction). The systematic observation of a lateral BIC at the $\Gamma$ point features structures with lateral symmetry as commented in the main text. 

\subsection{Expansion of complex eigenvalues in the vicinity of the $\Gamma$ point}\label{sec:expansion_complex}
Results of the expansion of complex eigenvalues of even modes (Eq. \eqref{eq:omega} with $\eta=1$) are given below:

\begin{itemize}
    \item Bright branch of the dispersion characteristics:
    \begin{equation}\label{eq:S21}
    \omega=\omega_{R}+\kappa_{d}+K+\kappa_c+2G \sin\psi_0-2iG(1+\cos\psi_0)+C\frac{k^2}{2}.
\end{equation}
$C=C_R+iC_I$ is the complex curvature (second derivative versus $k$) of the complex eigenvalue.
\begin{equation}\label{eq:curv_R}
   C_R=\frac{d^2\omega_R}{dk^2}=\frac{\frac{v^2}{|K+\kappa_c-U|^2}(K+\kappa_c+G \sin\psi_0)-\cos\psi_0\frac{\tau_{rad}}{\tau_{wg}}C_{rad}}{1-\cos\psi_0\frac{\tau_{rad}}{\tau_{wg}}},
\end{equation}
\begin{equation}\label{eq:curv_I}
   C_I=\frac{d^2\omega_I}{dk^2}=\frac{v^2}{|K+\kappa_c-U|^2}G(1+\cos\psi_0)-\sin\psi_0\frac{\tau_{rad}}{\tau_{wg}}\left(C_{rad}-\frac{d^2\omega_R}{dk^2}\right),
\end{equation}
with $U=iG(1+e^{i\psi_0})$ , $\psi_0=\psi(k=0)$ and where  $\tau_{wg}=\frac{1}{2G}$ and $\tau_{rad}=\frac{L_{opt}}{c}$, are respectively the average lifetime of photons in the wave-guided state before being emitted into the continuum, and in the radiated state during a one way trip between the two gratings. $C_{rad}=\frac{(\frac{c}{n})^2}{\omega_{R0}}$ , is named the radiated curvature: it is identical to the curvature of the dispersion characteristic of a Fabry Perot cavity, with n intra-cavity optical index and with perfect metallic reflectors, at the $\Gamma$ point around the resonance frequency $\omega_{R0}$. 
    \item Dark branch (with lateral BIC at the $\Gamma$ point) of the dispersion characteristics:
    \begin{equation}\label{eq:S24}
   \omega=\omega_R+i\omega_i\cong\omega_0+\kappa_{d}-K-\kappa_c+C\frac{k^2}{2},
\end{equation}
\begin{equation}\label{eq:S25}
   C_R=\frac{d^2\omega_R}{dk^2}=\frac{-v^2}{|K+\kappa_c-U|^2}(K+\kappa_c+G\sin\psi_0),
\end{equation}
\begin{equation}\label{eq:S26}
   C_I=\frac{d^2\omega_I}{dk^2}=\frac{-v^2}{|K+\kappa_c-U|^2}G(1+\cos\psi_0).
\end{equation}
\end{itemize}
As explained in section \ref{sssec:BandInversion} and expressed by Eqs.~\eqref{eq:S21}, \eqref{eq:S24}  and \eqref{eq:S25}, the real eigenvalues of the bright and the dark branches are degenerated at the $\Gamma$ point when $K+\kappa_c+G \sin\psi_0=0$, and the real dark branch is flat (zero real curvature).

\subsection{Imaginary dispersion characteristic around the wave vector $k_{BIC}$ of a simple transverse BIC; BIC merging processes: double transverse BIC from merging of 2 simple transverse BICs and triple BIC from merging of two simple transverse BICs and a lateral BIC}\label{sec:expansion_imaginary}

Expression of the imaginary dispersion characteristic around the wave-vector $k=k_{BIC}$ can be derived from expansion of Eqs.~\eqref{eq:omega} and \eqref{eq:psi}. It is given by the relation below, for the even mode:
\begin{equation}\label{eq:S11}
   \omega_I=-\frac{G}{2}\left(1\pm\frac{K+\kappa_c}{\sqrt{v^2k_{BIC}^2+(K+\kappa_c)^2}}\right)\left[\frac{\frac{L_{opt}^2}{(2p+1)\pi}\frac{k_{BIC}}{n^2}\left(1\pm\frac{\omega_{RBIC} v^2}{\frac{c^2}{n^2}\sqrt{v^2k_{BIC}^2+(K+\kappa_c)^2}}\right)}{1+\frac{L_{opt}^2}{(2p+1)\pi}\frac{\omega_{RBIC}}{c^2}G\left(1\pm\frac{K+\kappa_c}{\sqrt{v^2k_{BIC}^2+(K+\kappa_c)^2}}\right)}\right]^2\left(k-k_{BIC}\right)^2,
\end{equation}
with:
\begin{equation}\label{eq:S1_1}
  L_{opt}\sqrt{\left(\frac{\omega_{RBIC}}{c}\right)^2-\left(\frac{k_{BIC}}{n}\right)^2}=(2p+1)\pi.
\end{equation}
The sign + (-) corresponds to the air (dielectric) branch. If $K$>0 the air (dielectric) branch is bright (dark) at the $\Gamma$ point, and the other way around if $K$<0. 
The second derivative, or the curvature, of the imaginary dispersion characteristic is written as:
\begin{equation}\label{eq:S111}
\frac{d^2\omega_I}{dk^2}|_{(k=k_{BIC})}\cong-\frac{G}{2}\left(1\pm\frac{K+\kappa_c}{\sqrt{v^2k_{BIC}^2+(K+\kappa_c)^2}}\right)\left[\frac{\frac{L_{opt}^2}{(2p+1)\pi}\frac{k_{BIC}}{n^2}\left(1\pm\frac{\omega_{RBIC} v^2}{\frac{c^2}{n^2}\sqrt{v^2k_{BIC}^2+(K+\kappa_c)^2}}\right)}{1+\frac{L_{opt}^2}{(2p+1)\pi}\frac{\omega_{RBIC}}{c^2}G\left(1\pm\frac{K+\kappa_c}{\sqrt{v^2k_{BIC}^2+(K+\kappa_c)^2}}\right)}\right]^2.
\end{equation}
When the transverse BIC approaches the $\Gamma$ point, $k_{BIC}$ tends to zero and $\omega_{RBIC}$ tends to $\omega_{R0}$  , with $\frac{\omega_{R0}}{c}L_{opt}=(2p+1)\pi$ ; the curvature is then simply written as below (for example when the bright branch is air like, that is for $K$>0):

\begin{itemize}
    \item For the bright branch:
 \begin{equation}\label{eq:B12}
\frac{d^2\omega_I}{dk^2}\bigg|_{k=k_{BIC}}\cong-2G\left[\frac{\frac{L_{opt}}{n^2\frac{\omega_{R0}}{c}}\left(1+\frac{\omega_{R0}v^2}{\frac{c^2}{n^2}}(K+\kappa_c)\right)}{1+\frac{2L_{opt}}{c}G}\right]^2k_{BIC}^2.
\end{equation}  
The curvature varies like $k_{BIC}^2$  : this behaviour is a manifestation of the merging of two ordinary transverse BICs (belonging to the bright branch and occurring at $\pm|k_{BIC}|$ vectors), when $k_{BIC}$ approaches 0. We call this particular transverse BIC occurring at the $\Gamma$ point, double transverse BIC, which exhibits a flat imaginary dispersion characteristic. As a matter of fact, the $k_{BIC}^2$ dependence of the curvature expresses that the imaginary dispersion characteristic is like $k^4$ around the $\Gamma$ point.
\item For the dark branch:
\begin{equation}\label{eq:B13}
\frac{d^2\omega_I}{dk^2}\bigg|_{k=k_{BIC}}\cong-\frac{G}{2}\left[\frac{L_{opt}v}{n^2\frac{\omega_{R0}}{c}(K+\kappa_c)}\left(1-\frac{\omega_{R0}v^2}{\frac{c^2}{n^2}}(K+\kappa_c)\right)\right]^2k_{BIC}^4.
\end{equation}
\end{itemize}
The curvature varies like $k_{BIC}^4$ : this behaviour is a manifestation of the merging of two ordinary transverse BICs (belonging to the dark branch and occurring at $\pm|k_{BIC}|$ vectors) with the lateral BIC (systematically occurring at the $\Gamma$ point of a dark branch), when $k_{BIC}$ approaches 0. We call this particular transverse BIC, triple transverse BIC, which exhibits a very flat imaginary dispersion characteristic. As a matter of fact, the $k_{BIC}^4$ dependence of the curvature expresses that the imaginary dispersion characteristic is like $k^6$ around the $\Gamma$ point.

\subsection{Double Exceptional point}\label{sec:Double_EP}
A double exceptional point is formed when conditions for full degeneracy of the complex eigenvalues of the dark and bright branches is achieved  \cite{Zhen2015} (see section \ref{sssec:BandInversion}). 
In the case of the even branches, Eq.~\eqref{eq:omega} leads to the following eigenvalues:
\begin{equation}\label{eq:eigen-EP}
    \omega=\omega_{R}+i\omega_{i} = \omega_0+\kappa_{d}-iG(1+e^{i\psi})\pm\sqrt{v^2k^2+\left(K+\kappa_{c}-iG(1+e^{i\psi})\right)^2}.
\end{equation}
Degeneracy of the complex eigenvalues occurs when :
\begin{equation}\label{eq:eigen-EP2}
  v^2k_{EP}^2+\left(K+\kappa_{c}-iG(1+e^{i\psi(k_{EP})})\right)^2=0,
\end{equation}
where $k=k_{EP}$ is the wave-vector at the exceptional points.
It results:
\begin{equation}
  k_{EP}=\pm\frac{G\left(1+\cos{\psi(k_{EP})}\right)}{v},
  \label{eq:kexc}
\end{equation}
with the condition:
\begin{equation}
 K+\kappa_{c}+G\sin{\psi(k_{EP})}=0 ,
 \label{eq:psiexc}
\end{equation}
being to be held in addition.
Note that the previous condition can be met provided that $0\leq|K+\kappa_{c}|\leq G$. This can be adjusted, in practice, by an appropriate setting of the filling factor of the grating structure.
$\omega_R(k_{EP})$, $\omega_I(k_{EP})$ and $k_{EP}$ can be derived from Eqs.~\eqref{eq:omega}, \eqref{eq:kexc} and \eqref{eq:psiexc}.

Finally, it is reminded (see Eq.~\eqref{eq:psi}): 
\begin{equation}
    \psi(k_{EP})=\sqrt{\left[\frac{\omega(k_{EP})}{c}\right]^2 - \left(\frac{k_{EP}}{n}\right)^2}L_{opt},
    \label{S2_10}
\end{equation}
which is an additional condition for the double exceptional to be achieved. This can be adjusted in practice by an appropriate setting of $L_{opt}$, that is of the thickness of the grating structure.
Interestingly, for a given structure showing a double exceptional point, the real eigenvalues of the dark branch at the $\Gamma$ point (lateral BIC) and at the exceptional points coincide. Indeed, for the lateral BIC:
\begin{equation}\label{eq:S2_11}
    \omega_{R}(k=0) = \omega_0+\kappa_{d}-K-\kappa_c=\omega_0+\kappa_{d}+G\sin{\psi(k_{EP})}=\omega_R(k_{EP}).
\end{equation}
Finally, for wave-vector $k$ exceedind $k_{EP}$, the real dispersion characteristics of the dark and bright branches are linear $(\pm vk)$ and both branches equally share the loss rate, while, at the $\Gamma$ point, the bright branch takes the full part of the losses.

\subsection{Double transverse BIC with flat real dispersion characteristic}\label{sec:doubleBIC}
It is reminded that the second derivative $\frac{d^2\omega_R}{dk^2}$  or curvature $C_R$ of the real dispersion characteristic at the $\Gamma$ point can be written as (see Eq.~\eqref{eq:real_curvature_doubleBIC}):  
\begin{equation}\label{eq:real_curvature_doubleBIC_App}
    C_{R}=\frac{\frac{1}{2G}\frac{v^2}{K+\kappa_c}+\frac{L_{opt}}{c}\frac{\left(\frac{c}{n}\right)^2}{\omega_{R0}}}{\frac{1}{2G}+\frac{L_{opt}}{c}}=\frac{\tau_{wg}\times C_{wg} + \tau_{rad}\times C_{rad}} {\tau_{wg} + \tau_{rad}},
\end{equation}
where $C_{wg}=\frac{v^2}{K+\kappa_c}$ and $C_{rad}=\frac{\left(\frac{c}{n}\right)^2}{\omega_{R0}}$ are the guided and radiated curvatures respectively.
The radiated curvature is strictly positive, while the guided curvature may be positive or negative, as the parameter $K+\kappa_c$. These two cases have been illustrated in figure \ref{fig:band_inversion}, where the bright branch was shown to be either air like (positive curvature) or dielectric like (negative curvature), depending on the sign of $K+\kappa_c+G\sin{\psi_0}$, where $\psi_0=\psi(k=0)$. In the current discussion, the bright branch turns to be a transverse BIC at the $\Gamma$ point, therefore $\psi_0=\pi$ and $K+\kappa_c+G\sin{\psi_0}\equiv K+\kappa_c$. For $K+\kappa_c<0$, or $C_{wg}<0$, it is possible in principle to meet the condition $C_R=0$. Given that $v\cong\frac{c}{n}$, and that $G$ and $|K+\kappa_c|$ are in practice lower than $\omega_{R0}=\frac{2\pi c}{\lambda}$, the condition for a zero real curvature implies that $L_{opt}$ is significantly larger than $\lambda$. In other words, the radiated curvature $C_{rad}$ is rather small as compared to the guided curvature $C_{wg}$, and the former has to be over-weighted by a large $\tau_rad$ (that is large $L_{opt}$) with respect to the later in order to result in a total curvature $C_R=0$.

\subsection{Design rules of a Dirac point at a triple BIC}\label{sec:Rule_triple_BIC}
Two conditions have to be met simultaneously for the formation of a Dirac point at a triple BIC: first, the condition for degeneracy at the $\Gamma$ point of the frequency of the lateral BIC with the real part of the bright eigenfrequency at the $\Gamma$ point, which occurs when $K+\kappa_c+G\sin{\psi_0}=0$ ; second, the condition for the bright mode to turn into a double transverse BIC, when $\psi_0=\pi (mod2\pi)$  (see section \ref{sssec:DoubleTransverseBIC}). For a given index contrast between the gratings and the surrounding medium, the high index material filling factor (FF) and the effective optical thickness of the coupled grating structure are the two available “joysticks” to be handled simultaneously to achieve those two conditions. The degeneracy condition, which corresponds to the cancellation of the overall diffractive coupling processes between wave-guided resonances, can be viewed as the physical counterpart of the full transmission condition which occurs in a half wavelength Bragg mirror stack; this condition is met when the optical thicknesses of the high / low index layers are an integer of $\frac{\lambda}{2}$. For example this is simply achieved for a high index $FF=\frac{n_{low}}{n_{low}+n_{high}}$. For the wave-guided coupled grating structure, this condition is met for a different value of FF; for example, $FF<\frac{n_{low}}{n_{low}+n_{high}}$  when the two gratings are in contact. This is due to the fact that the phase change occurring at the reflection / transmission of the guided wave impinging the high / low index interface of the grating is different from 0 or $\pi$ $(mod2\pi)$, unlike the case of plane waves in a Bragg mirror. For a practical design, it is advisable to start with the value of FF which applies to a Bragg mirror stack (for example 0.25, if $n_{high}=3$ and $n_{low}=1$) and then adjust it, to compensate for the phase change difference and therefore, approach the condition requested for degeneracy between the lateral BIC and the bright mode frequency. Then, in order to convert the bright mode into a transverse double BIC, the effective optical thickness of the coupled grating structure must be adjusted: this can be achieved by adjusting the thickness $H$ of each grating and / or the distance D between them. A few iterations of those successive adjustments of FF and $h$, $D$ may be necessary to get close to the formation of the elusive perfect Dirac point at a triple BIC.

\section{Eigenvalues in structures with broken lateral symmetry}\label{sec:broken_lateral_sym}
Eigenvalues can be obtained from diagonalization of the Hamiltonian given in \ref{ssec:NonhermitianH} and expressed in the base formed by base vectors $(a_{1+},a_{1-},a_{2+},a_{2-})$. The Hamiltonian can rewritten in a new base formed by even and odd (along the transverse direction), forward and backward base vectors $(a_{e+},a_{e-},a_{o+},a_{o-})$, in the same way as for fully symmetrical structures (see section \ref{sec:eigenvalues}), given that the symmetry along the transverse direction is maintained. The new 4×4 Hamiltonien is as below:
\begin{equation}\label{eq:Hevenodd2}
      H=
    \left(\begin{matrix}
		\omega_0+vk+\kappa_{d}-iG(1+e^{i\psi}) & K+\kappa_{c}-iG(1+e^{i\psi})e^{-i\varphi} & 0 & 0\\
		K+\kappa_{c}-iG(1+e^{i\psi})e^{i\varphi} & \omega_0-vk-iG(1+e^{i\psi}) & 0 & 0\\ 
		0 & 0 & \omega_0+vk-\kappa_{d}-iG(1-e^{i\psi}) & 
		K-\kappa_{c}-iG(1-e^{i\psi})e^{-i\varphi}\\
		0 & 0 & K-\kappa_{c}-iG(1-e^{i\psi})e^{i\varphi} & \omega_0-vk-\kappa_{d}-iG(1-e^{i\psi})
	\end{matrix}\right),
\end{equation}
and can be rewritten as :
\begin{equation}\label{eq:Hevenoddcompact2}
      H=
    \left(\begin{matrix}
    H_e & 0 \\ 0 & H_o
    \end{matrix}\right),
\end{equation}
where $H_e$, $H_o$ are 2×2 Hamiltonians whose diagonalization provides, separately, the couples of even and odd eigenvalues. The two $H_e$, $H_o$ are fully bright, and do not result in any dark branch showing a lateral BIC at the $\Gamma$ point, unlike the case of fully symmetrical structures. This is a signature of the broken lateral symmetry, which expresses it-self by the phase $\varphi\neq0$ (mod$\pi$).

\section{Quasi transverse BICs at the $\Gamma$ point in transverse symmetry broken structures: the case of negligible near-field coupling between the gratings}\label{sec:quasi_transverse_BIC}

In section \ref{ssec:H3}, we show that, at $\Gamma$ point, 2 dark and 2 bright modes are obtained. We conclude in section \ref{ssec:SpecificPropertiesEig3} that transverse BIC cannot be formed at the $\Gamma$ point in transverse symmetry broken structures as a result of the finite phase $\varphi$. However, if the near field evanescent coupling is considered as negligible when the two gratings are far apart, then, in Eq.~\eqref{eq:H3}, $K_+\cong0$ and the 2x2 hamiltonian for the bright modes at the $\Gamma$ point is given by:

\begin{equation}\label{eq:H3-app}
    H=
    \left(\begin{matrix}
		\omega_1+K_{12}-iG_{12} & -i\sqrt{G_{12} G_{21}} e^{i(\psi-\varphi)} \\ 
		-i\sqrt{G_{12} G_{21}} e^{i(\psi+\varphi)}  & \omega_2+K_{21}-iG_{21} 
	\end{matrix}\right).
\end{equation}

It clearly results that the impact of the phase $\varphi$ is also negligible when it comes to diagonalize the 2×2 Hamiltonian of the bright modes. Therefore, the two corresponding eigenvalues can be made real when the phase $\psi$ is set at 0 (mod $\pi$). This is indeed possible provided that $\omega_1=\omega_2$. Then the 2 bright modes of the Hamiltonian turn into 2 quasi-BICS, but are not true BICS.

\section{"Fishbone" structure} \label{sec:fishbone}
\subsection{Aligned and half-period misalignment structures}\label{sec:Fishbone1}

The Hamiltonien given in section \ref{ssec:H4}, is rewritten along the more compact version below:
\begin{equation}\label{eq:E1}
    H=
    \left(\begin{matrix}
		A+vk & B_2 & C_2 & D\\
		B_1 & A-vk & D & C_1\\
		C_1 & D & A+vk & B_1\\
		D & C_2 & B_2 & A-vk
	\end{matrix}\right).
\end{equation}

Note that for aligned grating structures ($\phi=0$ (mod$2\pi$)) or with half period lateral off-set ($\phi=\pi$ (mod$2\pi$)), $B_1=B_2$ and $C_1=C_2$, which means that the compact expressions of the Hamiltonian are formally identical. Therefore for this two limit cases, the properties of the eigenvalues are also formally identical.
For half-period lateral off-set, we obtain the two couple of complex eigenvalues below:

\emph{Fundamental modes:}
\begin{equation}\label{eq:E2}
    \omega=\omega_{R}+i\omega_{i} = \omega_0+\kappa_{d}-iG(1-e^{i\psi})\pm\sqrt{v^2k^2+\left(K+\kappa_{c}+i|G|(1-e^{i\psi})\right)^2},
\end{equation}

\emph{Excited modes:}
\begin{equation}\label{eq:E3}
    \omega=\omega_{R}+i\omega_{i} = \omega_0-\kappa_{d}-iG(1+e^{i\psi})\pm\sqrt{v^2k^2+\left(K-\kappa_{c}+i|G|(1+e^{i\psi})\right)^2},
\end{equation}
with: $ K=- \kappa (1+\beta)$, $G=\gamma(1-\sqrt{\beta})^2$.

\subsection{Structures with arbitrary misalignment: expansion of complex eigenvalues in the vicinity of the $\Gamma$ point}\label{sec:FB2}
In general, complex eigenvalues $\omega$ are obtained from the diagonalization of the Hamiltonian given by Eq.~\ref{eq:H5}. The corresponding secular equation can be written in the compact form:
\begin{equation}\label{eq:S53}
    |H-\omega \mathbb{1}_{4} |=\left|
    \left(\begin{matrix}
		H_{B+} & H_V\\
		H_V & H_{B-}
	\end{matrix}\right)-\omega \mathbb{1}_{4}\right|,
\end{equation}
where $H_{B\pm}$ and $H_V$ are 2x2 matrices.
At the $\Gamma$ point, where $H_V=0$, the eigenvalues are the solution of 2 independent equations:

\begin{equation}\label{eq:E5}
   |H_{B_\pm}-\omega \mathbb{1}_{2}|=0.
\end{equation}

Complex in general, the eigenvalues become real if $\psi=0$ or $\pi$ and are the signatures of transverse BICs (see section \ref{ssec:SpecificPropertiesEig4}).
From expansion of Eq.~\eqref{eq:S53}, it is possible to derive an analytical expression of the curvature $C_R$ of the dispersion characteristics at the $\Gamma$ point. Considering a transverse BIC at the $\Gamma$ point in the vicinity of the eigenvalue numbered "3", corresponding with Eq. \eqref{eq:omega3} (section \ref{ssec:SpecificPropertiesEig4}), which is appropriate for the production of flat BIC, we get:

\begin{equation}
\label{eq:E6}
    C_R=\frac{\frac{1}{(1+\alpha)|G|}\frac{v^2}{K_{eff}}+\frac{L_{opt}}{c}\frac{\left(\frac{c}{n}\right)^2}{\omega_{R0}}}{\frac{1}{(1+\alpha)|G|}+\frac{L_{opt}}{c}},
\end{equation}
where:
\begin{equation}
\label{eq:E7}
   \alpha=\frac{\Re(\frac{\kappa_d+K}{G})}{\frac{\kappa_d+K}{G}},
\end{equation}

\begin{equation}
\label{eq:E8}
   \frac{1}{K_{eff}}=2\Re \left(  \frac{(|\kappa_d+K|-2i|G|-2\kappa_c)|\kappa_d+K|+\kappa_d^2-2i\kappa_d\Re (G)-|K|^2-2i\Re (KG^*)}{|\kappa_d+K|(\kappa_d-K-2iG)(\kappa_d-K^*-2iG^*)-(|\kappa_d+K|-2\kappa_c-2i|G|^2)}    \right).
\end{equation}

\end{document}